\documentclass[twocolumn]{aastex62}

\usepackage{multirow}
\usepackage{amsmath}

\usepackage{lmodern}
\usepackage[T1]{fontenc}
\usepackage{textcomp}
\usepackage{rotating}
\usepackage{multirow,bigdelim}

\newcommand{\GG}[1]{}

\graphicspath{{./}{./}}

\received{\today}
\submitjournal{ApJS}

\shorttitle{CME--CME Interactions as Sources of CME Geo-effectiveness}
\shortauthors{Scolini et al.}
\begin{document}

\title{CME--CME Interactions as Sources of CME Geo-effectiveness: \\
The Formation of the Complex Ejecta and Intense Geomagnetic Storm in Early September 2017}

\correspondingauthor{Camilla Scolini}
\email{camilla.scolini@kuleuven.be}

\author[0000-0002-5681-0526]{Camilla Scolini}
\affil{Centre for mathematical Plasma Astrophysics, Dept.\ of Mathematics, KU Leuven, 3001 Leuven, Belgium}
\affil{Solar--Terrestrial Centre of Excellence --- SIDC, Royal Observatory of Belgium, 1180 Brussels, Belgium}

\author[0000-0001-5953-0370]{Emmanuel Chan\'{e}}
\affil{Centre for mathematical Plasma Astrophysics, Dept.\ of Mathematics, KU Leuven, 3001 Leuven, Belgium}

\author[0000-0003-4867-7558]{Manuela Temmer}
\affil{Insitute of Physics, University of Graz, 8010 Graz, Austria}

\author[0000-0002-4489-8073]{Emilia K. J. Kilpua}
\affil{Department of Physics, University of Helsinki, FI-00014 Helsinki, Finland}

\author[0000-0001-5661-9759]{Karin Dissauer}
\affil{Insitute of Physics, University of Graz, 8010 Graz, Austria}

\author[0000-0003-2073-002X]{Astrid M. Veronig}
\affil{Insitute of Physics, University of Graz, 8010 Graz, Austria}
\affil{Kanzelh{\"o}he Observatory for Solar and Environmental Research, University of Graz, 9521 Treffen, Austria}

\author[0000-0001-6590-3479]{Erika Palmerio}
\affil{Department of Physics, University of Helsinki, FI-00014 Helsinki, Finland}

\author[0000-0003-1175-7124]{Jens Pomoell}
\affil{Department of Physics, University of Helsinki, FI-00014 Helsinki, Finland}

\author[0000-0002-8680-8267]{Mateja Dumbovi\'{c}}
\affil{Hvar Observatory, Faculty of Geodesy, University of Zagreb, 10000 Zagreb, Croatia}

\author[0000-0002-8707-076X]{Jingnan Guo}
\affil{University of Science and Technology of China, 230026 Hefei, China}

\author[0000-0002-6097-374X]{Luciano Rodriguez}
\affil{Solar--Terrestrial Centre of Excellence --- SIDC, Royal Observatory of Belgium, 1180 Brussels, Belgium}

\author[0000-0002-1743-0651]{Stefaan Poedts}
\affil{Centre for mathematical Plasma Astrophysics, Dept.\ of Mathematics, KU Leuven, 3001 Leuven, Belgium}
\affil{Institute of Physics, University of Maria Curie-Sk{\l}odowska, PL-20-031 Lublin, Poland}

%% Note that the \and command from previous versions of AASTeX is now
%% depreciated in this version as it is no longer necessary. AASTeX 
%% automatically takes care of all commas and "and"s between authors names.

%% AASTeX 6.2 has the new \collaboration and \nocollaboration commands to
%% provide the collaboration status of a group of authors. These commands 
%% can be used either before or after the list of corresponding authors. The
%% argument for \collaboration is the collaboration identifier. Authors are
%% encouraged to surround collaboration identifiers with ()s. The 
%% \nocollaboration command takes no argument and exists to indicate that
%% the nearby authors are not part of surrounding collaborations.

%
%______________________________________________________________
% ABSTRACT

%% Mark off the abstract in the ``abstract'' environment. 
\begin{abstract}
% context
Coronal mass ejections (CMEs) are the primary sources of intense disturbances at Earth, where their \textit{geo-}effectiveness is largely determined by their dynamic pressure and internal magnetic field, which can be significantly altered during interactions with other CMEs in interplanetary space.
We analyse three successive CMEs that erupted from the Sun during September~4--6, 2017, investigating the role of CME--CME interactions as source of the associated intense geomagnetic storm ($\mathrm{Dst}_\mathrm{min} = -142$~nT on September~7).
% aims
To quantify the impact of interactions on the (geo-)effectiveness of individual CMEs, we perform global heliospheric simulations with the EUHFORIA model, using observation-based initial parameters with the additional purpose of validating the predictive capabilities of the model for complex CME events.
% results
The simulations show that around 0.45~AU, the shock driven by the September~6 CME started compressing a preceding magnetic ejecta formed by the merging of two CMEs launched on September~4, significantly amplifying its $B_z$ until a maximum factor of $2.8$ around 0.9~AU. The following gradual conversion of magnetic energy into kinetic and thermal components reduced the $B_z$ amplification until its almost complete disappearance around 1.8~AU.
% conclusions
We conclude that a key factor at the origin of the intense storm triggered by the September~4--6, 2017 CMEs was their arrival at Earth during the phase of maximum $B_z$ amplification. Our analysis highlights how the amplification of the magnetic field of individual CMEs in space--time due to interaction processes can be characterised by a growth, a maximum, and a decay phase, suggesting that the time interval between the CME eruptions and their relative speeds are critical factors in determining the resulting impact of complex CMEs at various heliocentric distances (\textit{helio-}effectiveness).
\end{abstract}

%% Keywords should appear after the \end{abstract} command. 
%% See the online documentation for the full list of available subject
%% keywords and the rules for their use.

%% From the front matter, we move on to the body of the paper.
%% Sections are demarcated by \section and \subsection, respectively.
%% Observe the use of the LaTeX \label
%% command after the \subsection to give a symbolic KEY to the
%% subsection for cross-referencing in a \ref command.
%% You can use La=TeX's \ref and \label commands to keep track of
%% cross-references to sections, equations, tables, and figures.
%% That way, if you change the order of any elements, LaTeX will
%% automatically renumber them.
%%
%% We recommend that authors also use the natbib \citep
%% and \citet commands to identify citations.  The citations are
%% tied to the reference list via symbolic KEYs. The KEY corresponds
%% to the KEY in the \bibitem in the reference list below. 

%
%______________________________________________________________
% MAIN BODY

\section{Introduction} \label{sec:intro}

Coronal mass ejections (CMEs) are huge eruptions of plasma and magnetic fields from the Sun that propagate through the heliosphere and can eventually impact Earth and other planets and spacecraft. 
Considered to be the major drivers of strong space weather disturbances at Earth \citep{gosling:1991, gosling:1993, huttunen:2005, koskinen:2006, richardson:2012,kilpua:2017a}, CMEs and their related interplanetary structures \citep[i.e.\ CME-driven shocks, sheaths, and magnetic ejecta, e.g.][]{kilpua:2017b} have been found responsible for up to 90\% of all intense (Dst $<-100$~nT) geomagnetic storms \citep{zhang:2007}. 
These intense storms are primarily caused by the combination of long-lasting (typically over 3 hours), strongly southward (negative $B_z$) interplanetary magnetic fields and high dynamic pressure within magnetic ejecta \citep[see e.g.][]{tsurutani:1988, farrugia:1993}.

While the majority of these storms are driven by single CMEs (about 60\%), a significant fraction (about 27\%) is found to be caused by the passage of complex signatures generated by the interaction of individual CMEs with other transients, such as other CMEs and stream interaction regions (SIRs) \citep{zhang:2007,vennerstrom:2016}. 
While several studies established that CME--CME interactions are likely to increase the impact on Earth (\textit{geo-}effectiveness) of individual CMEs \citep[see][and references therein]{lugaz:2017}, the actual quantification of this amplification has been rarely investigated \citep[see e.g.][]{xiong:2007, xiong:2009, shen:2018}.

% the September 2017 events
Although the probability of CME--CME interactions in the corona and interplanetary space is higher during periods of maximum solar activity, when the CME occurrence can exceed the rate of $10$~CMEs/day \citep{yashiro:2004, robbrecht:2009}, intense geomagnetic storms caused by CME--CME interactions during activity minima might also occur, most likely in association with sympathetic and (quasi-)homologous CMEs that erupt from the same active region (AR) \citep{lugaz:2007, lugaz:2017}.
At the time of this study, the most recent intense geomagnetic storm related to interacting CMEs occurred in early September 2017, due to the intense negative $B_z$ generated as a consequence of the propagation of a CME-driven interplanetary shock through a preceding magnetic ejecta. This event was also associated with two of the four most intense X-class solar flares observed in Solar Cycle 24, originating from an AR (NOAA~AR~12763) that presented outstanding levels of CME and flare productivity \citep{chertok:2018, redmon:2018, bruno:2019}.
\cite{shen:2018} investigated the impact of these CMEs on Earth using remote-sensing and in-situ observations, and estimated an amplification of the geo-effectiveness of the individual CMEs by a factor of ${\sim}2$ due to CME--CME interactions close to 1~AU. The evolution of this amplification in space and time as the CMEs propagated from Sun to Earth, as well as its physical origin, remains unclear.

% helio-effectiveness
So far, very few studies have attempted to quantify the geo-effectiveness amplification (in terms of $B_z$ and/or other geomagnetic activity indices) by performing global Sun-to-Earth simulations of real CME events.
At the same time, the geo-effectiveness amplification at Earth can be expected to be the result of a gradual amplification developing in space--time as the CMEs involved propagate from Sun to Earth, as consequence of the various interaction phases. 
To the best of our knowledge, no previous study investigating the evolution of this amplification in space--time was ever performed. 
In order to address this previously uninvestigated aspect of CME--CME interactions, in this work we therefore introduce a new terminology to refer to the amplification of the potential impact of a given CME at a generic location in the heliosphere. 
Taking the Earth as example, we consider the magnitude of the north-south magnetic field component $B_z$ within CMEs as primary proxy for their potential impact at a generic location in the heliosphere, which we refer to as CME ``\textit{helio-}effectiveness''. In the following, the amplification of the CME helio-effectiveness at a generic location space will be quantified in terms of the amplification of the southward $B_z$ within a given CME as consequence of CME--CME interaction phenomena.

% modelling
Since the ultimate impact of CMEs on geospace is largely determined by their internal magnetic configuration at 1~AU, studies aiming to assess the helio-effectiveness of CMEs in space--time and their resulting geo-effectiveness at Earth require the use of global, physics-based models of the heliosphere capable of describing the 3D magnetic field structure of CMEs, usually by means of various classes of flux rope models.
Recent advances in the field \citep[see e.g.][for recent reviews of the available models]{green:2018, feng:2020} include the \textit{EUropean Heliospheric FORecasting Information Asset} \citep[EUHFORIA;][]{pomoell:2018}, which has been extended to model CMEs using a linear force-free spheromak model \citep{verbeke:2019b}. 
A first test of the predictive capability of this model, limited to non-interacting, single CMEs, has been performed by \citet{scolini:2019}, who also developed a basic methodological scheme to determine the complete set of CME kinematical, geometrical, and magnetic parameters from remote-sensing observations of CMEs in the solar corona. The modelling capabilities of the spheromak model in EUHFORIA in the case of complex, interacting CME events, however, have so far remained unexplored.
Clearly, in order to ultimately quantify the actual geo-effectiveness amplification resulting from the interaction of the CMEs and the terrestrial magnetosphere, heliospheric CME evolution models need to be further coupled to a model of the geospace. As the coupling between the EUHFORIA heliospheric model and physics-based global models of the magnetospheric-ionospheric environments is beyond the scope of this work, we leave the assessment of the capabilities of such a model chain for a future study.

% goal of this study
The double goal of this study is, therefore, (1) to quantify the increase of the helio-effectiveness (in terms of $B_z$ amplification) and the geo-effectiveness (in terms of $B_z$ and Dst index amplification) of individual CMEs due to interaction processes via global, physics-based heliospheric simulations of the specific CME event considered, and (2) to test the predictive performances of the EUHFORIA spheromak CME model for complex multi-CME events. 

% paper organisation
The paper is organised as follows:
in Section~\ref{sec:observations} we introduce the instruments and data used in this work and we present a complete Sun-to-Earth observational overview of the CMEs under study.
In Section~\ref{sec:methods} we describe the observation-based methods used to derive the CME geometric, kinematic, and magnetic parameters from remote-sensing observations of the CMEs close to the Sun.
In Section~\ref{sec:results} we introduce the simulations performed and we present a detailed analysis of the events comparing observational and modelling results.
Finally, in Section~\ref{sec:conclusions} we discuss the results and consider future improvements and applications.

%______________________________________________________________
\section{Observations}
\label{sec:observations}

In this section we describe the observational properties of a series of major CMEs that erupted from the Sun during September 4--6, 2017, and that resulted in a complex and geo-effective signature at Earth on September 6--9, 2017.
We start by analysing white-light coronagraph images of the CMEs taken by the \textit{Large Angle and Spectrometric Coronagraph} \citep[LASCO;][]{brueckner:1995} C2 and C3 instruments on-board the \textit{Solar and Heliospheric Observatory} \citep[SOHO;][]{domingo:1995}, 
and by the \textit{Sun Earth Connection Coronal and Heliospheric Investigation} \citep[SECCHI;][]{howard:2008} COR2 coronagraph on-board the \textit{Solar terrestrial Relations Observatory} \citep[STEREO;][]{kaiser:2008} - ahead (A) spacecraft.
We then discuss the global characteristics of their (common) source region as observed by the \textit{Helioseismic and Magnetic Imager} \citep[HMI;][]{scherrer:2012} and \textit{Atmospheric Imaging Assembly} \citep[AIA;][]{lemen:2012} instruments on-board the \textit{Solar Dynamics Observatory} \citep[SDO;][]{pesnell:2012}.
Finally, we present an overview of the in-situ measurements taken at the Sun--Earth Lagrange L1 point by the \textit{Magnetic Field Investigation} \citep[MFI;][]{lepping:1995} and the \textit{Solar Wind Experiment} \citep[SWE;][]{ogilvie:1995} instruments on-board the $Wind$ \citep{ogilvie:1997} spacecraft,
and by the \textit{Deep Space Climate Observatory} \citep[DSCOVR;][]{burt:2012} spacecraft, discussing the association of the complex in-situ signatures with the CME events observed at the Sun.

%__________________________________________________________________

\subsection{White-light CME Observations}
\label{subsec:corona}

%CME1 ==============================
The first CME (hereafter CME1) was first observed in the LASCO/C2 coronagraph on September 4 at 19:00~UT as a partial halo with a dominant propagation component towards the south-west.
It was associated with an M1.7 class flare (start: 18:46~UT -- peak: 19:37~UT -- end: 19:52~UT) localised in AR~12673.
The CME propagated in LASCO/C2 and LASCO/C3 with an average speed (projected in the plane of sky) of about 600~km~s$^{-1}$, exhibiting a slightly accelerating behaviour (from the \texttt{LASCO CME catalog}, \url{https://cdaw.gsfc.nasa.gov/CME_list/}). 
On the day of the first eruption, STEREO-A was separated from Earth by an angle of $128^\circ$, therefore Earth-directed CMEs could be well observed by the COR2 instrument on-board STEREO-A. In this instrument, the CME was visible starting from 19:39~UT, where it appeared to propagate towards the south-west.

%CME2 ==============================
When the leading edge of CME1 was at ${\sim}10$~solar radii ($R_{\odot}$) as seen by LASCO/C3, it was overtaken by a second, faster CME (hereafter CME2) that was first observed in LASCO/C2 at 20:36~UT.
This second CME appeared from Earth as a full halo having an intensified frontal part propagating towards the south, with an average projected speed of about 1420~km~s$^{-1}$.
CME2 was associated with an M5.5 class flare (start: 20:28 -- peak: 20:33 -- end: 20:37) also localised in AR~12673.
In STEREO/COR2-A, the CME was seen to propagate towards the south-west (first appearance: 20:39~UT), catching up with CME1 shortly after 21:00~UT.
By 21:42~UT, the leading edges of CME1 and CME2 had completely merged as seen by LASCO/C3 as well, so that the two structures became indistinguishable in both LASCO and COR2 images.
The two CMEs erupted from the same AR less than 2~hours apart and exhibited similar coronal signatures, suggesting a sympathetic, (quasi-)homologous nature \citep{zhang:2002, cheng:2006, wang:2013}.

%CME not modelled ==============================
The day after the eruption of CME1 and CME2 (i.e.\ September 5), little activity was observed in the LASCO and STEREO-A coronagraphs.
One faint partial halo CME, most probably also erupting from AR~12673, was visible starting from 17:36~UT in LASCO/C2, but it became too faint to be tracked in the LASCO/C3 field of view. It was at all times barely visible even in running-difference images from STEREO/COR2-A, where its front appeared to propagate predominantly below the ecliptic plane.
The faintness of this CME in LASCO and STEREO-A coronagraph images, combined with the limited eruption signatures visible in EUV images of the solar disk, makes the reconstruction of its kinematic, geometric and magnetic parameters using the techniques presented in Section~\ref{sec:methods} particularly complicated. Considering also its propagation direction below the ecliptic plane, we have neglected this event in the following analysis \citep[see also][for a similar modelling approach]{werner:2019}. However, we point out that this CME could have contributed to the complexity of the event observed at Earth by interacting with the preceding and following CMEs.

%CME3 ==============================
Finally, on September 6, a full halo CME, hereafter CME3, was observed entering the LASCO/C2 field of view at 12:24~UT.
This CME originated from the same AR as the previous ones and it was associated with a remarkably intense flare of class X9.3 (start: 11:53~UT -- peak: 12:02~UT -- end: 12:10~UT). 
The CME was observed to propagate towards the south-west with a projected speed of about 1570~km~s$^{-1}$ and its leading edge was characterised by a highly elliptical shape tilted by about $45^\circ$ with respect to the solar equator.
In STEREO/COR2-A the CME appeared as a full halo (first appearance: 12:24~UT) characterised by a south-west propagation direction (see Appendix~\ref{sec:appendixA} for additional details).

%__________________________________________________________________

\subsection{Source Region Observations}
\label{subsec:ar12673}

\begin{figure*}
\centering
{\includegraphics[width=\hsize,trim={110mm 160mm 110mm 160mm},clip]{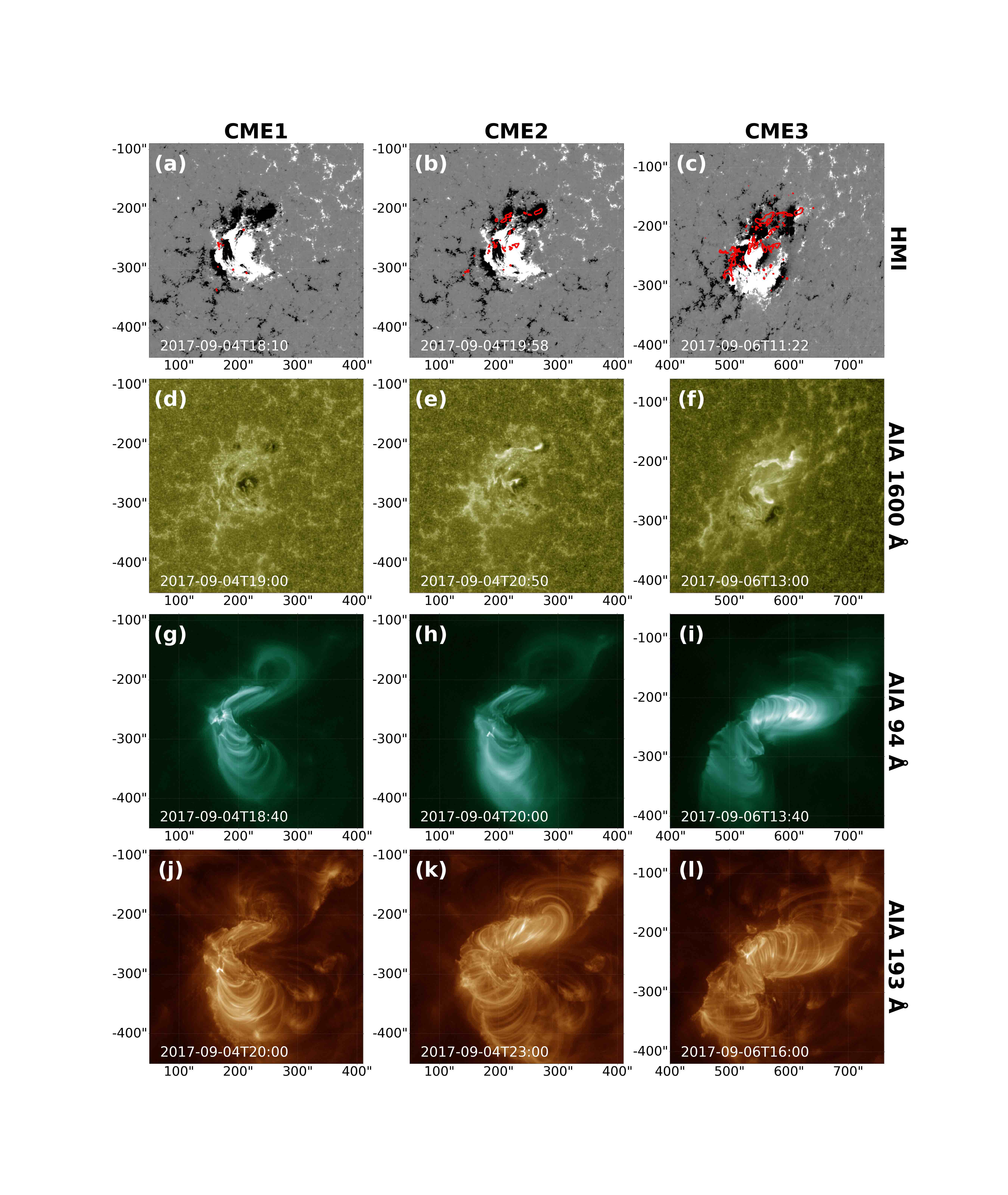}}
\caption{SDO observations of AR~12673 around the eruption times of the CMEs under study.
(a--c, top row) SDO/HMI line-of-sight magnetograms with red contours indicating the location of flare ribbons from SDO/AIA images in the 1600~{\AA} filter (d--f, second row).
(g--i, third row) SDO/AIA images in the 94~{\AA} filter.
(j--l, bottom row) SDO/AIA images in the 193~{\AA} filter.}
\label{fig:ar_hmi_aia} 
\end{figure*}

The source regions of the CMEs discussed above can all be located within AR~12673.
This AR presented outstanding levels of CME and flare productivity persisting for more than a full week \citep{chertok:2018, redmon:2018, bruno:2019}.
The region was first classified as a simple $\alpha$ region \citep{hale:1919, kunzel:1965} on August 30, when it was rotating toward the solar disk centre from the eastern limb.
It was then classified as $\beta \gamma$ on September 3, i.e.\ the day before the eruption of CME1.
The region then developed into a $\beta \gamma \delta$ configuration starting from September 5.
Figure~\ref{fig:ar_hmi_aia} shows SDO observations of the AR as observed by the HMI and AIA instruments.

% PILs
From September 4 onwards, photospheric magnetograms of the AR show the presence of a complex system of polarity inversion lines (PILs) that evolved and rotated over the days (Figure~\ref{fig:ar_hmi_aia}(a)--(c)). 
Two main PIL systems, one in the south-east part of the AR, characterised by an approximately north-south orientation, and one in the north-east part, exhibiting an approximately east-west direction, are visible.

% which CME associated with which PIL 
We use SDO/AIA 171~{\AA} and 1600~{\AA} images to pinpoint the location of the eruption of the three CMEs within the AR.
% CME1
CME1 erupted in the south-east part of the AR, as indicated by the development of flare ribbons (visible in 1600~{\AA}, Figure~\ref{fig:ar_hmi_aia}(d)) and by the southward expansion of coronal loops during the eruption (visible in 171~{\AA}, not shown here).
A post-eruptive arcade (PEA) (visible in Figure~\ref{fig:ar_hmi_aia}(j)) also formed after the first eruption until the onset of the second eruption. 
% CME2
Starting at 20:28~UT, associated with the eruption of CME2, more extended flare ribbons developed in the northern part of the AR (Figure~\ref{fig:ar_hmi_aia}(e)).
These observations suggest that the eruption of CME1 remained confined to the south-east part of the AR, while the eruption of CME2 developed through the whole PIL system up to its north-west end (Figure~\ref{fig:ar_hmi_aia}(b)).
The formation of a stable PEA is visible in the AIA 193~{\AA} filter (Figure~\ref{fig:ar_hmi_aia}(k)), confirming that the magnetic reconnection processes associated with the eruption extended over the whole PIL system elongation.
The short waiting time between CME1 and CME2 (less than $3$~hours) and their origin from the same AR strongly favour the scenario of quasi-homologous CMEs, where the second eruption is commonly interpreted as a consequence of the flux rope destabilisation caused by the rearrangement of coronal magnetic fields following the first eruption \citep{torok:2011, bemporad:2012, wang:2013, chatterjee:2013, liu:2017}.
% CME3
The eruption of CME3, occurring about 41~hours after CME2, originated from the vertical PIL located around the centre of the AR, where changes in the surface magnetic field in 45-sec HMI observations are visible starting from 11:54~UT \citep[see also][]{mitra:2018}. Bright flare ribbons visible in the AIA 1600~{\AA} line (Figure~\ref{fig:ar_hmi_aia}(f)), and a PEA visible in the AIA 193~{\AA} filter (Figure~\ref{fig:ar_hmi_aia}(l)) indicate that magnetic reconnection extended over the whole PIL elongation (Figure~\ref{fig:ar_hmi_aia}(c)).

%__________________________________________________________________

\subsection{In-situ Observations at Earth}
\label{subsec:insitu}

\begin{figure*}
\centering
{\includegraphics[width=0.98\hsize,trim={20mm 40mm 20mm 40mm},clip]{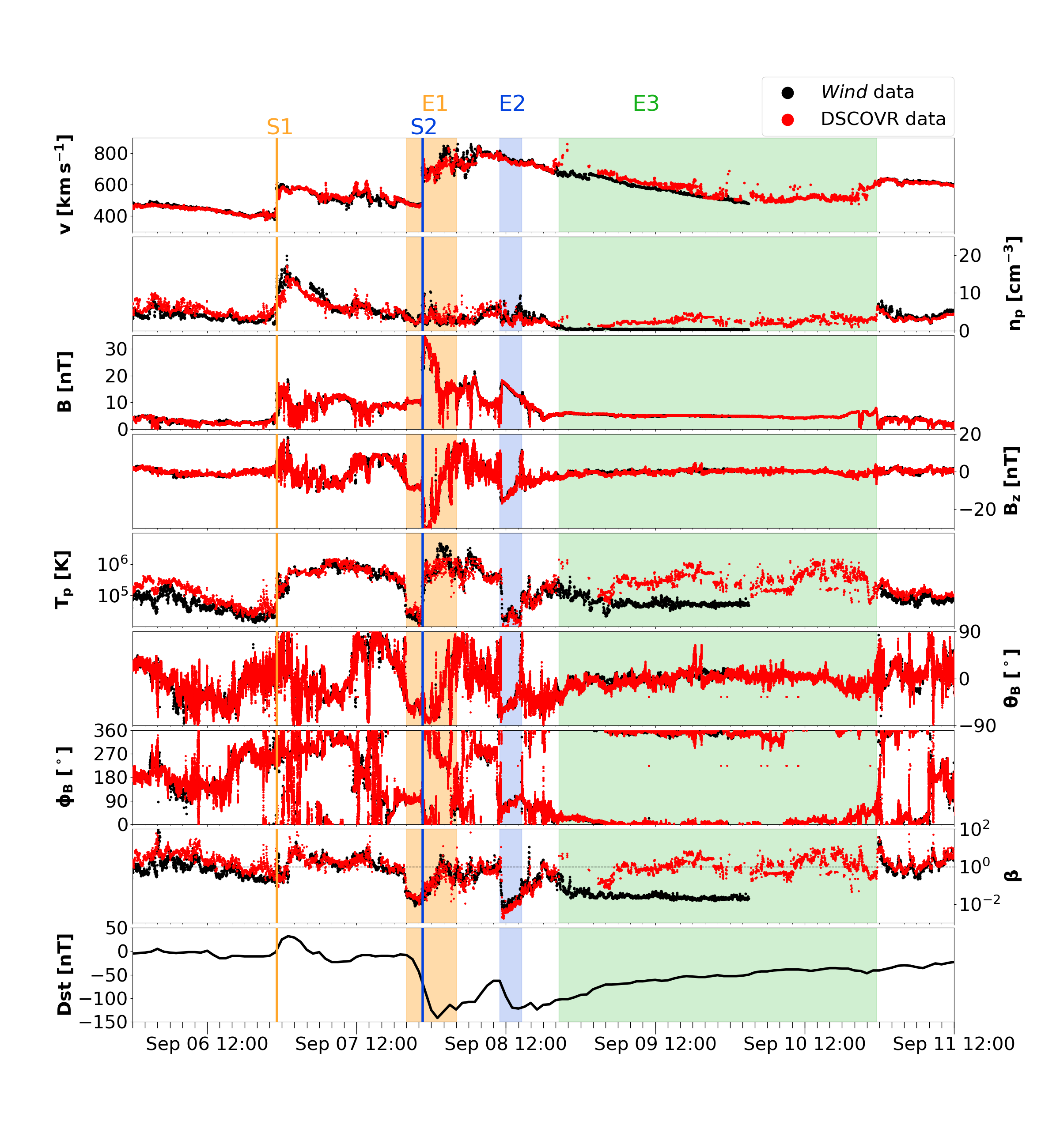}} 
\caption{1-min averaged solar wind magnetic field and plasma parameters from the \textit{Wind} (in black) and DSCOVR (in red) spacecraft at L1, between 6 and 11 September 2017. 
From top to bottom: 
plasma speed ($v$), 
proton number density ($n_p$), 
magnetic field magnitude ($B$), 
magnetic field $B_z$ component in Geocentric Solar Ecliptic (GSE) coordinates, 
proton temperature ($T_p$), 
magnetic field elevation ($\theta_B$), 
magnetic field azimuthal angle ($\phi_B$), 
proton plasma $\beta$.
The bottom panel shows the 1-hour Dst index. 
The vertical lines mark the interplanetary shocks S1 and S2, and the shaded areas mark the periods associated with magnetic ejecta E1, E2, and E3.}
\label{fig:insitu} 
\end{figure*}

Figure~\ref{fig:insitu} shows 1-min averaged in-situ measurements taken by the $Wind$ and DSCOVR spacecraft during the days following the eruptions, together with the 1-hour Dst index measured on ground (provided by the World Data Center for Geomagnetism, Kyoto; \url{http://wdc.kugi.kyoto-u.ac.jp/dstdir/}).
 
% CME1+CME2 =============
Since CME1 and CME2 are observed to merge into a single structure (hereafter CME1+CME2) in the fields of view of LASCO and SECCHI, i.e.\ already in the corona, it is reasonable to expect them to drive a single, common shock \citep[e.g.,][]{odstrcil:2003,xiong:2007,lugaz:2013} as they propagate in interplanetary space, most probably the one observed at $Wind$ at 23:13~UT on September 6 (hereafter S1).
S1 was followed by a prolonged sheath region (with a duration of $\sim$21~hours, corresponding to a thickness of $\sim$0.25~AU for a structure moving at $\sim$500~km~s$^{-1}$) characterised by a fluctuating magnetic field and relatively high density and temperature.
Such a structure can imply a spacecraft crossing through a thick sheath region formed by the merging of the CME1 and CME2 sheaths, whose formation is compatible with the early merging of the two CMEs.
On September 7 around 20:00~UT a region of low plasma $\beta$ and smooth magnetic field, compatible with a magnetic ejecta (hereafter E1), was observed. E1, most probably associated with the merged CME1+CME2 ejecta, was also listed in the Richardson \& Cane ICME list 
\citep[\url{http://www.srl.caltech.edu/ACE/ASC/DATA/level3/icmetable2.htm};][]{cane:2003, richardson:2010} with start time around 20:00~UT and end time around 04:00~UT on September 8, i.e.\ with a duration of 8 hours.
In the list, the ejecta was classified with a quality flag equal to 1, indicating that it exhibited only some of the typical characteristics of magnetic clouds, e.g.\ a low plasma $\beta$ and coherent magnetic field rotation, but lacked some other characteristics such as an enhanced magnetic field magnitude. 

Moreover, the ejecta was characterised by the presence of an interplanetary shock (hereafter S2) propagating through it. 
The shock was observed at 22:38~UT on September 7, i.e.\ 2.5 hours after the start of E1, and it was most likely driven by CME3.
S2 compressed the magnetic field of the E1 ejecta, resulting in a significant amplification of the southward $B_z$ from a pre-existing value of ${\sim}-10$~nT, to ${\sim}-30$~nT.
The negative $z$-component of the magnetic field in the shock upstream region triggered the beginning of a geomagnetic disturbance, marked by a decrease in the Dst index to about $-50$~nT. The further enhancement of the negative $B_z$ in the downstream region led to the development of the first and strongest dip in the Dst profile, reaching $-142$~nT around 1~UT on September 8. Overall S2 presented several characteristics typical of shocks propagating inside preceding ejecta, including a low $\beta$ in the upstream and downstream shock regions, and a magnetic field clock angle almost constant across the shock \citep[][]{lugaz:2015b, lugaz:2017}.

% CME3 =============
A second period of enhanced magnetic field and low $\beta$ was observed between 11:00~UT on September 8 and 20:00~UT on September 10.
This period was classified in the Richardson \& Cane ICME list as a single ejecta with a quality flag equal to 1.
However, we note the presence of a region of fluctuating fields and relatively high proton temperature and plasma $\beta$ between 14:30~UT and 20:30~UT on September 8, which suggests the passage of two separate ejecta regions (hereafter E2 and E3).
Both E2 and E3 exhibited low plasma $\beta$ and enhanced magnetic fields with different levels of rotation. 
The period marked as E2 is characterised by a rotating magnetic field, a typical characteristic of spacecraft crossings through the flux rope structure.
This ejecta period is most probably associated with the passage of CME3 that also drove S2.
E3 exhibits typical characteristics of leg encounters, as featured by the large $B_x$ component (not shown), the lack of magnetic field rotations, and a long duration \citep{marubashi:2007, moestl:2010, kilpua:2011, kilpua:2013, owens:2016}. 
In view of the coronal and in-situ CME/ICME observations, we consider most probable that E2 and E3 were associated with the same CME (i.e.\ CME3) at the Sun, and simply corresponded to crossings of the spacecraft through different parts of the flux rope.
In this picture, E2 would correspond to a passage closer to the apex of CME3, and E3 to the passage through its leg. 
Additional evidences are provided by the speed profile which decreases very coherently through and between E2 and E3, as well as by the coherent rotation of the magnetic field vectors.
This interpretation would require a bending or deformation of the flux rope global structure as a consequence of its interaction with the ambient solar wind or preceding ejecta \citep{crooker:1998, mulligan:1999, dasso:2007, marubashi:2007}.
The passage of E2 generated a second dip in the Dst index (minimum of $-124$~nT on September 8 around 13~UT) i.e.\ the complex ejecta investigated here resulted in a two-step geomagnetic storm  \citep{farrugia:2006, liu:2014b}. The magnetic field in E3 was low (about 5~nT) and pointed primarily to the north. As a consequence, the Dst recovered during E3 without further intensification. 

We note that in such high-activity periods, the identification of ICME signatures and their linking with the corresponding CMEs at the Sun becomes very complex due to interactions among the various structures in the corona and interplanetary space, and the elevated number of potential CME candidates. For this reason, we cannot rule out \textit{a priori} that E2 and E3 were associated with two different CMEs at the Sun, possibly involving the faint partial halo CME discussed in Section~\ref{subsec:corona}.  
However, in order to keep the assumptions as simple as possible at the beginning, in our simulations we consider both ejecta to be associated with CME3. This is a reasonable assumption as our primary focus concerns the investigation of the nature and evolution of the main geo-effective structures, i.e.\ the southward field in E1 due to the compression by S2, and the southward field in E2, rather than the origin of E3.
We also note that with this interpretation our in-situ analysis is consistent with the previous studies by \citet{shen:2018} and \citet{werner:2019}.

Between the end of E1 and the start of E2, a region characterised by plasma $\beta {\sim}1$, modest and fluctuating magnetic fields, and increasing density, suggests the occurrence of magnetic reconnection at the interaction surface between E1 and E2 \citep{maricic:2014}. These in-situ signatures exhibit several characteristics of ongoing CME--CME interactions, consistent with a picture in which the interaction of CME1+CME2 and CME3 was still at an early stage at 1~AU \citep{lugaz:2015b, lugaz:2017}.
%
%__________________________________________________________________

\section{Methods and models}
\label{sec:methods}

%--------------------------------------
\subsection{CME Kinematics and Geometry}
\label{subsec:gcs}

During the days of the CME eruptions, STEREO-A was located at a longitude of $-128^\circ$ in Stonyhurst coordinates \citep{thompson:2006}, providing a second vantage point to investigate the coronal evolution of the CMEs under study, in addition to the observations made along the Sun--Earth line. 
To constrain the kinematics and geometry of the CMEs in the corona, we perform a 3D fitting of the events from these two viewpoints (SOHO and STEREO-A) using the Graduated Cylindrical Shell \citep[GCS;][]{thernisien:2006, thernisien:2009} model.
The results from the GCS fitting are then used as input for the CME modelling using EUHFORIA (Section~\ref{subsec:euhforia}).

% Table
\begin{table*}
\centering
\begin{tabular}{llll}
\hline
\hline
\textbf{Parameter}              & \textbf{CME1}     & \textbf{CME2}         & \textbf{CME3} \\
\hline
$\theta_\mathrm{CME}$            &  0$^\circ$        & -25$^\circ$           & -11$^\circ$  \\
$\phi_\mathrm{CME}$              &  25$^\circ$       & 0$^\circ$           & 21$^\circ$  \\  
$\kappa_\mathrm{CME}$            &  0.38             & 0.50                  & 0.43   \\
$\alpha_\mathrm{CME}$            &  10$^\circ$       & 30$^\circ$            & 15$^\circ$  \\
$\omega_\mathrm{CME}/2$ (EO$/$AV$/$FO)   &  $22^\circ / 27^\circ / 32^\circ$     & $30^\circ / 45^\circ / 60^\circ$ & $25^\circ / 33^\circ  / 40^\circ$  \\
$\gamma_\mathrm{CME}$            &  0$^\circ$        & 0$^\circ$             & 40$^\circ$  \\
$v_\mathrm{CME}$                 &  960 km s$^{-1}$  & 1585 km s$^{-1}$      & 1910 km s$^{-1}$  \\
$v^\mathrm{rad}_\mathrm{CME}$           &  697 km s$^{-1}$      & 1057 km s$^{-1}$      & 1293 km s$^{-1}$  \\
$v^\mathrm{exp}_\mathrm{CME}$           &  263 km s$^{-1}$      & 528 km s$^{-1}$       &  617 km s$^{-1}$ \\
Time at 0.1~AU           & 2017-09-04T23:00     & 2017-09-04T22:44    &    2017-09-06T14:11               \\
 \hline
 \hline
\end{tabular}
\caption{Results from the GCS fitting of the three CMEs under study. 
EO = edge-on, FO = face-on, AV = average.}
\label{tab:gcs}
\end{table*}

The GCS fitting provides as output the following parameters (in Stonyhurst coordinates):
CME latitude ($\theta_\mathrm{CME}$), 
longitude ($\phi_\mathrm{CME}$), 
front height ($h_\mathrm{CME}$), 
aspect ratio ($\kappa_\mathrm{CME}$), 
half angle ($\alpha_\mathrm{CME}$), 
and tilt ($\gamma_\mathrm{CME}$).
For $\alpha_\mathrm{CME} \neq 0^\circ$, the shell of the GCS model corresponds to a croissant-like shape. From the aspect ratio and half angle we derive the edge-on (EO), face-on (FO) and average (AV) CME half widths ($\omega_\mathrm{CME}/2$) \citep{thernisien:2009}. From a modelling perspective, the determination of a single value for the CME half width from raw GCS outputs is critical as in EUHFORIA CMEs are initialised with spherical shapes, i.e.\ their cross sections are symmetric in all directions.
From the fitting of sequential images, we also derive the deprojected (3D) speed of the CME apex ($v_\mathrm{CME}$) as well as 
the radial/translational CME speed ($v^\mathrm{rad}_\mathrm{CME}$, corresponding to the speed of the centre of the croissant tube), and the CME expansion speed ($v^\mathrm{exp}_\mathrm{CME}$, corresponding to the rate of increase of the croissant cross section radius) \citep[][see also Appendix~\ref{sec:appendixA} for the analytical derivation]{scolini:2019}.
Each speed was determined by performing a linear fitting of instantaneous results from LASCO/C2--C3 and STEREO/COR2-A images 
(CME1: between 19:54~UT and 20:39~UT on September 4, for a total 3 images;
CME2: between 20:39~UT and 21:54~UT on September 4, for a total 5 images;
CME3: between 12:39~UT and 13:54~UT on September 6, for a total 5 images).

Table~\ref{tab:gcs} lists the results of the GCS fitting for the three events under study. 
Snapshots of the GCS fitting results overplotted on LASCO/C3 and STEREO/COR2-A images are provided in Appendix~\ref{sec:appendixA}.
We note that from the extrapolation of the times at which the CME leading edges reached 0.1~AU, we obtain that CME2 reached this distance about 15~minutes earlier than CME1. 
This indicates that the interaction between CME1 and CME2 occurred at heliocentric distances close to or slightly lower than 0.1~AU.

%--------------------------------------
\subsection{CME Magnetic Parameters}
\label{subsec:cme_magnetic_parameters}

In preparation for the heliospheric CME simulations with the spheromak flux rope model in EUHFORIA (discussed in Section~\ref{subsec:euhforia}), we discuss in the following subsections the observational derivation of three key parameters characterising the magnetic (flux rope) structure of the CMEs under study: their chirality, tilt and the amount of magnetic flux reconnected during the eruption.

% -----------------------------------------
\subsubsection{Chirality and Tilt of the Flux Ropes}
\label{subsubsec:chirality_tilt}

% chirality
\paragraph{Chirality of the Flux Ropes}
Observationally, the magnetic helicity sign (or \textit{chirality}) of ARs can be inferred from different morphological features \citep[e.g.][]{demoulin:2009, palmerio:2017}. In the particular case under study, images in the extreme ultraviolet SDO/AIA filter at 94~{\AA} (Figure~\ref{fig:ar_hmi_aia}(g)--(i)) suggest that AR~12673 was characterised by a negative chirality as indicated by the presence of a reverse-S sigmoid in its northern part, which is also consistent with the recent analyses by \citet{mitra:2018}, \citet{yan:2018} and \citet{price:2019}. Although cases of inconsistency between the chirality of the source region and that of the erupted CME have been observed \citep[e.g.][]{chandra:2010}, for most of the events the two are found to match \citep{bothmer:1998, palmerio:2018}. For this reason, in the following analysis we start by assuming the erupted structures to be characterised by a negative chirality as their source region.

\begin{figure}
\centering
{\includegraphics[width=0.90\hsize]{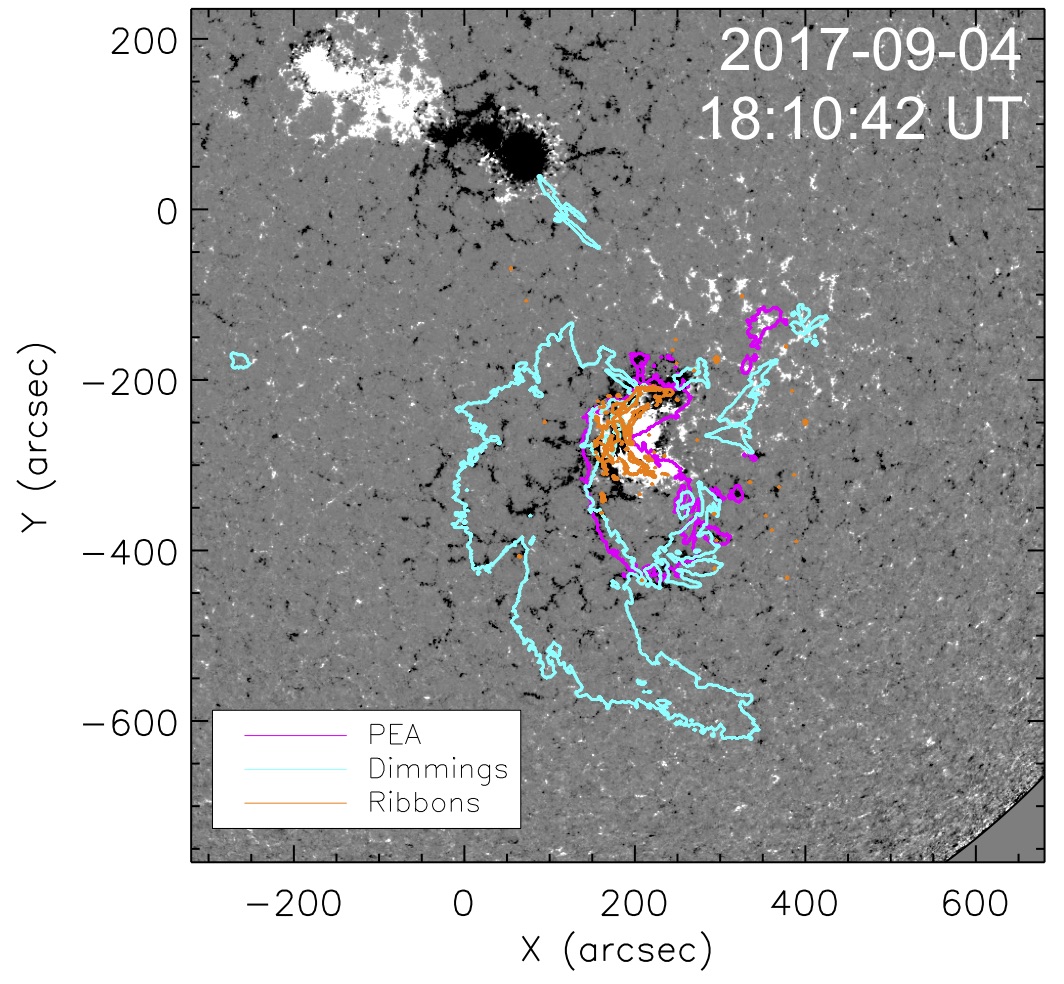} \\
\includegraphics[width=0.90\hsize]{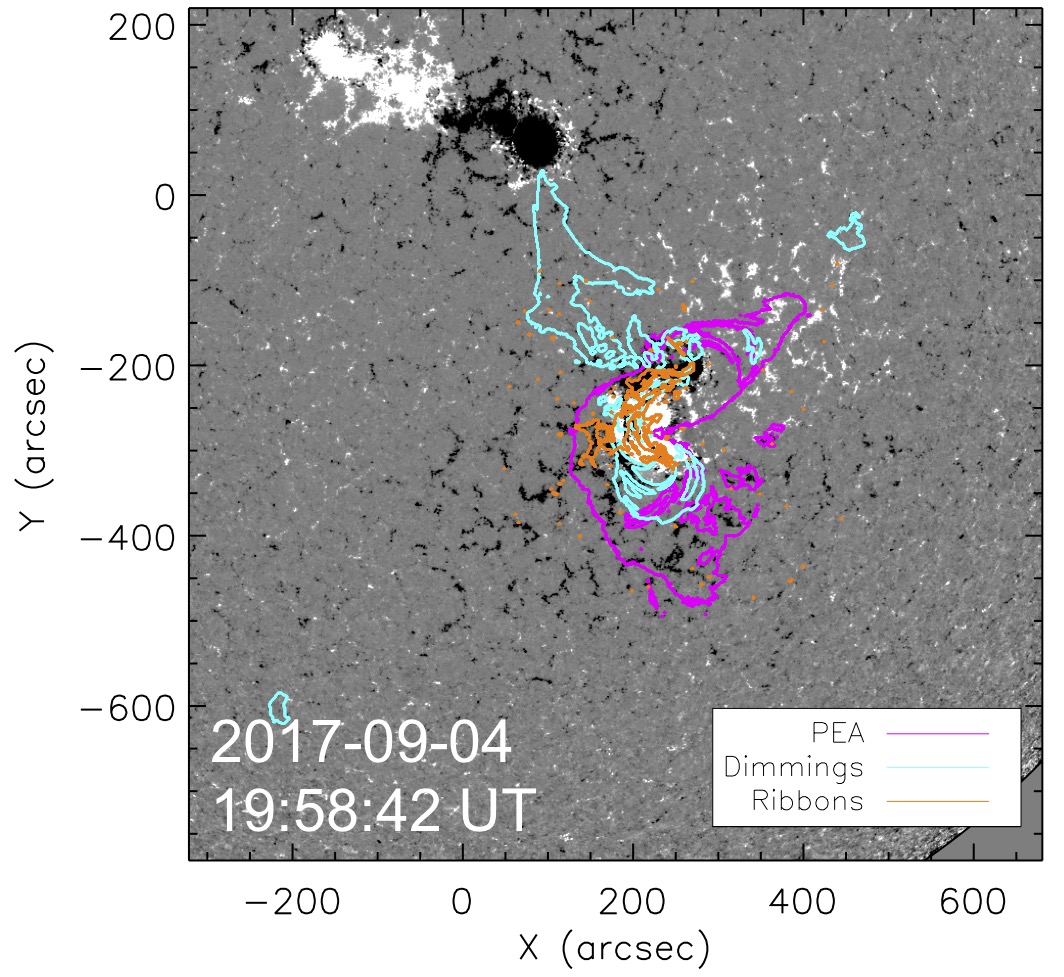} \\
\includegraphics[width=0.90\hsize]{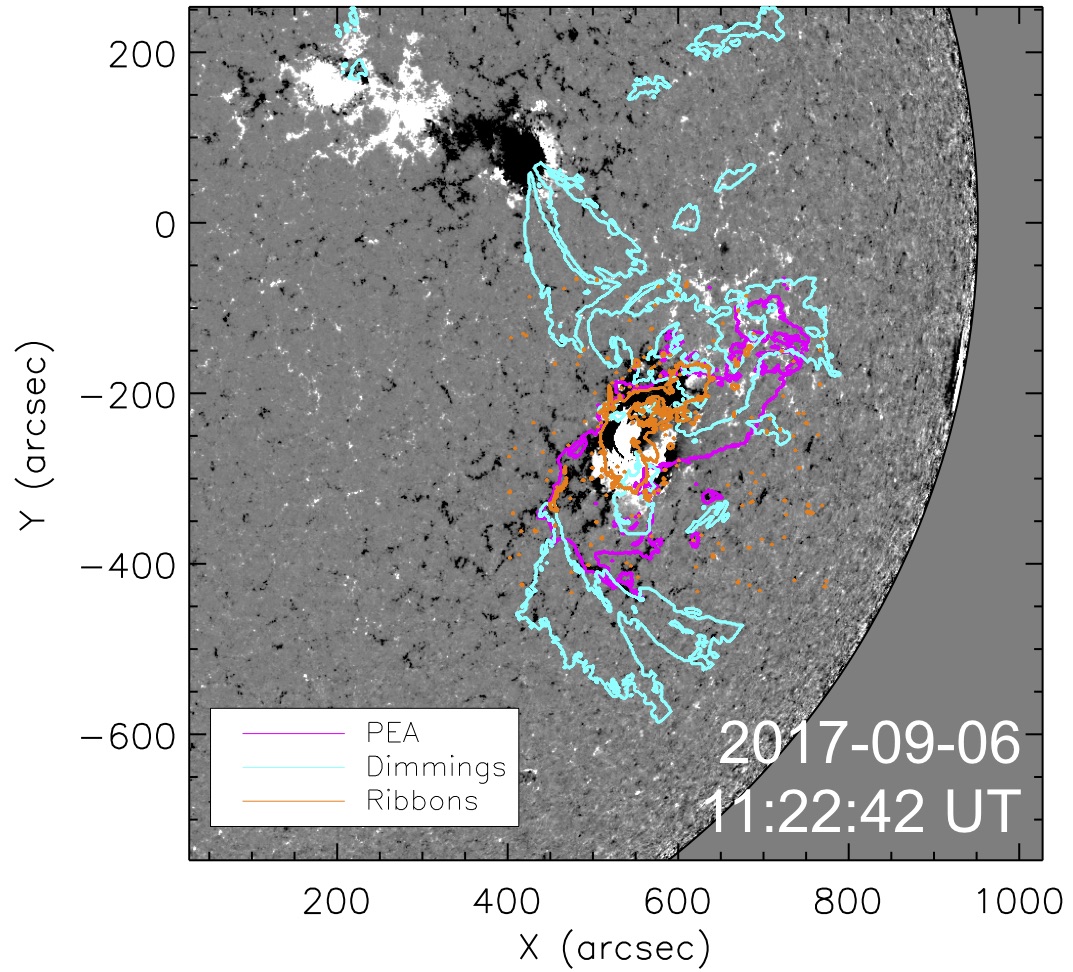}}
\caption{
Location of flare ribbons (orange curves), coronal dimmings (cyan curves), and PEAs (magenta curves) associated with the eruption of CME1 (top), CME2 (middle), and CME3 (bottom). The greyscale backgrounds show the HMI line-of-sight magnetic fields on 
September 4 at 18:16~UT (top) and 19:58~UT (middle), and September 6 at 11:23~UT (bottom), saturated to $\pm 100$~G with black and white representing the negative and positive polarities, respectively.
}
\label{fig:hmi_masks} 
\end{figure}

% tilt at the source region
\paragraph{Tilt of the Flux Ropes}
In order to estimate the orientation of the flux ropes at the Sun, we use proxies based on the orientation of PEAs and PILs \citep{moestl:2008, palmerio:2017, palmerio:2018}. As shown in Figure~\ref{fig:ar_hmi_aia}, the PEA forming after the eruption of CME1 was confined to the southern portion of the AR/PIL, and exhibited an approximately north-south orientation. 
For CME2 and CME3, we observe PEAs developing along the whole PIL structure. 
Although a global direction from south-east to north-west can be identified, the shape of such PEAs appears to be bent in a reverse-S shape. This reflects the complexity of the underlying PIL system, and makes the determination of an unambiguous flux rope tilt based on such observations extremely difficult.
Similar conclusions about the initial flux rope tilts can be obtained by considering the locations of coronal dimmings and flare ribbons (Figure~\ref{fig:hmi_masks}).
Combining these tilts with the information about flux rope chirality and magnetic polarity regions from HMI magnetograms, we recover an ENW flux rope type for CME1 and intermediate ENW--NWS flux rope types for CME2 and CME3 in the lower corona \citep[using the same classification as][]{bothmer:1998, mulligan:1998, palmerio:2018}.

% tilt in the upper corona
To constrain the orientation of the flux ropes in the upper corona (${\sim}5$~$R_{\odot}$ to ${\sim}20$~$R_{\odot}$), we consider the results from the GCS fittings. The derived tilts suggest the axial magnetic field of CME1 and CME2 to be oriented parallel to the solar equatorial plane ($\gamma_\mathrm{CME} = 0^\circ$), while that of CME3 to have an inclination of ${\sim}40^\circ$ with respect to the solar equator. 
According to reconstruction of the heliospheric current sheet (HCS) by potential-field source-surface \citep[PFSS;][]{altschuler:1969} models (from the GONG network: \url{https://gong.nso.edu/data/magmap/pfss.html}), the derived CME tilts are (quasi-) aligned to the HCS \citep{yurchyshyn:2008, isavnin:2014}.
Using the GCS tilt estimates as given in Table~\ref{tab:gcs} and assuming the flux rope chirality at the source region to be preserved in space--time, two possible flux rope types are possible for each CME.
CME1 and CME2 are most probably either SEN flux rope types, or NWS flux rope types.
CME3 can be associated either with an intermediate ENW--NWS flux rope, or with an intermediate WSE--SEN flux rope.
We note, however, that the $\gamma_\mathrm{CME}$ parameter is associated with the highest uncertainties \citep{thernisien:2009}, and it is known to be very sensitive to the subjectivity involved in performing the GCS fitting \citep[see e.g.\ Figure~5 in][for an alternative fitting of the CMEs using the GCS model, leading to a quite different interpretation of the tilt angles in the corona]{shen:2018}.
For this reason, we consider its reconstruction as particularly uncertain.

% comparison with tilt in AR
Overall, the comparison of the CME flux rope types recovered in the lower and upper corona suggests that CME1 and CME2 underwent considerable rotations (rotation $\ge 90^\circ$ for CME1, $\ge 45^\circ$ for CME2, and $\ge 0^\circ$ for CME3). Similar conclusions are reached when considering the GCS fitting performed by \citet{shen:2018}. Such estimates are in the upper range of reported values \citep{lynch:2009, vourlidas:2011, isavnin:2014, kay:2015}, but are not surprising given the sympathetic nature of CME1 and CME2, consistent with a scenario of early interactions and strong magnetic forces that may have led to significant CME rotations \citep{kay:2015}.

% association at 1 AU
\paragraph{Association to Interplanetary Structures}
In order to further confirm the associations between the CMEs and their interplanetary counterparts described in Section~\ref{subsec:insitu}, we also compare the helicity sign and the flux rope types of the corresponding ejecta at the Sun with the magnetic structures recovered in situ. 

The magnetic field rotations observed in association with E1 (most probably associated with the structure resulting from the merging of CME1 and CME2) indicate that the ejecta can be described as a left-handed flux rope characterised by a ${\sim}45^\circ$ inclination between a SEN- and an ENW-type at 1~AU, although the presence of shock S2 propagating through E1 makes the interpretation of the flux rope type particularly problematic because of the disturbed properties in the downstream region.
The magnetic field rotations within E2 indicate that the second ejecta can be described as a left-handed low-inclination flux rope of SEN type at 1~AU. The lack of magnetic field rotations within E3 makes it difficult to determine the flux rope type of this structure. 
Overall, the chiralities of the flux ropes recovered from in-situ observation at Earth are consistent with the chiralities recovered from source region images. 
On the other hand, occurrence of significant rotations in the corona and/or interplanetary space is needed to explain the different flux-rope orientations recovered at the source region, in the upper corona, and at 1~AU. Although deviating from typical scenarios, this aspect can be interpreted as a consequence of the multiple coronal and interplanetary interactions that occurred among the CMEs under study, as well as with other heliospheric structures (e.g.\ the HCS).

% -----------------------------------------
\subsubsection{Reconnected Magnetic Fluxes}
\label{subsubsec:phir}

In order to perform EUHFORIA simulations with the spheromak CME model, next to observational estimations of the flux rope tilts and chiralities, we also need an estimate of the amount of magnetic flux contained within the magnetic structure. As a proxy, we use the flux ($\varphi_{r}$) that reconnected in association with each CME eruption. To have a robust estimate of $\varphi_{r}$, we compare the results obtained from the analysis of a variety of post-eruptive signatures (such as flare ribbons, coronal dimmings, and PEAs) in the AR involved in the eruptions. 

% Table
\begin{table*}
\centering
\begin{tabular}{lllll}
\hline
\hline
\textbf{Parameter}              & \textbf{CME1}     & \textbf{CME2}     & \textbf{CME3} & \\
\hline
$\varphi_{r}$ \citep[based on][]{kazachenko:2017}     
                                    & $2.3$   
                                    & $4.9$            
                                    & $30$ & \rdelim\}{5}{2mm}[statistical] \\
$\varphi_{r}$ \citep[based on][]{tschernitz:2018}     
                                    & $2.8$  
                                    & $5.5$            
                                    & $28$ & \\
$\varphi_r$ \citep[based on][]{dissauer:2018b}    
                                    & $1.9$   
                                    & $3.1$            
                                    & $9.9$ & \\
$\varphi_r$ \citep[based on][]{pal:2018}    
                                    & $4.8$   
                                    & $9.9$            
                                    & $13$ & \\
$\boldsymbol{\varphi_\mathrm{r}}$ \textbf{(average)}           
                                    & $\boldsymbol{3.0}$       
                                    & $\boldsymbol{5.9}$            
                                    & $\boldsymbol{17}$ & \\
$\varphi_{r}$ \citep[based on][]{kazachenko:2017}       
                                    & $0.81$          
                                    & $0.78$                      
                                    & $3.9$ & \rdelim\}{4}{2mm}[single-event] \\
$\varphi_r$ \citep[based on][]{dissauer:2018b}                
                                    & $4.9$       
                                    & $3.4$           
                                    & $7.6$ & \\
$\varphi_\mathrm{r}$ \citep[based on][]{gopalswamy:2017}              
                                    & $8.2$       
                                    & $8.7$            
                                    & $10$ & \\
$\boldsymbol{\varphi_\mathrm{r}}$ \textbf{(average)}              
                                    & $\boldsymbol{4.6}$       
                                    & $\boldsymbol{4.3}$            
                                    & $\boldsymbol{7.2}$ & \\ 
 \hline
 \hline
\end{tabular}
\caption{$\varphi_r$ (in units of $10^{21}$~Mx) for the three CMEs under study as recovered from statistical and single-event analyses.}
\label{tab:phir}
\end{table*}

\paragraph{Statistical Relations}
We first estimate the reconnected fluxes using CME--flare statistical relations proposed in previous works. 
Among all, we focus on the relations between the flare peak intensity in soft X-rays and the reconnected flux derived from flare ribbons and coronal dimmings \citep[e.g.][]{tschernitz:2018, kazachenko:2017, dissauer:2018b}, and on the relations between the CME speed and the reconnected flux derived from PEAs \citep{pal:2018}.
Major advantages in using statistical relations instead of an in-depth analysis of single events are the applicability to a larger set of events (not restricted to eruptions originated close to the disk centre and characterised by specific post-eruptive signatures), and the simplicity of use which makes them potentially suitable for operational forecasting applications, as they can be used to routinely initialise the parameters used by physics-based flux rope CME models running in forecasting mode. 

% flare ribbons -----------------
\cite{tschernitz:2018} studied a set of 51 flares ranging between B3 and X17 in GOES class,
reporting a very tight correlation (Pearson correlation coefficient $r_P =0.92$ in log--log space) between the flare peak intensity $I_\mathrm{SXR}$ (in units of W~m$^{-2}$) and the reconnected flux $\varphi_{r}$ (in units of Mx) estimated from flare ribbons:
\begin{equation}
   \log(\varphi_{r}) = 24.21 + 0.58 \log(I_\mathrm{SXR}).
\label{eqn:fl_rcf_T2018}
\end{equation}
Considering a larger sample of about 3000 flares ranging from C1 to X5 in GOES class (\texttt{RibbonDB} catalog, \url{http://solarmuri.ssl.berkeley.edu/~kazachenko/RibbonDB/}), \citet{kazachenko:2017} reported a correlation of
\begin{equation}
    \log(\varphi_{r}) = 24.72 + 0.64 \log(I_\mathrm{SXR}),
\label{eqn:fl_rcf_K2017}
\end{equation}
with Spearman's rank correlation coefficient $r_S = 0.66$.
Correcting for the different definitions of $\varphi_{r}$ used by \cite{kazachenko:2017} ($\varphi_{r}$ = total (unsigned) magnetic flux) compared to \cite{tschernitz:2018} ($\varphi_{r}$ = average of the positive and negative fluxes), the relation in Equation~\ref{eqn:fl_rcf_K2017} becomes
\begin{equation}
    \log(\varphi_{r}) = 24.42 + 0.64 \log(I_\mathrm{SXR}),
\label{eqn:fl_rcf_K2017b}
\end{equation}    
where $\varphi_{r}$ is now defined to be consistent with the definition used by \citet{tschernitz:2018}.
Considering the large span of the flare GOES classes associated with the three CMEs under study (M1.7 to X9.3) 
and the higher correlation coefficients reported by \cite{tschernitz:2018} and \cite{kazachenko:2017} compared to other studies, in the following we use Equations~\ref{eqn:fl_rcf_T2018} and \ref{eqn:fl_rcf_K2017b} to identify the most probable range of flare ribbon reconnected flux values $\varphi_r$ associated with each CME event.

% coronal dimmings -----------------
In addition to flare ribbons, we consider coronal dimmings as a secondary signature to estimate the reconnected flux during the eruptions under study based on flare peak intensities.
\cite{dissauer:2018b} performed a statistical analysis based on coronal dimming regions observed in association with a set of 62 CME events, reporting a correlation between the flare peak intensity and reconnected flux estimated from coronal dimmings equal to
\begin{equation}
    \log(\varphi_{r}) = 23.26 + 0.42 \log(I_\mathrm{SXR}),
\label{eqn:fl_rcf_D2018}
\end{equation}
with $r_P = 0.62$ (in log--log space).

Applying Equations~\ref{eqn:fl_rcf_T2018},~\ref{eqn:fl_rcf_K2017b},~and~\ref{eqn:fl_rcf_D2018} to the flare peak intensities observed in association with the three CME events under study (obtained from the NOAA SWPC data achive, \url{ftp.swpc.noaa.gov/pub/warehouse}), i.e.\ 
$I_\mathrm{SXR} = 1.7 \times 10^{-5}$~W~m$^{-2}$ (CME1), 
$I_\mathrm{SXR} = 5.5 \times 10^{-5}$~W~m$^{-2}$ (CME2), 
and $I_\mathrm{SXR} = 9.3 \times 10^{-4}$~W~m$^{-2}$ (CME3), 
we obtain an estimate of the reconnected fluxes based on statistical relations as listed in Table~\ref{tab:phir}.

% PEAs -----------------
In a recent study, \citet{pal:2018} derived statistical relations linking the reconnected fluxes obtained from the Flux Rope from Eruption Data \citep[FRED;][]{gopalswamy:2017} method, using PEAs as a primary signature to calculate the reconnected flux, and the CME 3D speed $v_\mathrm{3D}$ (in units of km~s$^{-1}$) in the corona estimated by applying the GCS fitting technique.
Based on 33 CME events, they reported a correlation of ($r_P = 0.66$):
\begin{equation}
    v_\mathrm{3D} = 327 \, \varphi_{r}^{0.69}.
\label{eqn:pea_rcf_P2018}
\end{equation}
Inverted, the relation above is
\begin{equation}
    \varphi_{r} = \left ( \frac{v_\mathrm{3D}}{327} \right )^{1/0.69},
\label{eqn:pea_rcf_P2018b}
\end{equation}
which allows to estimate the reconnected flux once the 3D speed of the CME is known.
Using as input the 3D speeds reconstructed from the GCS fitting ($v_\mathrm{CME}$) listed in Table~\ref{tab:gcs},
the reconnected fluxes for the three CMEs obtained from Equation~\ref{eqn:pea_rcf_P2018b} are provided in Table~\ref{tab:phir}.

\paragraph{Reconnected Fluxes from Single-event Analysis}
In order to obtain an event-based estimate of the reconnected fluxes and to assess the performance of statistical relations in the specific case of the events considered, we complement the $\varphi_r$ values recovered from statistical relations with results from the single-event analysis of each of the three eruptions under study. 
In order to consider a broad spectrum of CME-related signatures, 
we estimate the reconnected flux involved in each CME eruption using the following methods: 
\begin{enumerate}
\item 
A method to identify flare ribbon areas based on the work by \cite{kazachenko:2017}.
Flare ribbons are detected in SDO/AIA 1600~{\AA} images with 24-sec cadence by applying a cutoff threshold based on the median image intensity. Images are taken between 30~minutes before and 3~hours after the start of the flare associated with each CME under study.
\item 
A method to identify coronal dimming areas based on the work by \cite{dissauer:2018a}. 
Coronal dimmings are detected based on a thresholding method that is applied to logarithmic base-ratio SDO/AIA 211~{\AA} images. Similar as for the flare ribbons, dimming pixels are detected between 30~minutes before the flare and up to 3~hours after the flare, respectively.
\item 
A method to identify PEA areas based on the FRED method described by \cite{gopalswamy:2017}.
For each CME under study, we use SDO/AIA 193~{\AA} taken around the moment of maximal extension of the PEAs. We note that instead of performing a manual identification of the PEA areas as in \citep{gopalswamy:2017}, we here employ an automatic identification algorithm using a cutoff threshold based on the median image intensity.
\end{enumerate}
To recover $\varphi_r$ based on the areas identified with the methods above, we use full-disk SDO/HMI line-of-sight magnetograms (\texttt{hmi.m\textunderscore720s}) taken about 30 minutes before the flare start times (shown in Figure~\ref{fig:hmi_masks}). 

% variability of the results and large uncertainties
Results from the single-event analyses are listed in Table~\ref{tab:phir}, while the location of the various EUV signatures overplotted on HMI magnetograms for the three events is shown in Figure~\ref{fig:hmi_masks}. 
The large spread in the recovered $\varphi_r$ values from single-event analyses reflects the different areas covered by the signatures considered (ribbons, dimmings, and PEAs), and we note that large uncertainties affect the estimation of $\varphi_{r}$, i.e.\ up to $\pm50$\% of the measured value, as reported by various studies \citep{qiu:2007, temmer:2017, tschernitz:2018, dissauer:2018b, gopalswamy:2017, pal:2017}. 
Despite the scatter, the different values recovered can therefore be considered to be consistent within the (large) error bars.
At the same time, these results highlight that using this methodology one should aim to recover an order of magnitude for the reconnected flux, rather than a precise estimate of it.

When averaging out the variability of the $\varphi_{r}$ results obtained from the different methods, we recover very similar results between the single-event analyses and the statistical analyses of CME1 and CME2 (rows in bold in Table~\ref{tab:phir}).
The results from the two approaches for CME3, on the other hand, appear somehow less consistent. We note that a contributing factor to this variability comes from the fact that the CME erupted about $40^\circ$ away from the solar disk center, which is close to the limit of the applicability of the single-event analysis methods due to the increased projection effects when moving away from the disk centre \citep[e.g.][]{dissauer:2018a, gopalswamy:2017}. This most probably resulted in higher uncertainties affecting the reconstruction of $\varphi_r$, i.e. the results from single-event and statistical analyses are still consistent among each other due to the larger error bars.
Overall, these results indicate that for the specific events considered, using different statistical methods to quantify the reconnected fluxes provides results that are on average consistent with those of more sophisticated single-event analysis methods. 
Such statistical methods are fast and easy to apply as they only require as input easy-to-use data products, such as the peak intensity of the CME associated flares, or the 3D speed of the CME in the corona as recovered from the GCS fitting or other reconstruction methods. In the context of operational forecasting operations, these results therefore highlight how statistical methods could represent promising potential alternatives to otherwise time-consuming single-event analysis methods.

%__________________________________________________________________

\subsection{The EUHFORIA Model}
\label{subsec:euhforia}

Magnetohydrodynamic (MHD) simulations of the evolution of magnetised CMEs in the heliosphere are powerful complementary tools to observations, as they can provide information on the evolution of CME structures in 3D space that is often difficult to infer from remote sensing or in-situ observational data alone---particularly in cases limited by single-spacecraft in-situ measurements as here. 
In this work, we investigate the heliospheric propagation and interaction of the three successive CMEs under study using the EUHFORIA model.

EUHFORIA is a recently-developed 3D MHD model of the inner heliosphere \citep{pomoell:2018}
that allows the modelling of the background solar wind and CME events using a linear force-free spheromak flux rope model \citep{verbeke:2019b}.
The model is composed by
(1) a semi-empirical Wang--Sheeley--Arge-like \citep[WSA;][]{arge:2004} global coronal model
that provides the background solar wind parameters at the heliocentric radial distance of $0.1$~AU, 
starting from synoptic maps of the photospheric magnetic field, 
and (2) a time-dependent 3D MHD model of the inner heliosphere (between 0.1~AU and 2~AU).
In the heliosphere, it is possible to model solely the ambient solar wind or to also include CMEs, which are inserted in to the heliospheric domain via time-dependent boundary conditions at 0.1~AU.
In this work we use a computational domain for the heliospheric model between $\pm 180^\circ$ in longitude ($\phi$), $\pm 80^\circ$ in latitude ($\theta$), and from 0.1~AU to 2~AU in the radial direction ($D$). 
The use of a sufficiently high spatial resolution is particularly necessary to better resolve shock structures in the simulation domain, which are extremely important in the context of global CME--CME interactions, as discussed in Section~\ref{sec:results}. For this reason, our simulations are performed using a homogeneous grid with {1024} cells in the radial direction (corresponding to a radial resolution of $\Delta D={0.00186}$~AU~$=0.4$~$R_{\odot}$ per cell), and with $2^\circ$ resolution in the longitudinal and latitudinal directions. As input for the coronal model we use the magnetogram synoptic map generated by the Global Oscillation Network Group (GONG) on September 4, 2017 at 00:04~UT \url{https://gong.nso.edu/data/magmap/QR/bqs/201709/mrbqs170904/mrbqs170904t0004c2194_055.fits.gz}.
All simulations are carried out with EUHFORIA version 1.0.4.
In the following, all coordinates are given in the Heliocentric Earth Equatorial (HEEQ) coordinate system,
unless specified otherwise.

%--------------------------------------
\subsubsection{CME Modelling}

% spheromak CMEs
In this work, we initialise spheromak CMEs at 0.1~AU using the following observation-based parameters recovered from the GCS fitting:
longitude ($\theta_\mathrm{CME}$), 
latitude ($\phi_\mathrm{CME}$), 
and half width ($\omega_\mathrm{CME}/2$, average of the values provided in Table~\ref{tab:gcs}).
Moreover, the speeds of the inserted CMEs are set using 
the CME radial speed $v_\mathrm{CME}^\mathrm{rad}$ derived from the GCS fitting, 
as discussed in detail by \citet{scolini:2019}.
Due to the more limited observational constraints available, 
two additional parameters, the CME mass density and temperature, 
are set to default values ($\rho_\mathrm{CME} = 10^{-18}$~kg~m$^{-3}$ and $T_\mathrm{CME} = 0.8 \times 10^{6}$~K, respectively).
The speed, density and temperature are set to be homogeneous within the CME body during the insertion in the heliosphere.
% magnetic parameters
The three parameters that describe the magnetic structure of spheromak CMEs 
are partially derived from observations.
% chirality
The magnetic chirality is set equal to $-1$ (negative, indicating a left-handed flux rope) for all CMEs, as provided by the low-coronal observations of the source region.
% tilt
Due to the large uncertainties affecting the reconstructed orientation ($\gamma_\mathrm{CME}$) of the CME magnetic structures at $0.1$~AU (i.e.\ because of observational limitations in white-light images, subjectivity in the GCS fitting, as well as strong CME rotations, as discussed in Section~\ref{subsec:cme_magnetic_parameters}), we test several tilt angles $\tau$ for the spheromak configurations in EUHFORIA. Among all, an initial tilt corresponding to a WSE flux rope type for all three CMEs provides the best $B_z$ predictions compared to in situ observations.
% magnetic flux
We set the toroidal magnetic flux $\varphi_t$ of each spheromak CME 
based on the estimated reconnected flux $\varphi_{r}$ derived from statistical and single-event studies (Table~\ref{tab:phir}) \citep[using the same methodology as][]{scolini:2019},
and under the assumption that the reconnected flux only contributes to the poloidal flux of the flux rope \citep[i.e.\ $\varphi_{r} \approx \varphi_{p}$;][]{qiu:2007, moestl:2008, gopalswamy:2017}.
The results (rounded to the closest integer) calculated from the $\varphi_{r}$ estimates are:
$\varphi_{t} = 5 \times 10^{21}$~Mx (CME1), 
$\varphi_{t} = 5 \times 10^{21}$~Mx (CME2),
$\varphi_{t} = 1 \times 10^{22}$~Mx (CME3).

% CME simulations - labelling
We perform a total of 5 simulations, labelled according to the following format: ``XX-XX-XX'' is a generic simulation with ``XX'' being the label of individual CMEs. ``00'' means that a given CME is not modelled, while ``01'' means that the CME was modelled.
We start by performing one simulation (run 00-00-00) without any CMEs inserted, in order to characterise the ambient solar wind through which the three CMEs propagated.
We then perform a set of three runs where we progressively add one CME at a time in our simulations, i.e.\ first modelling only CME1 (run 01-00-00), then including also CME2 (run 01-01-00), and finally adding CME3 (run 01-01-01). This is done to see how the modelling results change by consecutively adding CMEs in the simulations, in order to better isolate the contribution of each CME to the final modelling results and the effect of the CME--CME interactions on the propagation of CME1 and CME2.
Finally, we also perform a simulation with CME3 alone (run 00-00-01), which allows us to compare the propagation of CME3 with or without the presence of the preceding CMEs.
The complete list of simulations is provided in Table~\ref{tab:euhforia_runs}.
%
% Table
\begin{table}
\centering
\begin{tabular}{c|ccc}
\hline
\hline
\textbf{Run number}             & \textbf{CME1} & \textbf{CME2} & \textbf{CME3}\\
\hline
00-00-00     &  --              & --            & --    \\
01-00-00     &  spheromak       & --            & --    \\
01-01-00     &  spheromak       & spheromak     & --    \\
01-01-01     &  spheromak       & spheromak     & spheromak \\
00-00-01     &  --              & --   & spheromak  \\
 \hline
 \hline
\end{tabular}
\caption{Summary of the EUHFORIA simulations performed in this study.}
\label{tab:euhforia_runs}
\end{table}
Full 3D simulations outputs of the whole heliospheric domain are extracted with a 1-hour cadence, while time series of all MHD variables at Earth and selected virtual spacecraft are produced with a 10-minute cadence. 
Similarly to \citet{scolini:2019}, we place in our simulation domain an array of virtual spacecraft in the surroundings of Earth, separated by an angular distance of $\Delta \sigma =  5^\circ$ and $\Delta \sigma = 10^\circ$ (different combinations of longitudes and latitudes are considered), in order to assess the spatial variability of the results in the vicinity of Earth.

%--------------------------------------
\subsection{Geo-effectiveness Predictions and Dst Index}
\label{subsec:dst}

In order to quantify the resulting geo-effectiveness of the CME events as predicted by EUHFORIA, 
we calculate the predicted Dst index from the modelled time series at Earth after conversion into Geocentric Solar Magnetospheric (GSM) coordinates, using the AK2 model proposed by \citet{obrien:2000a, obrien:2000b}. Such predictions are compared with hourly Dst values from the World Data Center for Geomagnetism, Kyoto (\url{http://wdc.kugi.kyoto-u.ac.jp/dstdir/}) and with predictions obtained by applying the same coupling function to 1-min solar wind time series provided by $Wind$.

The quality of the CME geo-effectiveness predictions is then quantified by comparing the minimum Dst predicted by EUHFORIA, with predictions obtained from $Wind$ measurements and with actual observations.

%__________________________________________________________________

\section{Results and Discussion}
\label{sec:results}

In this section we present the results of the simulations performed, discussing the evolution of CME--CME interactions and their impact on the helio- and geo-effectiveness of individual CMEs.

%------------------------------------------------------------------
\subsection{Overview}
\label{subsec:results_overview}

\begin{figure}
\centering
\includegraphics[width=0.98\hsize]{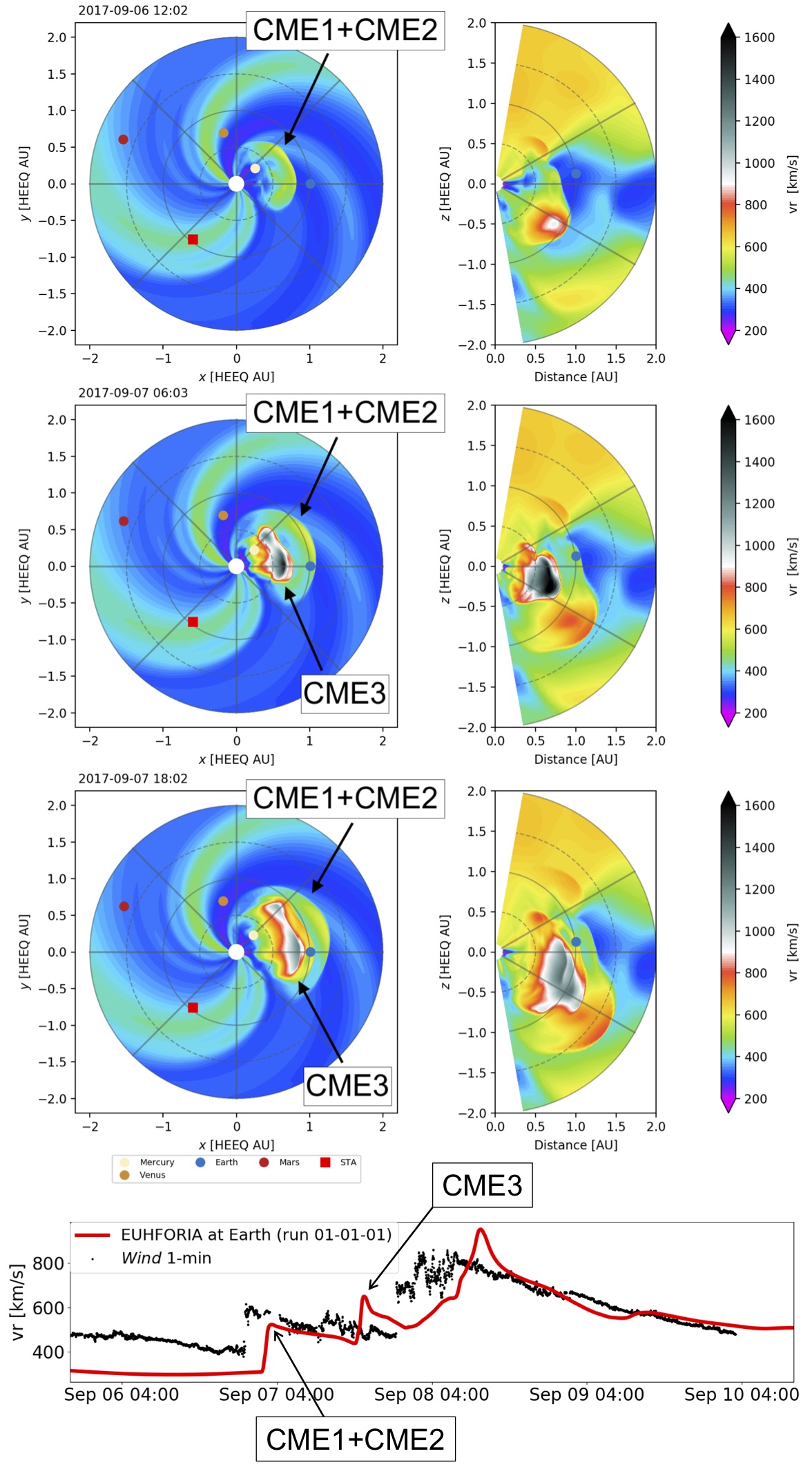} 
\caption{Top: propagation of the three CMEs in EUHFORIA simulations: 
snapshot of the radial speed $v_r$ from run 01-01-01 on 
September 6, 2017 at $\sim$12:00~UT (top),
September 7, 2017 at $\sim$06:00~UT (middle), and 
September 7, 2017 at $\sim$18:00~UT (bottom),
in the heliographic equatorial plane (left) and in the meridional plane (right) that includes Earth (which is indicated by solid blue circles). 
The fronts of CME1, CME2, and CME3 are indicated by the black arrows.
An animation of this figure is available. The animation runs from 00:03~UT on September 2, to 12:03~UT on September 10.
Bottom: comparison of EUHFORIA time series (red) with in situ measurements from $Wind$ (black).
(An animation of this figure is available.)
}
\label{fig:euhforia_vr} 
\end{figure}

\begin{figure*}
\centering
{ 
\includegraphics[width=0.99\hsize]{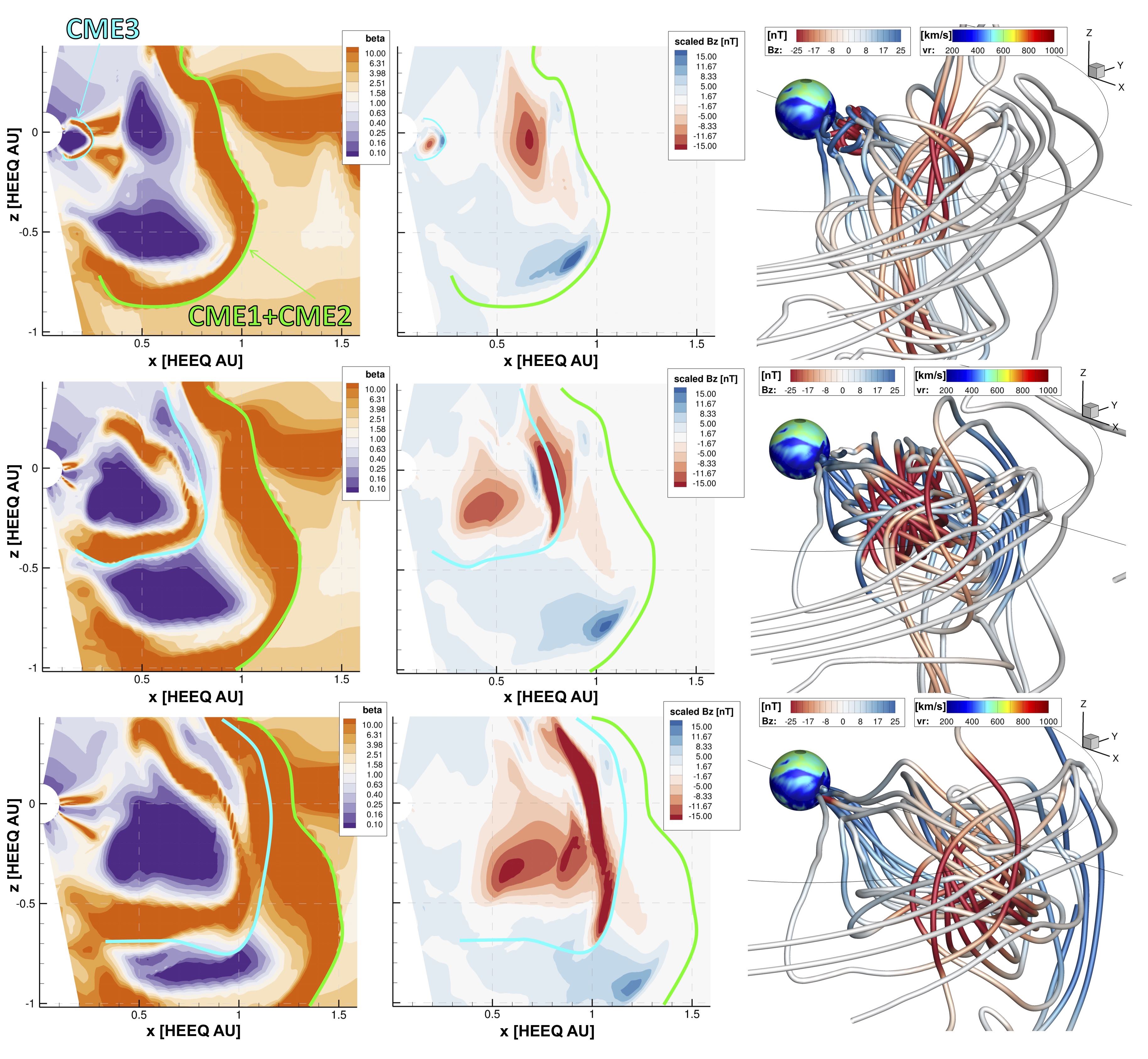}}
\caption{
Interaction between CME1+CME2 and CME3 in EUHFORIA (run 01-01-01) at three different times:
September 6, 2017 at 18:00~UT (top), September 7, 2017 at 06:00~UT (middle), and September 7, 2017 at 18:00~UT (bottom).
Left: 2D view of the meridional plane containing Earth showing the different plasma $\beta$ regions 
(purple: $\beta<1$, marking magnetic ejecta, orange: $\beta>1$, marking shock and sheath regions).
The shocks associated with CME1+CME2 and CME3 are marked in green and cyan, respectively.
Centre: 2D view of the meridional plane containing Earth showing the scaled $B_z$ 
($B_z (D/1~\mathrm{AU})^2$) polarity regions (blue: marking positive, non-geo-effective regions, red: marking negative, helio-effective regions).
Right: 3D view of the magnetic field lines in the heliospheric domain. 
The spherical contour shows the inner boundary ($D=0.1$~AU), coloured based on the radial speed $v_r$. The magnetic field lines are coloured based on $B_z$. The 1~AU distance is marked by the black circle, the Sun-Earth line by the black straight line, and the position of Earth is at the intersection of the two.
}
\label{fig:euhforia_interaction_2D} 
\end{figure*}

An overview of the event in the heliosphere as obtained from EUHFORIA run 01-01-01 is presented in the top panels of Figure~\ref{fig:euhforia_vr}, showing the radial speed ($v_r$) in the ecliptic plane and in the meridional plane containing Earth, at three different times in the simulation (the radial speed, number density and co-latitudinal magnetic field plots for all the runs performed are provided in Appendix~\ref{sec:appendixB}).
The position of the leading edges of the CMEs as marked by their interplanetary shocks at various times are indicated by the black arrows. The bottom panel of Figure~\ref{fig:euhforia_vr} shows the $v_r$ prediction at Earth compared to in-situ observations from $Wind$ (a comparison of the EUHFORIA time series at and around Earth for all runs performed, and in-situ observations from $Wind$, is included in Appendix~\ref{sec:appendixC}).
The evolution and interactions of the CMEs in terms of 3D topology of their magnetic field lines, and of the plasma $\beta$ and $B_z (D/1 \,\, \mathrm{AU})^2$ in the meridional plane containing Earth, are provided in Figure~\ref{fig:euhforia_interaction_2D}. 
The simulation results show that CME1 and CME2 interacted already during the insertion in the heliospheric domain, as expected given the very close insertion times derived from the GCS reconstruction (listed in Table~\ref{tab:euhforia_runs}) and from coronagraph images. 
In the heliosphere, the two CMEs propagate as a merged structure (hereafter CME1+CME2) all the way from 0.1~AU to 2~AU. 
As further discussed in Section~\ref{subsec:results_1AU}, its predicted arrival time at Earth supports the interpretation of S1 as the interplanetary shock driven by the CME1+CME2 merged structure. Figure~\ref{fig:euhforia_vr} also shows CME3 first propagating through the perturbed solar wind in the wake of CME1+CME2, and then interacting with them. The interaction with CME1+CME2 appears to be still ongoing at the time CME3 reaches 1~AU, as indicated by the shocks of CME1+CME2 and CME3 still distinct as predicted at Earth, also supporting the interpretation of S2 being the interplanetary shock driven by CME3.

%------------------------------------------------------------------
\subsection{CME--CME Interactions in the Heliosphere}
\label{subsec:results_interaction_shock}

\begin{figure}
\centering
{ 
\includegraphics[width=0.95\hsize, trim={0mm 10mm 0mm 30mm},clip]{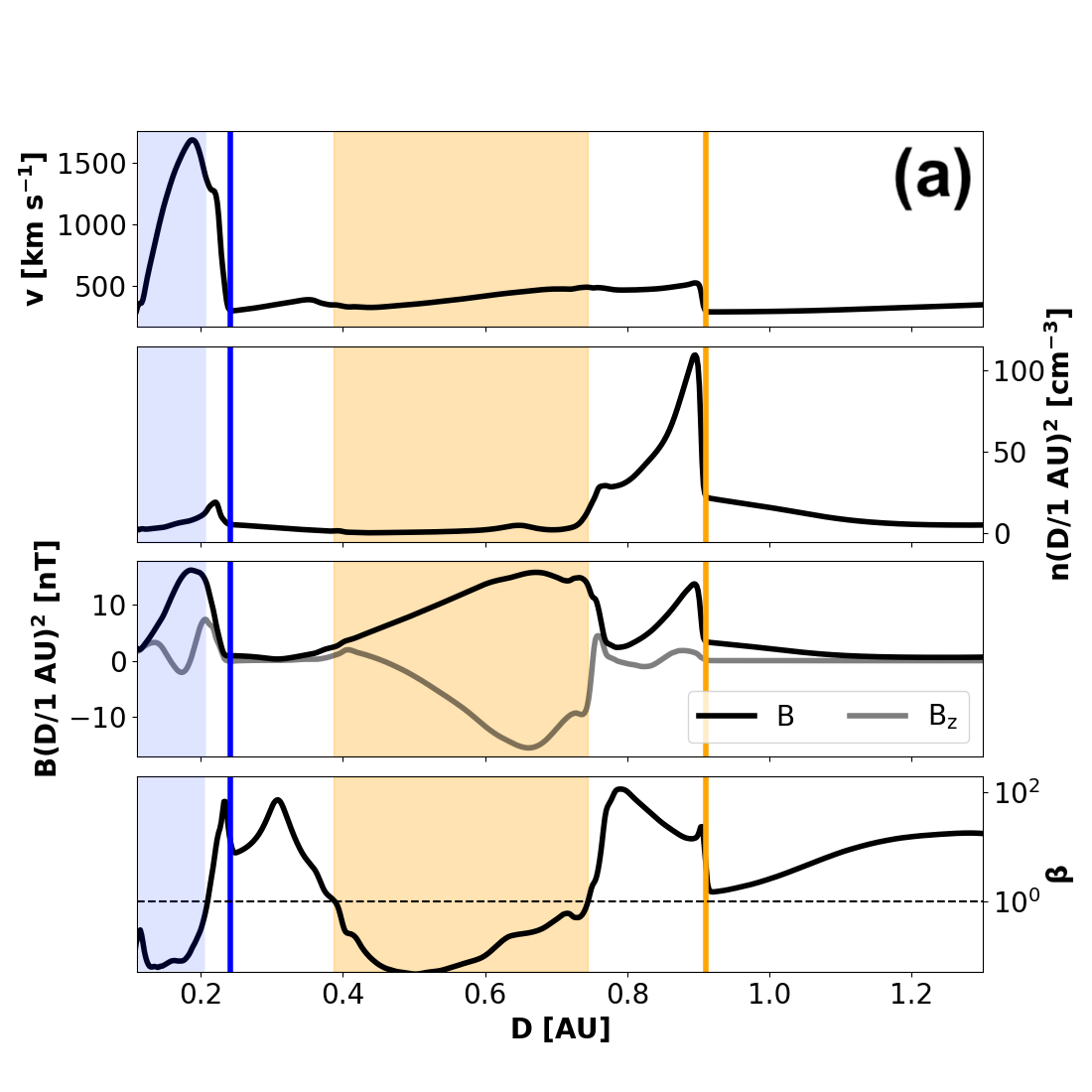} \\
\includegraphics[width=0.95\hsize, trim={0mm 10mm 0mm 30mm},clip]{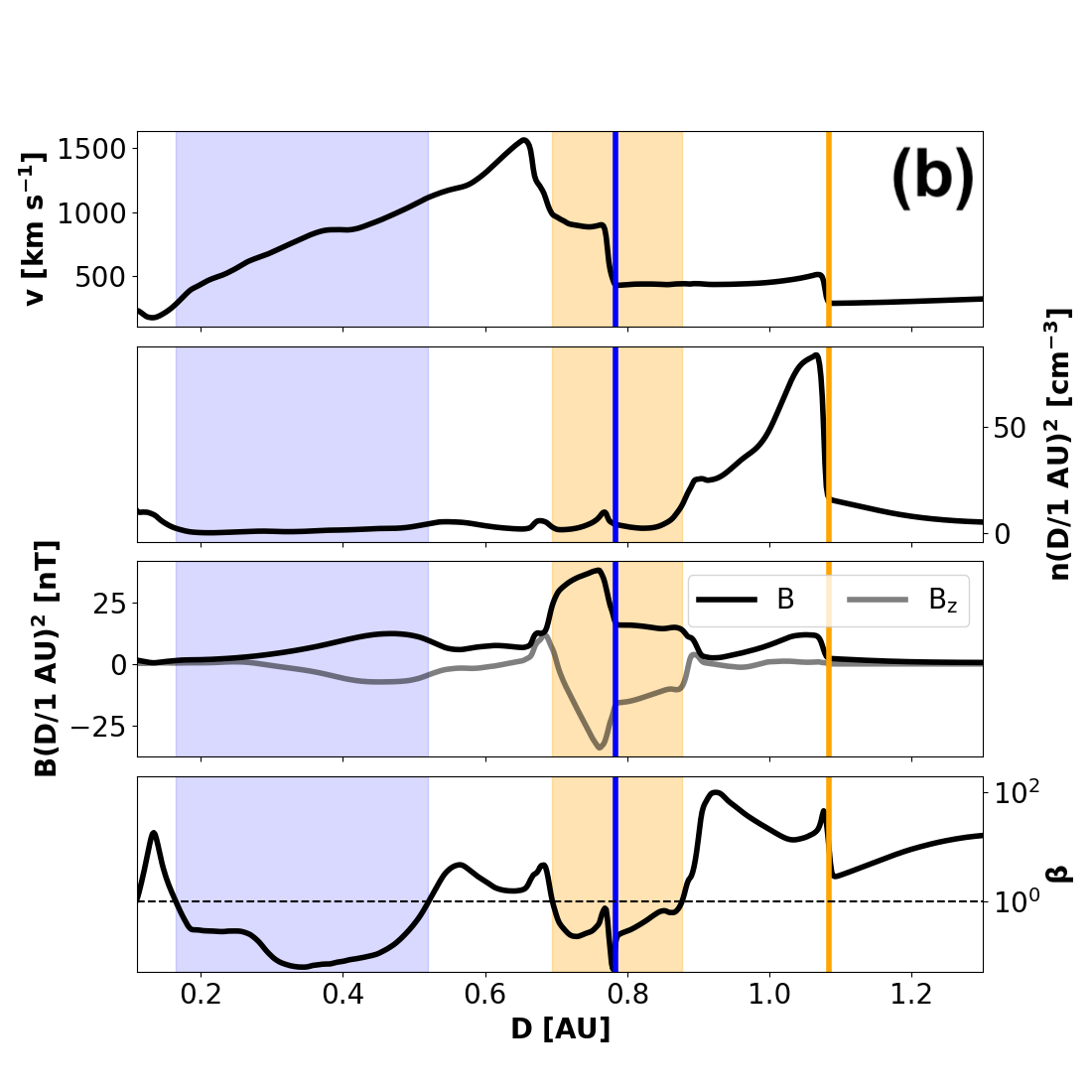} \\ 
\includegraphics[width=0.95\hsize, trim={0mm 10mm 0mm 30mm},clip]{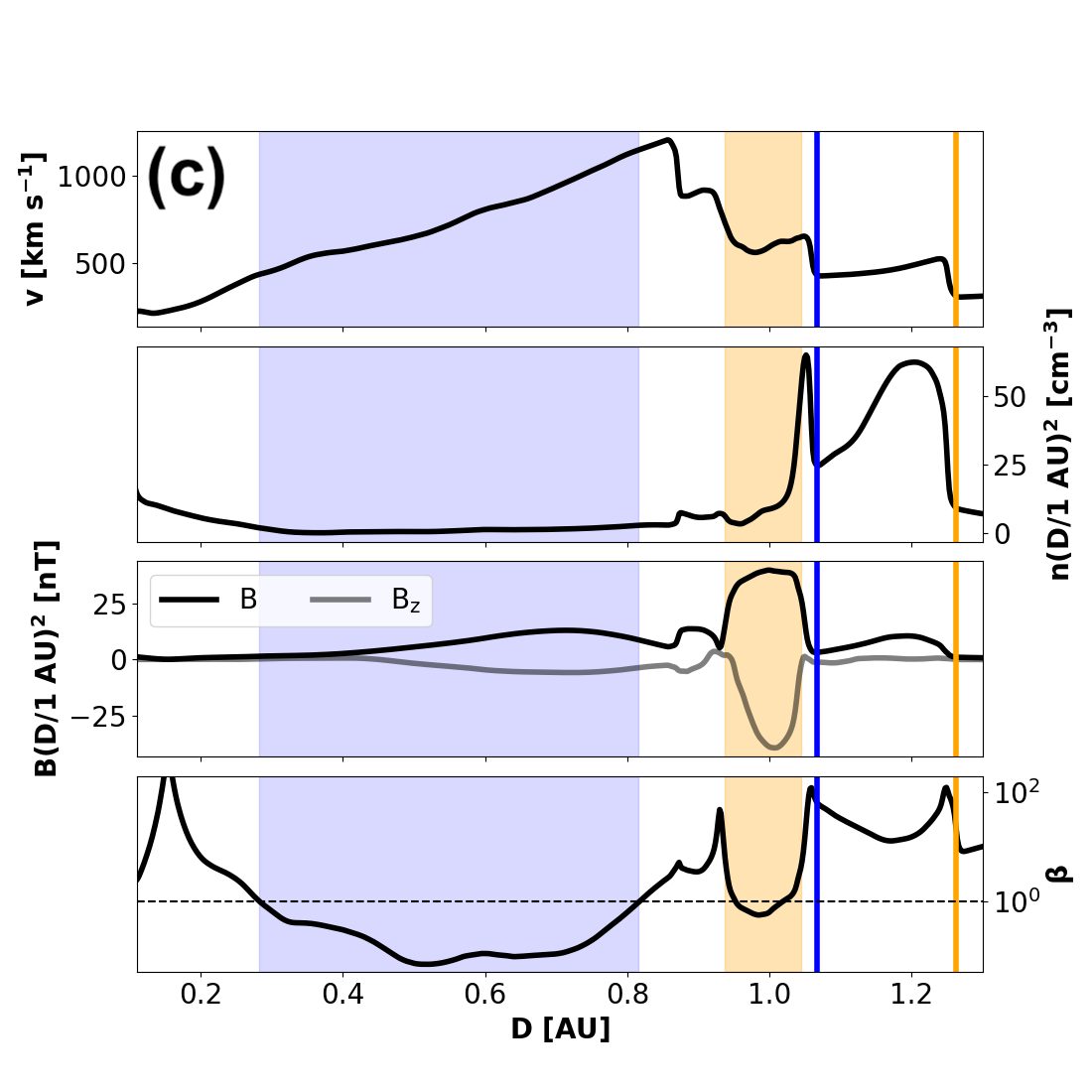}}
\caption{
Interaction along the Sun--Earth line in EUHFORIA run 01-01-01, at the same times as in Figure~\ref{fig:euhforia_interaction_2D}.
From top to bottom: speed ($v$), scaled number density ($n \, (D / 1 \, \mathrm{AU})^2$),
scaled magnetic field magnitude ($B \, (D / 1 \, \mathrm{AU})^2$) and north--south component ($B_z \, (D / 1 \, \mathrm{AU})^2$), and plasma $\beta$. 
The orange and blue vertical lines indicate the location of the shocks driven by CME1+CME2 and CME3.
The orange and blue shaded regions indicate the ejecta associated with CME1+CME2 and CME3.}
\label{fig:euhforia_interaction_phases_1D} 
\end{figure}

Because of the intrinsic difficulties in identifying the boundaries of the various ejecta and related structures in 3D, to characterise the phases of the interaction between CME1+CME2 and CME3 as they propagate in interplanetary space, we apply an approach based on 1D cuts taken along the Sun--Earth line (i.e.\ approximately the direction of propagation of the structures eventually arriving at Earth) at various times in our simulation (run 01-01-01). At each time, we identify the location of the leading and trailing shocks (driven by CME1+CME2 and CME3, respectively) and the boundaries of the two respective ejecta by considering a low-$\beta$ region to correspond to an ejecta. We also characterise the position of each ejecta in terms of their geometrical center.  Figure~\ref{fig:euhforia_interaction_phases_1D} illustrates the main MHD parameters along the Sun--Earth line, together with the location of the various shock and ejecta structures, at three different times in the simulation, which clearly associate with three of the four typical phases of CME--CME interactions \citep[as defined by][]{lugaz:2005a}. 
In particular, Figure~\ref{fig:euhforia_interaction_phases_1D}(a) associates with phase 1, corresponding to a period before the start of the interaction. Figure~\ref{fig:euhforia_interaction_phases_1D}(b) associates with phase 2, corresponding to the shock-ejecta interaction phase, and Figure~\ref{fig:euhforia_interaction_phases_1D}(c) with phase 3, corresponding to the shock-sheath interaction phase. Phase 4, corresponding to the shock-shock interaction phase, is not present in Figure~\ref{fig:euhforia_interaction_phases_1D} as this phase occurs after the CMEs have already left the simulation domain. 
A more detailed investigation of each interaction phase is given in the following paragraphs.

\begin{figure}
\centering
{ 
\includegraphics[width=0.98\hsize]{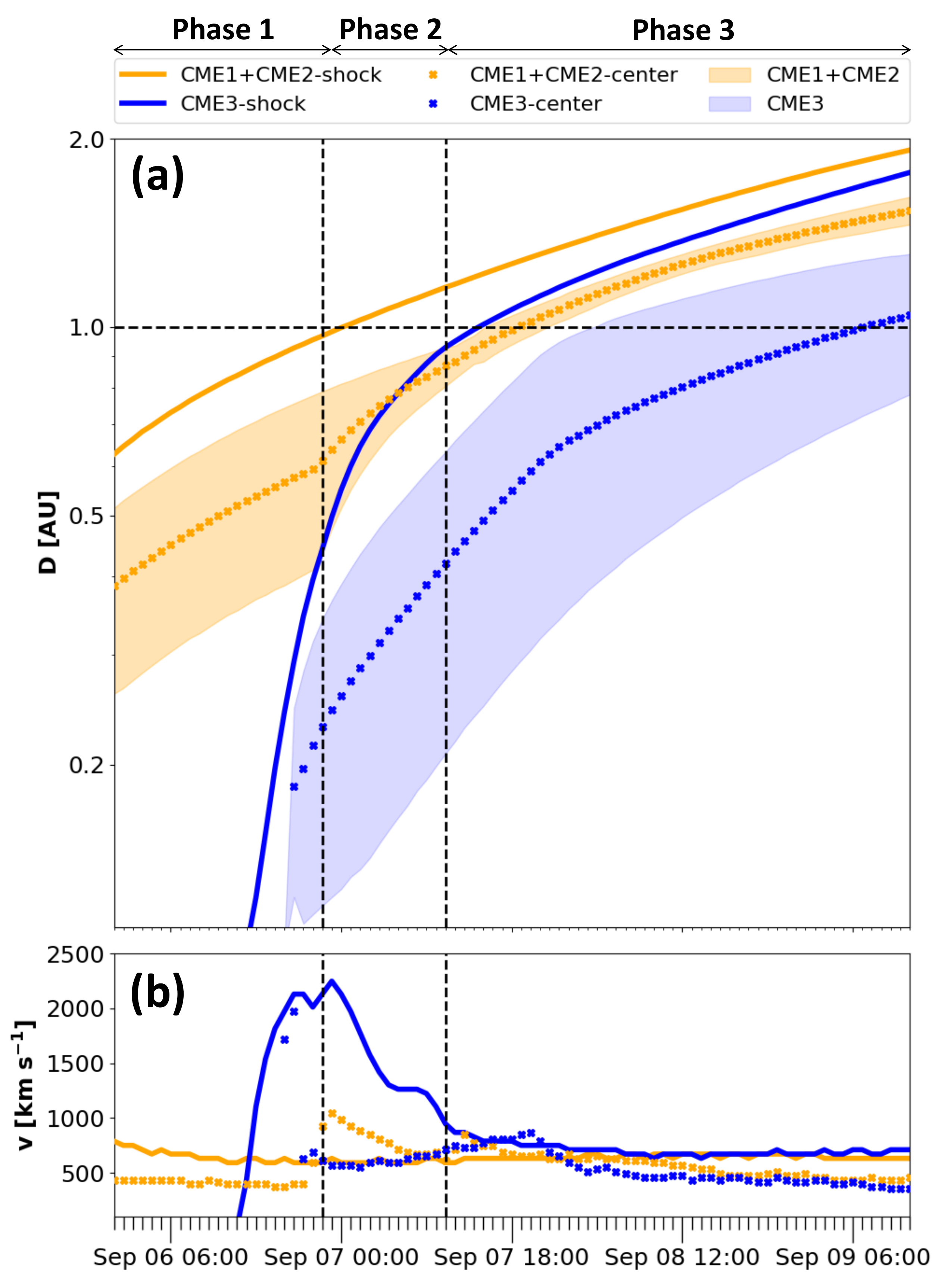}}
\caption{
Propagation of the shocks and ejecta along the Sun--Earth line in EUHFORIA run 01-01-01. 
(a): time--distance plot of the shocks and ejecta associated with CME1+CME2 (in orange) and CME3 (in blue) between 0.1 AU and 2.0 AU in EUHFORIA. The solid lines indicate the location of the shocks. The shaded regions indicate the extension of the ejecta. The crosses mark the geometrical centre of the ejecta. 
The horizontal dashed lines marks the 1~AU distance.
The vertical dashed lines mark the boundaries of the various interaction phases.
(b): time--speed plot for the shocks and ejecta centres.}
\label{fig:euhforia_propagation} 
\end{figure}

We also calculate the speed of the shocks and ejecta centres in the reference frame of the Sun. 
The propagation of each structure in terms of time--distance and time--speed profiles is shown in Figure~\ref{fig:euhforia_propagation}. 

\begin{figure*}
\centering
{\includegraphics[width=0.471\hsize]{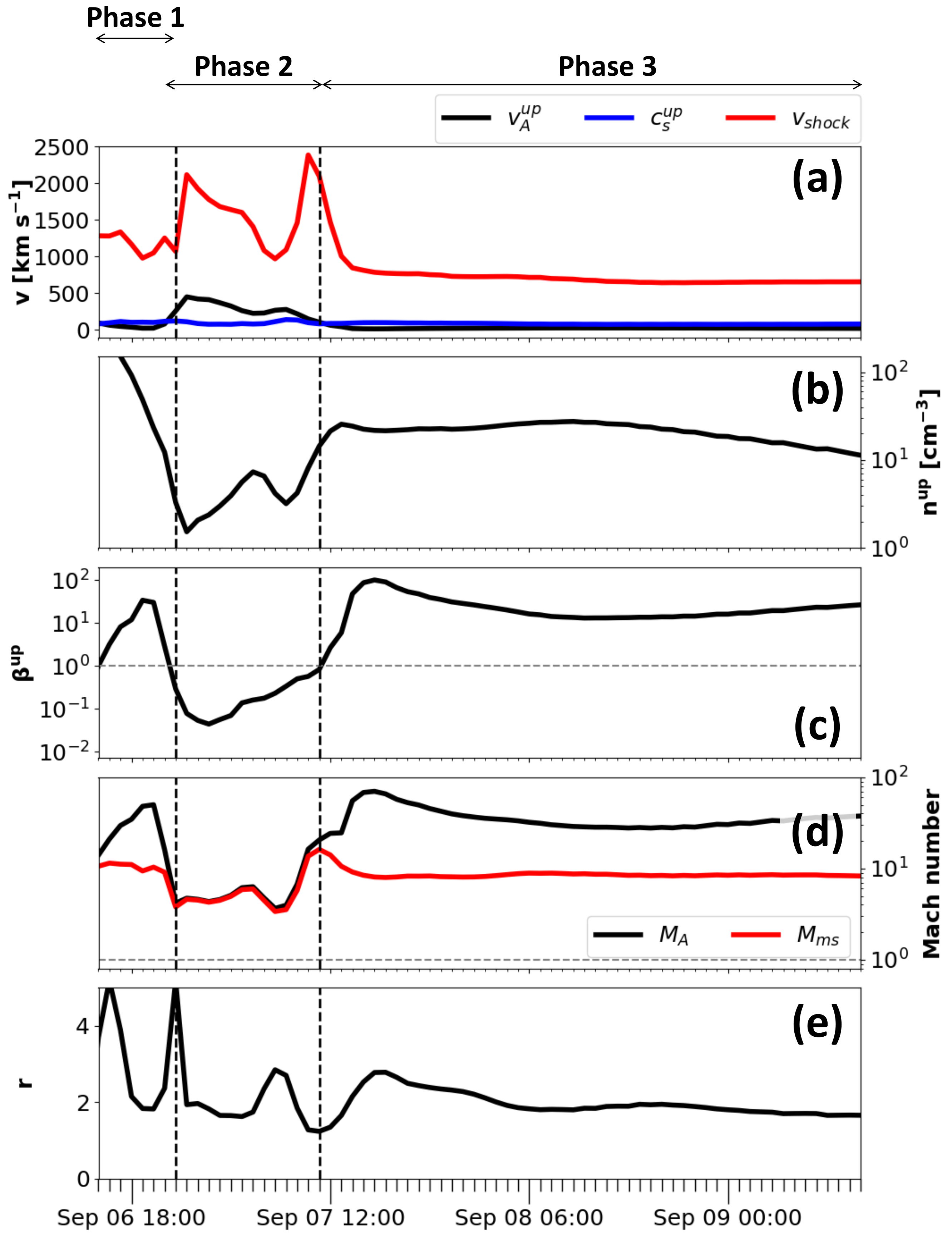}
\includegraphics[width=0.48\hsize]{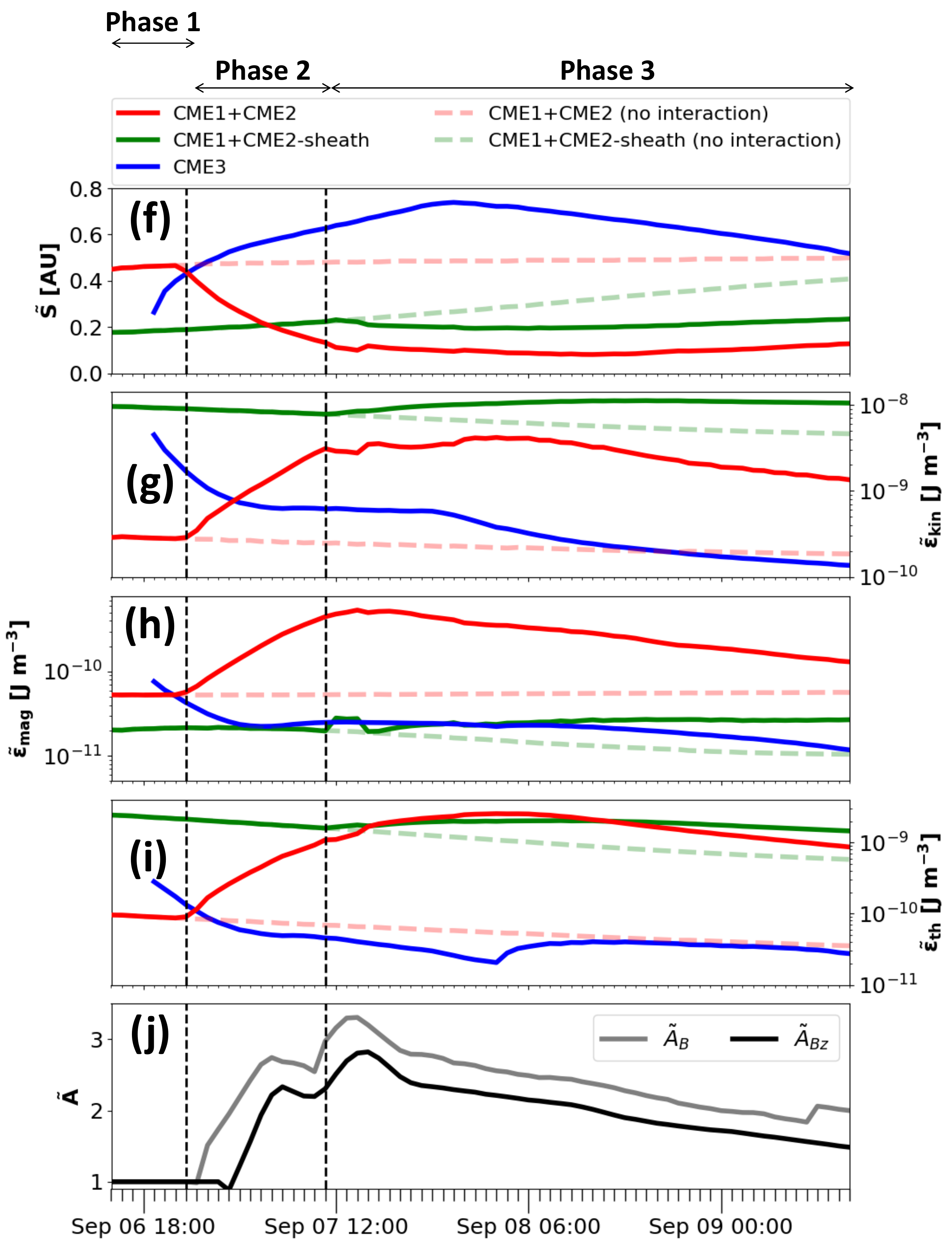}}
\caption{
Left: panels (a)--(e) show the conditions as observed just upstream of the shock driven by CME3 during its propagation in interplanetary space.
Right: panels (f)--(i) show the scaled radial size and the kinetic, magnetic, and thermal energy densities of CME1+CME2 (red), its sheath (green), and CME3 (blue).
Panel (j) shows the geo-effectiveness amplification factors ($\tilde{A}$) of CME1+CME2 in terms of $\tilde{B}$ ($\tilde{A}_B$) and $\tilde{B}_z$ ($\tilde{A}_{Bz}$) due to the interaction with the shock driven by CME3.
The vertical dashed lines mark the boundaries of the various interaction phases.}
\label{fig:euhforia_evolution} 
\end{figure*}

At each time available, we characterise the properties of the trailing (CME3) shock by determining its speed in the reference frame moving with the upstream plasma by applying the Rankine--Hugoniot relations (assuming the 1D cut to be parallel to the shock normal) as: 
\begin{equation}
    v_\mathrm{shock} = \frac{v^\mathrm{down} \, \rho^\mathrm{down} - v^\mathrm{up} \, \rho^\mathrm{up}}{\rho^\mathrm{down} - \rho^\mathrm{up}}
\end{equation}
where ``down'' and ``up'' refer to quantities calculated in the downstream an upstream shock regions, respectively.
In addition to the shock speed, we also calculate the Alfv\'{e}n speed ($v^\mathrm{up}_\mathrm{A}$), sound speed ($c^\mathrm{up}_\mathrm{s}$), plasma $\beta$ ($\beta^\mathrm{up}$) in the upstream region, and the shock Alfv\'{e}n and magnetosonic (fast) Mach numbers ($M_\mathrm{A}$, $M_\mathrm{ms}$), together with the density compression ratio $r = \rho^\mathrm{down} / \rho^\mathrm{up}$.
The evolution in space--time of all these quantities is shown in Figure~\ref{fig:euhforia_evolution}(a)--(e).

To clarify the relation between the evolution of the shock driven by CME3 and the evolution of the two ejecta, we also calculate the average scaled kinetic ($\tilde{\epsilon}_\mathrm{kin}$), magnetic ($\tilde{\epsilon}_\mathrm{mag}$), and thermal ($\tilde{\epsilon}_\mathrm{therm}$) energy densities within the two ejecta as:
\begin{equation}
\begin{cases}
    \tilde{\epsilon}_\mathrm{kin} = \big\langle \frac{1}{2} \tilde{\rho} \tilde{v}^2 \big\rangle \\
    \tilde{\epsilon}_\mathrm{mag} = \big\langle \frac{\tilde{B}^2}{2 \mu_0} \big\rangle \\
    \tilde{\epsilon}_\mathrm{therm} = \big\langle \frac{ 2 \tilde{n} k_B \tilde{T}}{\gamma-1} \big\rangle \\
\end{cases}
\end{equation}
where $\gamma =1.5$ is the adiabatic index \citep[consistent with][]{pomoell:2018}, 
$k_\mathrm{B}$ is the Boltzmann constant, 
$\langle{x}\rangle$ indicates the average along the radial coordinate taken over the ejecta extension calculated above, 
and $\tilde{x}$ indicates a scaled quantity.
Scaled quantities account for the radial evolution of the various CME parameters, and are employed in order to better compare the energy densities at different times in the same run (corresponding to different distances from the Sun) and at same times in different runs. This is needed because the propagation, i.e.\ the radial distance, of the leading ejecta is greatly affected by the presence (run 01-01-01) or lack (run 01-01-00) of the interaction with CME3, and to compare its properties at the same time in the two runs, a correction for the different distance from the Sun is needed.
In particular, we use the following scaling relations:
$\tilde{v} = v$, $\tilde{n} = n \, (D/1 \,\,  \mathrm{AU})^{2.32}$, $\tilde{\rho} = \rho \, (D/1 \,\,  \mathrm{AU})^{2.32}$, $\tilde{B} = B \, (D/1 \,\,  \mathrm{AU})^{1.85}$, $\tilde{T} = T \, (D/1 \,\,  \mathrm{AU})^{0.32}$, $\tilde{P} = P \, (D/1 \,\, \mathrm{AU})^{2.64}$, with the exponents derived from statistical studies of CMEs in the inner heliosphere \citep[from][]{liu:2005, gulisano:2010}. 
Together with the scaled energy densities, we also calculate the scaled radial size of the ejecta as $\tilde{S} = S \, (D/1 \,\, \mathrm{AU})^{0.45}$, where the scaling factor is taken as the lower limit of the range obtained by \citet{gulisano:2010}, which proves to well scale the decrease of the radial size of CME1+CME2 in the case without interaction (dashed red line in Figure~\ref{fig:euhforia_evolution}(f)).
We also apply the same approach to calculate the average energy densities in the region between the leading edge of the CME1+CME2 ejecta and its shock, corresponding to the sheath ahead of CME1+CME2. The evolution in time of the sizes and scaled energy densities associated with these three structures is shown in Figure~\ref{fig:euhforia_evolution}(f)--(i).

Finally, we calculate the amplification of the helio-effectiveness of CME1+CME2 due to its interaction with CME3 in terms of the magnetic field compression caused by the propagation of the shock driven by CME3 through the CME1+CME2 ejecta. We therefore compute the maximum $\tilde{B}$ ($\tilde{B}^\mathrm{max}$) and minimum $\tilde{B_z}$ ($\tilde{B}_z^\mathrm{min}$) within the boundaries of the ejecta, where $\tilde{B}_z = B_z \, (D/1 \,\, \mathrm{AU})^{1.85}$ assuming that the magnetic field components in a CME scale with the same behaviour as that of the magnitude $B$. The helio-effectiveness amplification factors ($\tilde{A}$) due to the interaction are then calculated as 
$\tilde{A}_B = \tilde{B}^\mathrm{max}_{010101}/\tilde{B}_{010100}^\mathrm{max}$ and 
$\tilde{A}_{Bz} =\tilde{B}_{z,010101}^\mathrm{min}/\tilde{B}_{z,010100}^\mathrm{min}$, i.e.\ taking the ratio of the values from run 01-01-01 and run 01-01-00. Results are shown in Figure~\ref{fig:euhforia_evolution}(j).

\paragraph{Phase 1: Pre-interaction} %%%%%%%%%%%%%%%%%%%%%%%%%%%%%%%%%%%%%%
In Figure~\ref{fig:euhforia_interaction_phases_1D}(a), the shock and ejecta associated with CME3 propagate through a high-$\beta$ ($\beta \sim 10$) solar wind perturbed by the earlier passage of the preceding CME1+CME2 structure. The rear edge of CME1+CME2 is still unaffected by the presence of CME3, indicating no direct interaction has yet occurred. This is also visible from the time-distance profile in Figure~\ref{fig:euhforia_propagation}(a). At this time the shock driven by CME3 is propagating with a speed of ${\sim}2130$~km~s$^{-1}$ and its ejecta with a speed of ${\sim}650$~km~s$^{-1}$, significantly higher than the shock and ejecta associated with CME1+CME2 (both moving at ${\sim}600$~km~s$^{-1}$). CME3 is progressively approaching CME1+CME2, as is clearly visible in Figure~\ref{fig:euhforia_propagation}(a).

During this phase, the scaled energy densities of CME1+CME2 are approximately constant in space--time (Figure~\ref{fig:euhforia_evolution}(g)--(i)), implying that the (non-scaled) energy densities are decreasing with radial distance due to the known interplay of expansion (which converts magnetic and thermal energy into kinetic energy), and drag (which slows down the CME reducing its kinetic energy) \citep{cargill:2004, vrsnak:2013}. 
Between 18:00~UT and 21:00~UT on September 6, CME3 is still being inserted in the domain, and it expands (Figure~\ref{fig:euhforia_propagation}(a), Figure~\ref{fig:euhforia_evolution}(f)) due to the propagation through the rarefied perturbed solar wind \citep{temmer:2015, temmer:2017, liu:2014b}, as indicated by its rapidly growing radial size and the rapid decrease of $\tilde{\epsilon}_\mathrm{kin}$, $\tilde{\epsilon}_\mathrm{therm}$, and $\tilde{\epsilon}_\mathrm{mag}$ (Figure~\ref{fig:euhforia_evolution}(g)--(i)).

\paragraph{Phase 2: Shock--Ejecta Interaction} %%%%%%%%%%%%%%%%%%%%%%%%%%%%%%
On September 6 at 22:00~UT, around $0.45$~AU, the shock driven by CME3 starts interacting with the preceding CME1+CME2 ejecta (see Figure~\ref{fig:euhforia_interaction_phases_1D}(b) and Figure~\ref{fig:euhforia_propagation}(a)). 
Figure~\ref{fig:euhforia_evolution}(a)--(e) shows that as the shock enters the preceding ejecta, the upstream plasma $\beta$ decreases by a factor of 100, the upstream density decreases by a factor of 8, the Alfv\'{e}n speed increases by a factor of 12, while the sound speed remains almost the same. The reduced density and higher magnetic field contribute to increase $v_\mathrm{A}$ and to lower $M_\mathrm{A}$, hence the shock quickly accelerates from a speed of ${\sim}1300$~km~s$^{-1}$ to a speed of ${\sim}2120$~km~s$^{-1}$ on September 6 at 23:00~UT while the density compression ratio of the shock decreases from $r \simeq 5$ (with 5 equal to the theoretical maximum for ideal MHD with $\gamma = 1.5$) to a value of about 1.9. 
After reaching the speed of ${\sim}2120$~km~s$^{-1}$ , the shock decelerates to a speed of ${\sim}970$~km~s$^{-1}$ when reaching the core (densest part) of the CME1+CME2 ejecta.
The compression ratio is found to increase again to 2.9 a few hours later.
The shock remains at all times a fast shock, with a minimum $M_\mathrm{ms}$ of 4.3 in the frame moving with the upstream plasma.
Following the passage through the denser ejecta core, the shock propagates through a rarefied region where it accelerates up to ${\sim}2390$~km~s$^{-1}$ on September 7 at 10:00~UT, right before exiting the leading edge of the ejecta.
 
During this phase, the shock--ejecta interaction is most efficient in amplifying the helio-effectiveness of the CME1+CME2 ejecta, as visible in the $\tilde{A}_B$ and $\tilde{A}_{Bz}$ amplification factors in Figure~\ref{fig:euhforia_evolution}(j). The shock driven by CME3 compresses, accelerates, and heats the plasma in the preceding ejecta as it propagates through it, enhancing the density, speed, temperature and magnetic field in the downstream region. This is visible as an increase in $\tilde{\epsilon}_\mathrm{kin}$, $\tilde{\epsilon}_\mathrm{mag}$, and $\tilde{\epsilon}_\mathrm{therm}$ of CME1+CME2 compared to the simulation without interaction (i.e.\ red continuous and dashed lines in Figure~\ref{fig:euhforia_evolution}(g)--(i)). The acceleration of CME1+CME2 is also clearly visible in the increased speed of the ejecta centre in Figure~\ref{fig:euhforia_propagation}(b). The radial size of CME1+CME2 also rapidly decreases (Figure~\ref{fig:euhforia_propagation}(a) and Figure~\ref{fig:euhforia_evolution}(f)), consistent with a radial compression by the trailing shock and ejecta \citep{vandas:1997, schmidt:2004, lugaz:2005a, xiong:2006a}. During this phase we observe a steady increase of the $\tilde{A}_B$ and $\tilde{A}_{Bz}$ amplification factors (``growth phase''), until maximum amplification factors of 3.3 for $\tilde{A}_B$ and 2.8 for $\tilde{A}_{Bz}$ are reached around September 7 at 14:00~UT, i.e.\ after the end of this phase (``maximum phase''). The maximum helio-effectiveness amplification occurs when the CME1+CME2 ejecta is around 0.9~AU in simulation 01-01-01, i.e.\ close to the location of Earth.

\paragraph{Phase 3: Shock--Sheath Interaction} %%%%%%%%%%%%%%%%%%%%%%%%%%%%%%
Eventually, the shock driven by CME3 reaches the front of the CME1+CME2 ejecta and starts interacting with the dense sheath of plasma ahead of it (Figure~\ref{fig:euhforia_interaction_phases_1D}(c) and Figure~\ref{fig:euhforia_propagation}(a)). In our simulations, this occurs around September 7 at 11:00~UT, i.e.\ close to the moment when it passes Earth/1~AU.
We note that this phase starts at smaller heliocentric distances than in the observed in-situ data by the spacecraft at L1, where the shock appears still to be fully embedded in the ejecta. 
This is most probably due to a combination of two factors:
(1) an over-estimated drag in our simulations (lower speed and higher density in the solar wind ahead of CME1+CME2 compared to observations, see Appendix~\ref{sec:appendixC}) --- this particularly contributed to slowing down CME1+CME2 (postponing its arrival time), 
and (2) uncertainties in the CME initial speeds based on the GCS and $\varphi_r$ reconstructions, which may have contributed to predicting an early arrival time of the shock of CME3 compared to in-situ observations (bottom panel in Figure~\ref{fig:euhforia_vr}).
During this phase the shock driven by CME3 enters again a high~$\beta$ ($\beta > 50$) environment, characterised by a low Alfv\'{e}n speed ($\sim$15~km~s$^{-1}$) and a sound speed comparable to that in the ejecta ($\sim$95~km~s$^{-1}$). The density in the sheath is about 2 orders of magnitudes higher than in the ejecta.
As the shock enters this new region, its speed with respect to the upstream plasma drops below $850$~km~s$^{-1}$, due to the increased upstream density. Its radial speed is still slightly larger than the speed of the leading shock driven by CME1+CME2 ($\sim$ 800~km~s$^{-1}$ compared to $\sim$630~km~s$^{-1}$), so the CME3-driven shock continues to get closer to the shock ahead, but at a slower rate. We expect them eventually to merge, but the rate of approach is so low that this would take place only beyond the outer boundary of the simulation domain (i.e.\ beyond 2~AU).

Analysing the energy of each substructure in this phase, we observe that the CME3 shock starts heating and compressing the CME1+CME2 sheath ahead, as indicated by the decrease in its radial size and by the increase in the associated $\tilde{\epsilon}_\mathrm{kin}$, $\tilde{\epsilon}_\mathrm{mag}$, and $\tilde{\epsilon}_\mathrm{therm}$ compared to the simulation without interaction (green continuous and dashed lines in Figure~\ref{fig:euhforia_evolution}(f)--(i)). 
At the same time, the scaled radial size of CME1+CME2 (CME3) slowly decreases (increases) until September 7 at 23:00~UT, when the trends invert.
Most notably, in this phase we also observe that the $\tilde{A}_B$ and $\tilde{A}_{Bz}$ amplification factors follow a decreasing trend starting from September 7 at 14:00~UT. 
We explain this by noting that while $\tilde{\epsilon}_\mathrm{mag}$ in CME1+CME2 starts decreasing from September 7 at 14:00~UT, its $\tilde{\epsilon}_\mathrm{kin}$ and $\tilde{\epsilon}_\mathrm{therm}$ continue increasing until September 8 at 03:00~UT. The earlier peak of $\tilde{\epsilon}_\mathrm{mag}$ compared to those of $\tilde{\epsilon}_\mathrm{kin}$ and $\tilde{\epsilon}_\mathrm{therm}$ suggests a conversion of the magnetic energy accumulated by CME1+CME2 during the shock--ejecta interaction phase, into kinetic and thermal energy of the same structure. The acceleration of CME1+CME2 due to this energy conversion is particularly visible in Figure~\ref{fig:euhforia_evolution}(b). During this energy conversion phase, the radial size of CME1+CME2 slightly increases as indicated by the constant scaled radial size (Figure~\ref{fig:euhforia_evolution}(g)), most probably due to the presence of CME3 at the back of CME1+CME2 preventing a further expansion. At the end of this phase CME1+CME2 and CME3 move at almost the same speed ($\sim$670~km~s$^{-1}$, Figure~\ref{fig:euhforia_evolution}(b)). The low expansion of CME1+CME2 is therefore consistent with numerical and observational studies \citep{xiong:2006a, xiong:2007, gulisano:2010, lugaz:2012}, which highlighted how the expanding behaviour of the leading ejecta after the end of the main interaction phase depends on the delicate interplay between the natural tendency for the CME to expand and the compressing action of the trailing ejecta.

The most relevant consequence of the expansion and conversion of magnetic energy into acceleration and heating, is that CME1+CME2 progressively ``forgets'' the amplification of $B_z$ caused by the interaction with CME3, and it slowly returns to $B_z$ conditions similar to the case without interaction (``decay phase'').
This suggests that one of the key factors at the origin of the intense geomagnetic storm triggered by the September 4--6, 2017 CMEs was their arrival at Earth during the phase of maximum helio-effectiveness amplification due to the interaction of CME1+CME2 with CME3.

%------------------------------------------------------------------
\subsection{Effect of the Interactions at 1~AU}
\label{subsec:results_1AU}

\subsubsection{$B_z$ and Arrival Time Predictions} %%%%%%%%%%%%%%%%%%%%%%%%%%%

Together with the intensity and orientation of their internal magnetic field at Earth \citep{savani:2015, savani:2017, palmerio:2018}, another key CME parameter that the community has long tried to predict is the arrival time \citep{moestl:2014, riley:2018, verbeke:2019a}. Current estimates of prediction uncertainties for this parameter are about $\pm 10$~hours depending on the exact metric considered and on the model used \citep{riley:2018, wold:2018}, with numbers increasing for more complex events. In the context of interacting CMEs, the arrival time of individual CMEs can be significantly affected by two different effects impacting the CME kinematics: (1) the preconditioning of the ambient solar wind due to the passage of previous CMEs \citep{temmer:2017}, and (2) direct CME--CME interactions \citep[\textit{CME--CME collisions}; see e.g.][]{liu:2014a, shen:2017}. 
Given the relevance of CME--CME interactions in affecting the kinematics of individual CMEs and, consequently, the prediction of their arrival times at Earth, we address here how much the $B_z$ and arrival times at Earth of the shocks and ejecta associated with CME1+CME2 and CME3 were affected by their interaction. 
To do so, we start from the identification of the shocks and ejecta boundaries performed in Section~\ref{subsec:results_interaction_shock}, then we extract and compare times at which the shock and leading edge of the ejecta associated with CME3 arrived at 1~AU, in simulations with (run 01-01-01) and without (run 00-00-01) the presence of the preceding CMEs. To assess the impact of the interaction on the arrival time of CME1+CME2, we similarly extract and compare the position in time of the shock and leading edge of the ejecta associated with CME1+CME2 obtained from run 01-01-01 and run 01-01-00.

\begin{figure*}
\centering
{\includegraphics[width=0.55\hsize]{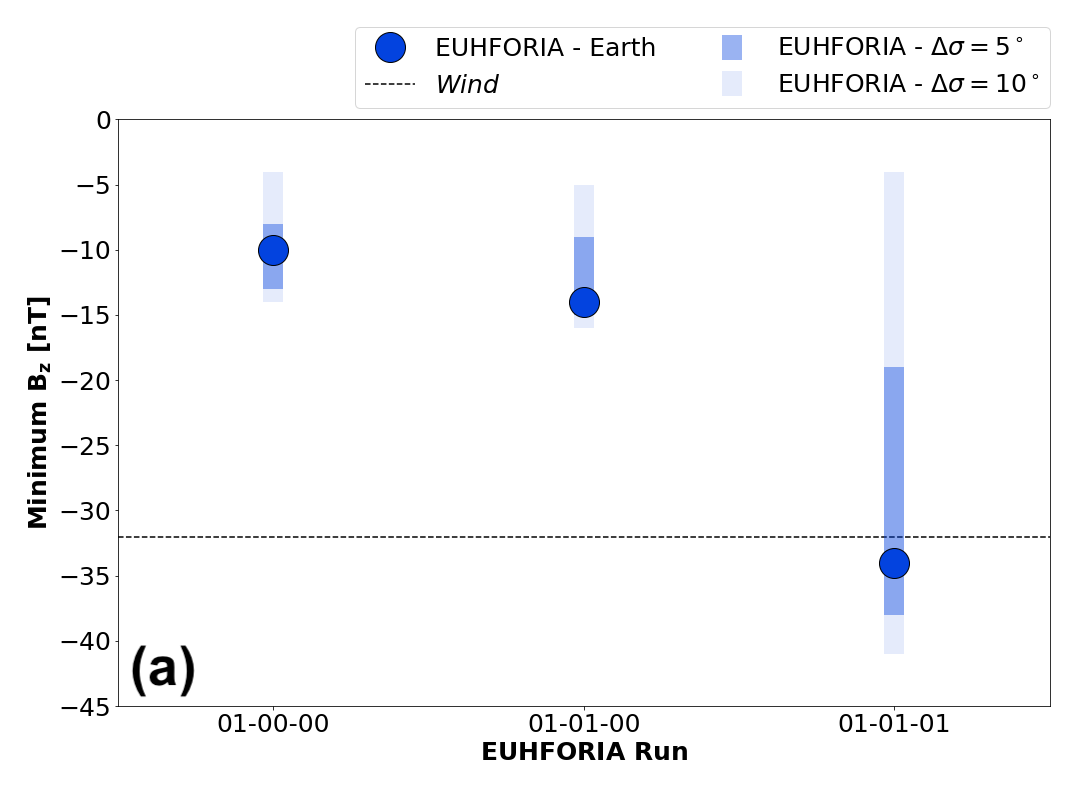} 
\includegraphics[width=0.44\hsize]{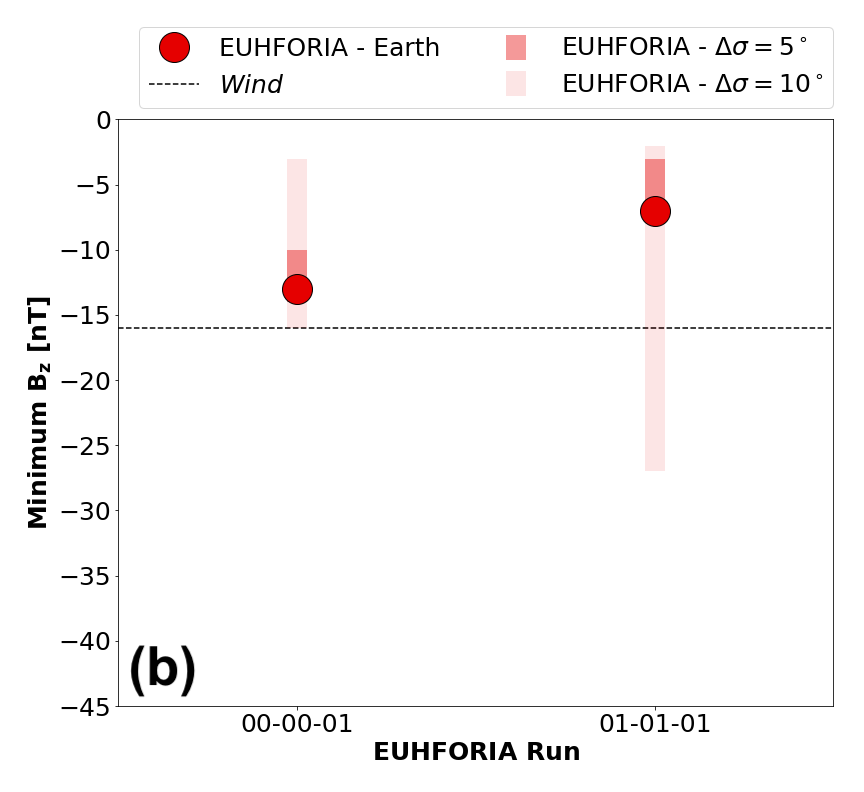}}
\caption{Scatter plots summarising the minimum $B_z$ predicted from the various EUHFORIA runs for CME1+CME2 (a) and CME3 (b), compared to $Wind$ in-situ measurements of the minimum $B_z$ associated with E1 (a) and E2 (b) (black dashed lines). Predictions at Earth are indicated with coloured dots, while predictions at virtual spacecraft separated by $\Delta \sigma = 5^\circ$ and $\Delta \sigma = 10^\circ$ from Earth are indicated as coloured bars. }
\label{fig:euhforia_Bz_scatter} 
\end{figure*}

% ejecta E1 ===================================
\paragraph{Effect on CME1+CME2}
In Figure~\ref{fig:euhforia_Bz_scatter}(a) we plot the predicted minimum $B_z$ associated with CME1+CME2 from the EUHFORIA time series at Earth and the surrounding virtual spacecraft, together with the observed minimum $B_z$ associated with E1 (see Figure~\ref{fig:insitu}). 
We observe that the simulation of CME1 alone (run 01-00-00) predicts a modest minimum $B_z$ of $-10$~nT ($-13$~nT to $-8$~nT considering spacecraft separated by an angular distance $\Delta \sigma = 5^\circ$ from Earth, and $-14$~nT to $-4$~nT considering spacecraft separated by $\Delta \sigma = 10^\circ$). While the strength of the magnetic field within CME1 at Earth is similar to the unshocked region in E1, due to its relatively low initial speed compared to CME2 (Table~\ref{tab:gcs}), the shock driven by CME1 without the inclusion of CME2 in our simulations is predicted to arrive at Earth on September 7 at 09:33~UT, i.e.\ about 10 hours later than the observed S1.
% 01-01-00
When adding CME2 to the simulations (run 01-01-00), the early interaction of CME1 and CME2 results in a merged structure that propagates in the heliosphere with a speed close to the speed of the fastest CME involved, i.e.\ CME2, arriving at Earth on September 7 at 02:23~UT, i.e.\ only 3 hours later than the actual arrival time of S1. The predicted minimum $B_z$ of this combined structure is $-14$~nT ($-15$~nT to $-9$~nT considering spacecraft separated by $\Delta \sigma = 5^\circ$ from Earth, and $-16$~nT to $-5$~nT considering spacecraft separated by $\Delta \sigma = 10^\circ$), less than $5$~nT lower than the unshocked (upstream) region observed in E1. 
The start of the ejecta associated with CME1+CME2 is predicted on September 7 at 23:13~UT, only 3 hours later than the starting time of ejecta E1 based on $Wind$ in-situ observations.
% 01-01-01
When including all three CMEs (run 01-01-01), the predicted minimum $B_z$ of CME1+CME2 at Earth drops to $-35$~nT, i.e.\ very close to the minimum observed value of $-32$~nT. When accounting for uncertainties related to the initial CME directions of propagation, the predicted value varies between $-38$~nT and $-19$~nT for a $\Delta \sigma = 5^\circ$ separation from Earth, and between $-41$~nT and $-4$~nT for a $\Delta \sigma = 10^\circ$ separation from Earth.
Overall, we also observe a good level of agreement between the predicted and the observed $B$ and $B_z$ time series at and around Earth (see Appendix~\ref{sec:appendixC}).
% arrival times
While the predicted arrival time of the shock driven by CME1+CME2 is unaffected by the inclusion of CME3 in our simulations, the ejecta associated with CME1+CME2 arrives about 5 hours earlier, i.e. on September 7 at 17:43~UT. The ejecta is therefore predicted to arrive about 2 hours earlier than the starting time of ejecta E1 based on $Wind$ in-situ observations. Overall, the very close match between the modelled and observed arrival times and minimum $B_z$ in run 01-01-01 provides strong evidence that S1 and E1 were indeed the interplanetary counterparts of CME1+CME2.

% conclusions
By comparing the minimum $B_z$ prediction from runs 01-01-00 and 01-01-01 at and around Earth, we conclude that the presence of S2 and E2 contributed to an increase in the minimum $B_z$ associated with E1 by a factor of 2.5 (ranging between 2.1 and 2.5 for a $\Delta \sigma = 5^\circ$ separation from Earth, and between 0.8 and 2.6 for a $\Delta \sigma = 10^\circ$ separation from Earth), a value consistent with the results obtained by \citet{shen:2018} based on observational arguments and with the analysis in Section~\ref{subsec:results_interaction_shock}.

% ejecta E2 ===================================
\paragraph{Effect on CME3}

For completeness, in Figure~\ref{fig:euhforia_Bz_scatter}(b) we plot the predicted minimum $B_z$ at and around Earth associated with CME3 in a similar way as already done for CME1+CME2. Simulation results at Earth from run 00-00-01 are consistent with the observed minimum $B_z$ within E2 ($-13$~nT compared to $-16$~nT), while inclusion of the preceding CMEs predicts a minimum $B_z$ of $-6$~nT. By considering virtual spacecraft in the vicinity of Earth, however, we observe that predictions vary between $-6$~nT and $-3$~nT for $\Delta \sigma = 5^\circ$, and between $-27$~nT and $-2$~nT for $\Delta \sigma = 10^\circ$. Therefore, although run 00-00-01 gives a slightly better prediction for the minimum $B_z$ within E2 at Earth, results in the vicinty of Earth from both simulations are overall consistent with the value measured in situ. The larger spread in the minimum $B_z$ in the vicinity of Earth predicted in run 01-01-01 compared to run 00-00-01 may reflect the development of finer structures within CME3, or potential deflections in its trajectory, as a consequence of its interaction with CME1+CME2 (or a combination of the two).

Considering its arrival time at Earth, run 00-00-01 predicts the shock of CME3 to arrive at Earth on September 8 at 08:23~UT, followed by the ejecta starting on September 9 at 01:43~UT. As expected, the inclusion of CME1+CME2 ahead of CME3 (run 01-01-01) caused an earlier arrival of the shock, i.e. on September 7 at 17:03~UT, about 15~hours earlier, and of the ejecta, i.e. on to September 8 at 09:43~UT, about 16~hours earlier, because of the solar wind preconditioning induced by CME1+CME2 (which also contributed to the early start of the interaction).
These final predictions match well the observed arrival time of S2 (September 7 at 22:38~UT, about 6 hours later than predicted) and the starting time of E2 (September 8 at 11:00~UT, about 1 hour later than predicted) at $Wind$, supporting their association with CME3 at the Sun. At the same time, we note how the predicted long duration of the ejecta associated with CME3 in run 01-01-01 suggests that it was associated with both the E2 and E3 (see Figure~\ref{fig:insitu}), supporting the interpretation that they were indeed both associated with the same CME at the Sun.

% cone model
%For comparison, the predicted minimum $B_z$ obtained using the cone CME model \citep{scolini:2018b} is approximately the same for all runs and it never goes below $-5$~nT as consequence of the lack of internal magnetic field structure in the CMEs. 

\subsubsection{Dst Predictions and Geo-effectiveness} %%%%%%%%%%%%%%%%%%%%%%%%%%%

\begin{figure*}
\centering
{\includegraphics[width=0.55\hsize]{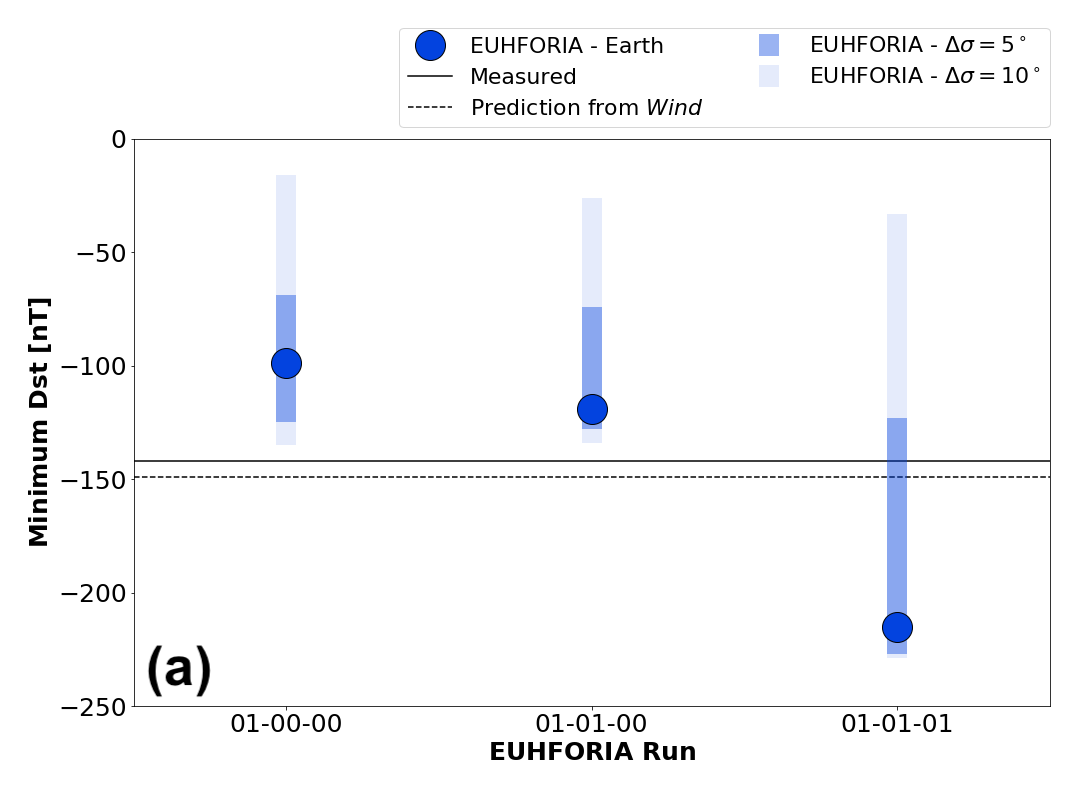} 
\includegraphics[width=0.44\hsize]{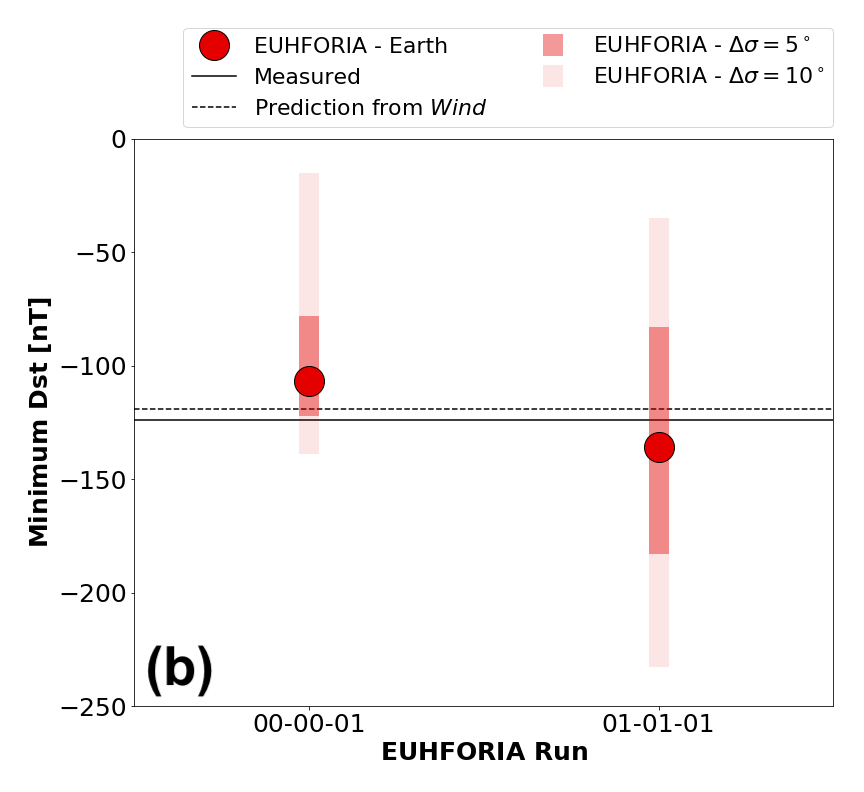}}
\caption{Scatter plots summarising the minimum Dst predicted from the EUHFORIA runs at Earth for CME1+CME2 (a) and CME3 (b), compared to predictions from $Wind$ in-situ measurements associated with E1 and E2 (black dashed lines), and the minimum Dst measured at Earth (black solid lines). EUHFORIA predictions at Earth are indicated with coloured dots, while predictions at virtual spacecraft separated by $\Delta \sigma = 5^\circ$ and $\Delta \sigma = 10^\circ$ from Earth are indicated as coloured bars.}
\label{fig:euhforia_Dst_scatter} 
\end{figure*}

To quantify the effect of CME--CME interactions on the actual impact of the merged CME1+CME2 structure on Earth (i.e. its geo-effectiveness), we apply the method presented in Section~\ref{subsec:dst} to the EUHFORIA time series extracted at and around the location of Earth. To better quantify by how much EUHFORIA-based Dst predictions are off due to mispredictions of the CME impact parameters, and how much of the discrepancy is actually due to limitations of the specific coupling function used, we also apply the coupling function to $Wind$ in-situ measurements. As previously done for the minimum $B_z$, we separately discuss the predictions obtained for CME1+CME2 (Figure~\ref{fig:euhforia_Dst_scatter}(a)), and for CME3 (Figure~\ref{fig:euhforia_Dst_scatter}(b)).

% ejecta E1 ===================================
\paragraph{Minimum Dst Associated with CME1+CME2}
As already found for the $B_z$ prediction, Dst predictions associated with CME1+CME2 are highly dependent on the simulated CMEs in different runs (Figure~\ref{fig:euhforia_Dst_scatter}(a)). 
% 01-00-00
While CME1 alone (run 01-00-00) predicts a minimum Dst of $-99$~nT ($-125$~nT to $-69$~nT considering spacecraft separated by $\Delta \sigma = 5^\circ$ from Earth, and $-135$~nT to $-16$~nT considering spacecraft separated by $\Delta \sigma = 10^\circ$),
% 01-01-00
the addition of CME2 (run 01-01-00) leads to a predicted minimum Dst of $-119$~nT ($-128$~nT to $>0$~nT considering spacecraft separated by $\Delta \sigma = 5^\circ$ from Earth, and $-134$~nT to $>0$~nT considering spacecraft separated by $\Delta \sigma = 10^\circ$) due to the early interaction between CME1 and CME2.
% 01-01-01
The simulation including all three CMEs (run 01-01-01) finally provides us with an estimate of the contribution of CME--CME interaction processes between CME1+CME2 and CME3 to the geo-effectiveness of CME1+CME2.
In particular, we note that the minimum Dst predicted at Earth is ${\sim}-215$~nT. When accounting for uncertainties related to the initial CME directions of propagation, the predicted value varies between $-227$~nT and $-123$~nT for a $\Delta \sigma = 5^\circ$ separation from Earth, and between $-229$~nT and $-33$~nT for a $\Delta \sigma = 10^\circ$ separation from Earth.
% conclusions
By comparing the minimum Dst prediction from runs 01-01-00 and 01-01-01 at and around Earth, we conclude that the presence of S2 and E2 significantly enhanced the geo-effectiveness of E1 by enhancing the minimum Dst by a factor of 1.8 \citep[ranging between 1.7 and 1.8 for a $\Delta \sigma = 5^\circ$ separation from Earth, and between 1.7 and 1.3 for a $\Delta \sigma = 10^\circ$ separation from Earth; consistent with][]{shen:2018}.
% actual measurements
In terms of absolute values, the minimum Dst recorded on ground was $-142$~nT, while the predicted minimum based on $Wind$ in-situ measurements was $-149$~nT, i.e.\ the two values are very close. Therefore, we observe EUHFORIA run 01-01-01 tends to overestimate the Dst at Earth by a factor of $1.5$ compared to actual observations (1.4 when compared to $Wind$ predictions). The over-prediction is significantly higher than the one observed in $B_z$ ($1.1$ compared to $Wind$ observations), most probably due to the larger density \citep[entering the calculation of Dst via the solar wind dynamic pressure ($P_{dyn}$), see Equation~3 in][]{obrien:2000b} predicted by EUHFORIA (see Appendix~\ref{sec:appendixC}).

% ejecta E2 ===================================
\paragraph{Minimum Dst Associated with CME3}
For completeness, in Figure~\ref{fig:euhforia_Dst_scatter}(b) we plot the predicted minimum Dst at and around Earth associated with CME3 in a similar way as already done for CME1+CME2. Dst estimates reflect what is already found for $B_z$, i.e.\ that the magnetic field strength within CME3 was not significantly altered by the interaction process, as the minimum Dst predicted in runs with (run 00-00-01) and without (run 01-01-01) the preceding CME1+CME2 are consistent with each other and with actual observations and $Wind$-based predictions. We also observe an anti-correlation between the minimum $B_z$ and the minimum Dst predicted in runs 00-00-01 and 01-01-01: while the minimum $B_z$ increases in the immediate surroundings of Earth, the minimum Dst decreases. This result is most probably due to the combination of the higher density and speed associated with CME3 in run 01-01-01 compared to run 00-00-01 \citep[contributions due to dynamic pressure ($P_\mathrm{dyn}$) and the dawn-to-dusk component of the electric field ($V B_s$), see Equations~2 and 3 in][]{obrien:2000b}, and the fact that in run 01-01-01 the Dst dip caused by CME3 started from a condition of highly-disturbed Dst already \citep[see also][]{kamide:1998, vennerstrom:2016}.

% cone model
%For comparison, minimum Dst predictions from cone model CME simulations never go below $-40$~nT (i.e.\ a minor storm).

%------------------------------------------------------------------
\subsection{Implications for Space Weather Events at Other Locations in the Heliosphere}
\label{subsec:discussion}

From Section~\ref{subsec:results_interaction_shock} we concluded that one of the key factors at the origin of the intense storm triggered by the September 4--6, 2017 CMEs was their arrival at Earth during the phase of maximum amplification of the southward $B_z$ (and consequently, of their helio-/geo-effectiveness) due to the interaction of CME1+CME2 with CME3.
Moreover, Figure~\ref{fig:euhforia_evolution}(j) highlighted the existence of a correlation between the evolutionary phase of CME--CME interactions and the amplification of the helio-effectiveness of the leading ejecta involved in the interaction, at least for the specific CME events and the specific (Sun-to-Earth) direction under study. 
 
In general, for two generic CMEs launched in approximately the same direction, the spatial/temporal windows of the various interaction phases depend on three main parameters: (1) the ambient solar wind through which the CMEs (particularly the preceding one) are propagating, (2) the time interval between the eruptions of the individual CMEs involved, and (3) their relative speed. Therefore, we note that different combinations of such parameters will lead the ejecta to reach Earth or any other location in the heliosphere at different evolutionary phases, and hence during different phases of helio-effectiveness amplification. An extensive exploration of the parameter space was performed by \citet{xiong:2007} via 2.5D simulations of interacting CMEs varying the time interval between the eruptions (from 10~to 44~hours), and their relative speeds (from $50$~km~s$^{-1}$ to $800$~km~s$^{-1}$).
By comparing the resulting geo-effectiveness amplification at Earth in the case of interactions that reached 1~AU at different evolutionary stages, they similarly suggested that the evolution stage may be a dominant factor in determining the ultimate geo-effectiveness of interacting CMEs, although a comparison of model results with in-situ observations of real CME events was not presented. 
They also suggested that the exact evolution profile in space--time of the helio-effectiveness amplification may depend on the impact angle between the spacecraft crossing and the CME apex, with spacecraft locations close to the CME nose more likely to maintain the helio-effectiveness amplification due to the persistent push of the trailing ejecta on the leading ejecta (preventing further expansion of the latter), and spacecraft locations close to the CME flanks exhibiting a decay of the helio-effectiveness amplification due to the narrower extension of the trailing magnetic ejecta compared to its driven shock (which induced the compression). An in-depth analysis of such angular dependencies in EUHFORIA simulations is left for future studies.

The existence of a decay phase in the helio-effectiveness amplification starting after the end of the shock--ejecta interaction clearly has strong implications on the impact of CME--CME interactions at various locations in the heliosphere. In particular, each CME--CME interaction event may be associated with a ``helio-effectiveness amplification zone'', corresponding to the heliocentric distances associated with the maximum amplification phase for a given combination of CME waiting times and relative speeds. The amplification of the helio-effectiveness of individual CMEs will be null for spacecraft locations closer to the Sun than the distance at which the interaction starts, it will progressively increase for spacecraft locations between the start and the end of the shock--ejecta interaction (growth phase), 
it will be maximal for spacecraft locations at the outer edge of this distance range (maximum phase), and it will progressively decrease for spacecraft locations farther away from the Sun (decay phase). Although a more extensive study of this impact needs to be addressed in a future work, we speculate that the non-uniform probability distribution of the CME waiting times and relative speeds \citep{wang:2013} may also result in higher probabilities of having helio-effective CME--CME interaction events at specific heliocentric distances than at others.

%__________________________________________________________________

\section{Summary and conclusions}
\label{sec:conclusions}

In this work we have performed a comprehensive Sun-to-Earth analysis of three successive CMEs that erupted from AR~12673 during a remarkably active week in early September 2017, and which resulted in an intense two-step geomagnetic storm (main dip: $\mathrm{Dst}_\mathrm{min} = -142$~nT, secondary dip: $\mathrm{Dst}_\mathrm{min} = -124$~nT) driven by the interplanetary interactions occurring among the CMEs involved. Together with the analysis of the CME-related signatures at the source region, in the corona, and at L1, we have also performed global simulations in the heliosphere using the spheromak CME model in EUHFORIA initialised with observation-based kinematic, geometric, and magnetic parameters for the CMEs.

Remote-sensing observations show that the first two CMEs (CME1 and CME2) were sympathetic events that erupted less than 3~hours apart, with CME2 being faster than CME1 by $\sim$500~km~s$^{-1}$. They interacted already in the upper corona (around $20$~$R_{\odot}$), and they subsequently propagated through the heliosphere as a merged structure. CME3 erupted about 2~days later, with a speed in the corona of $\sim$2000~km~s$^{-1}$, i.e.\ $\sim$500~km~s$^{-1}$ faster than CME2, eventually catching up with the two preceding CMEs in the heliosphere.

Modelling results allowed us to associate the interplanetary shock driven by CME1+CME2 with the shock observed at L1 on September~6 at 23:13~UT (S1), and the CME1+CME2 structure with the magnetic ejecta observed starting on September~7 at 20:00~UT (E1). The interplanetary shock on September~7 at 22:38~UT (S2) was most likely driven by the following CME3, and it was propagating through the preceding CME1+CME2 ejecta. Simulation results also supported the interpretation that both the E2 and E3 observed in situ at L1 were associated with CME3 at the Sun. 

% FR helicity 
By comparing EUV observations of the source region and in-situ observations at L1 we also found that the chirality of the flux ropes in the source AR was consistent with the chirality of the flux rope inferred from in-situ observations at Earth, providing additional support to our linking of structures at the Sun with their interplanetary counterparts. 
% FR tilt 
On the other hand, we found significant rotations between the flux rope orientations at the source region, in the corona, and at 1~AU, which are most probably due to the interaction processes occurring among the three CMEs involved at various stages during their propagation.
Because of the difficulties in constraining the flux rope orientations at 0.1~AU, i.e.\ at the distance of the inner boundary of our heliospheric simulation domain, we tested CME simulations using different orientations, ultimately finding that the initialisation of CMEs as ESW flux rope types at 0.1~AU generated the best predictions at Earth.

% reconnected flux / FR toroidal flux
To initialise the toroidal magnetic flux $\varphi_t$ of the spheromak CMEs in our simulations, we tested a combination of observational methods to determine the reconnected flux $\varphi_r$ associated with each eruption in a more robust way. The results from the application of statistical relations between the main flare and CME properties (such as the flare peak intensity and the CME 3D speed) and different post-eruptive signatures were found to be, on average, compatible with the results from more sophisticated single-event analyses. This result is particularly relevant for space weather forecasting purposes, as it suggests a quick and easy-to-apply method to initialise the magnetic field strength in flux rope CME models that can be potentially applied routinely by forecasters or even via automatic algorithms.

% interaction phases
An analysis of the interaction of CME1+CME2 and CME3 based on 3D simulation results show that the interaction between the shock driven by CME3 and the preceding magnetic ejecta formed by the merging of CME1 and CME2 started around 0.45~AU (on September 6 at 22:00~UT). 
Analysing the impact of the shock--ejecta interaction on the amplification of the $B_z$ magnetic field (i.e.\ \textit{helio-}effectiveness) of CME1+CME2 at various times/heliocentric distances along the Sun--Earth line, we found it could be characterised by a growth phase, a maximum phase, and a decay phase. For the particular event considered, a maximum helio-effectiveness amplification of 2.8 for the minimum $B_z$ was reached near 0.9~AU, i.e.\ close to Earth's location. This amplification phase was also found to be associated with the compression, acceleration, and heating of CME1+CME2 by the shock driven by CME3. The growth and maximum phases were followed by a slow decay at larger heliocentric distances, which was associated with the conversion of magnetic energy into kinetic and thermal energy of CME1+CME2. The helio-effectiveness amplification had almost completely disappeared by the time the merged CME1+CME2+CME3 structure reached 1.8~AU.

% geo-effectiveness amplification at Earth
The simulation results showed that the impact of CME1+CME2 on Earth (\textit{geo-}effectiveness) was amplified by the interaction with CME3 by a factor of 2.5 for the minimum $B_z$, and by a factor of 1.8 for the minimum Dst index, consistent with the recent observational study of this event by \citet{shen:2018}. Moreover, while impacting Earth the system was found to be close to the maximum helio-effectiveness amplification reached at the end of the shock--ejecta interaction phase. We therefore concluded that one of the key factors for causing the event to result in the intense storm on September 7--8, 2017 was the arrival of the CMEs at 1~AU during this evolutionary phase. 
Also, CME3 arrived about 15~hours earlier because of the interaction with the preceding ejecta, i.e.\ the interaction significantly impacted the arrival time prediction for the trailing ejecta. 

% overall performance of EUHFORIA
Overall, the simulation of the three CMEs was able to reproduce the main features in the speed, density and magnetic field observed profiles at Earth/1~AU with a good level of agreement. In terms of CME geo-effectiveness, the model predicted a minimum $B_z$ of $-35$~nT in association with E1, matching the value observed by $Wind$ at L1 (i.e.\ $-32$~nT, in association with E1) remarkably well. The predicted minimum Dst index resulted a factor 1.5 higher than the minimum observed value (1.4 when compared to Dst index predictions based on $Wind$ solar wind measurements), most probably as consequence of an over-estimated dynamic pressure, but still consistent with observations within the error bars given by the virtual spacecraft located in the surroundings of Earth.

% helio-effectiveness amplification zone
Significantly advancing our previous knowledge of CME--CME interactions and their influence on the geo-effectiveness of individual CMEs depending on the interaction phase, this work shows evidence, for the first time, of the space--time evolution of the helio-effectiveness amplification of a real CME event using 3D simulations in a realistic set up.
In general, the exact location in space--time of each of such phases is primarily determined by the time interval between the successive eruptions, and by the relative speed of the individual CMEs involved (in addition to the solar wind conditions ahead of the first CME launched), which ultimately constrain the helicentric distances of the ``helio-effectiveness amplification zone'', i.e.\ corresponding to the maximum helio-effectiveness amplification. This is expected to be maximal for spacecraft/planet locations impacted by the CMEs close to the end of the shock--ejecta interaction phase, and its location will vary depending on the CME waiting times and relative speeds of the specific event considered.

% implications for other spacecraft locations - probabilities
Although a more detailed investigation of this impact needs to be addressed in a future study,
because of the non-uniform probability distribution of the CME waiting times and relative speeds, we speculate that higher probabilities of having helio-effective CME--CME interaction events may be found at given heliocentric distances rather than others, i.e.\ there could be a range of heliocentric distances where the impact of interaction phenomena on the individual CME helio-effectiveness can be expected to be higher than others. 
This may potentially be of great relevance for current and future explorations of new regions of the inner and outer Solar System, e.g.\ from Parker Solar Probe and Solar Orbiter, to the Voyager missions, and for predictions of space weather events at other planets.

%
%______________________________________________________________
% ACKNOWLEDGMENTS

%% If you wish to include an acknowledgments section in your paper,
%% separate it off from the body of the text using the \acknowledgments
%% command.
\acknowledgments
CS and EC acknowledge funding by the Research Foundation -- Flanders (FWO) (grants No 1S42817N and 12M0119N).
MT, KD, and AMV acknowledge funding by the Austrian Space Applications Programme of the Austrian Research Promotion Agency FFG (ASAP-11 4900217 CORDIM, ASAP-13 859729 SWAMI, and ASAP-14 865972 SSCME). 
EKJK acknowledges the Academy of Finland project SMASH 310445. EKJK and JP acknowledge the SolMAG project (ERC-COG 724391) funded by the European Research Council (ERC) in the framework of the Horizon 2020 Research and Innovation Programme, the BRAIN-be project CCSOM. The work of EKJK, EP, and JP has been achieved under the framework of the Finnish Centre of Excellence in Research of Sustainable Space (Academy of Finland grant number 312390).
MD acknowledges support by the Croatian Science Foundation under the project 7549 (MSOC) and EU H2020 grant agreement No 824135 (project SOLARNET). 
JG is supported by the Key Research Program of the Chinese Academy of Sciences (grants No XDPB11 and QYZDB-SSW-DQC015).
LR thanks the European Space Agency (ESA) and the Belgian Federal Science Policy Office (BELSPO) for their support in the framework of the PRODEX Programme. This work was made possible thanks to the Solar--Terrestrial Centre of Excellence, a collaborative framework funded by the Belgian Science Policy Office.
SP was funded by KU Leuven and via the projects C14/19/089 (Internal Funds KU Leuven), G.0A23.16N (FWO) and C~90347 (ESA PRODEX).
EUHFORIA is developed as a joint effort between  the  University  of  Helsinki  and  KU Leuven. The full validation of the solar wind and CME modelling is being performed within the CCSOM project (\url{http://www.sidc.be/ccsom/}). The EUHFORIA website, including a repository for simulation results and the possibility for users to access the code, is currently under construction; please contact the authors for full 3D simulation outputs and model details. The simulations were carried out at the VSC -- Flemish Supercomputer Center, funded by the  Hercules foundation and the Flemish Government -- Department EWI.

%
%______________________________________________________________
% FACILITIES

%% To help institutions obtain information on the effectiveness of their 
%% telescopes the AAS Journals has created a group of keywords for telescope 
%% facilities.
%
%% Following the acknowledgments section, use the following syntax and the
%% \facility{} or \facilities{} macros to list the keywords of facilities used 
%% in the research for the paper.  Each keyword is check against the master 
%% list during copy editing.  Individual instruments can be provided in 
%% parentheses, after the keyword, but they are not verified.

\vspace{10mm}
\facilities{SDO (AIA, HMI); SOHO (LASCO); STEREO (SECCHI); $Wind$ (MFI, SWE); DSCOVR (PlasMag).}

%% Similar to \facility{}, there is the optional \software command to allow 
%% authors a place to specify which programs were used during the creation of 
%% the manusscript. Authors should list each code and include either a
%% citation or url to the code inside ()s when available.

\software{
          EUHFORIA \citep{pomoell:2018}; 
          Python SunPy \citep{sunpy:2015};
          IDL SolarSoft \citep{freeland:1998}; 
          ESA JHelioviewer \citep{muller:2017}.
          }

%
%______________________________________________________________
% APPENDIX

%% Appendix material should be preceded with a single \appendix command.
%% There should be a \section command for each appendix. Mark appendix
%% subsections with the same markup you use in the main body of the paper.

%% Each Appendix (indicated with \section) will be lettered A, B, C, etc.
%% The equation counter will reset when it encounters the \appendix
%% command and will number appendix equations (A1), (A2), etc. The
%% Figure and Table counter will not reset.

\appendix

\section{Graduated Cylindrical Shell analysis}
\label{sec:appendixA}

\begin{figure*}
\centering
{\includegraphics[width=0.99\textwidth]{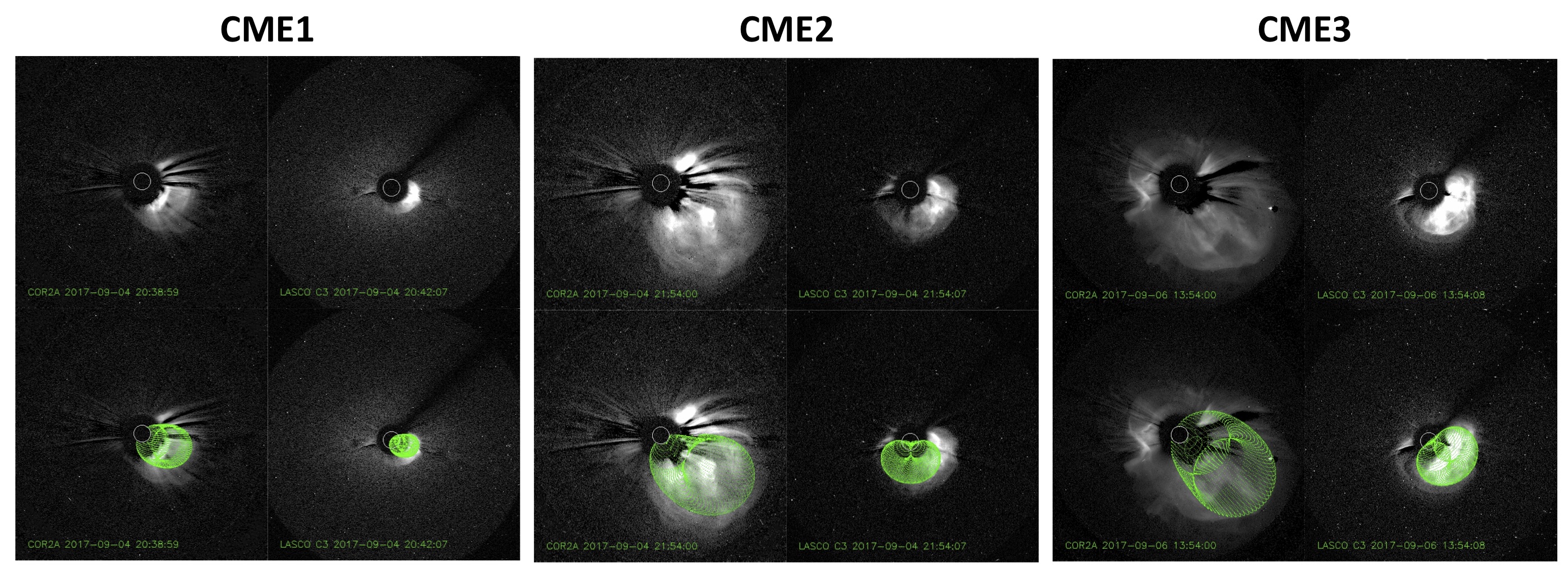}}
\caption{LASCO/C3 and STEREO/COR2-A pre-event background-subtracted intensity images of CME1 (left; on September 4, 2017 around 20:39~UT), CME2 (middle; on September 4, 2017 at 21:54~UT) and CME3 (right; on September 6, 2017 at 13:54~UT), with and without the GCS model wireframe (in green).}
\label{fig:gcs} 
\end{figure*}

\begin{figure}
\centering
{\includegraphics[width=0.7\textwidth]{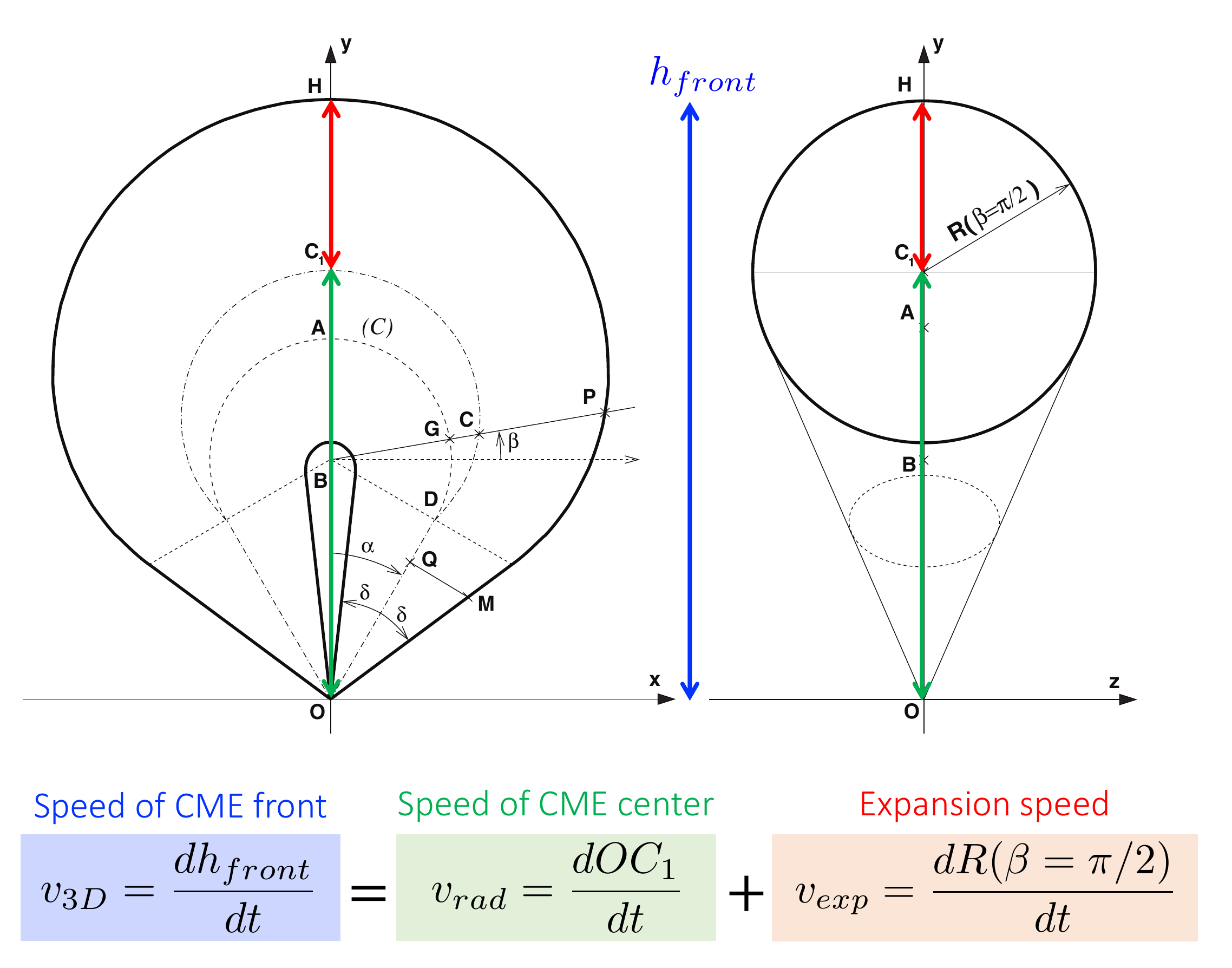}}\\
\caption{
Schematic of the GCS model, adapted from \citet{scolini:2019} and \citet{thernisien:2011}: 
face-on (left) and edge-on (right) representations. In the case $\alpha = 0$, the face-on and edge-on views coincide.
The blue double arrow marks the height of the CME front, the green double arrow marks the height of the CME centre, and the red double arrow marks the CME radius.
Their variation in time can be used to estimate the total ($v_\mathrm{3D} = v_\mathrm{rad} + v_\mathrm{exp}$), 
radial ($v_\mathrm{rad}$) and expansion speed ($v_\mathrm{exp}$), as described by the relations in the coloured boxes.}
\label{fig:gcs_sketch} 
\end{figure}

In the following we describe the method used to recover the radial and expansion speed from the GCS fitting of the CMEs under study (shown in Figure~\ref{fig:gcs}), first described by \citet{scolini:2019}. Using the same notation as \citet{thernisien:2011}, the heliocentric distance of the CME front at its apex, $h_\mathrm{front}$, is defined as 
\begin{equation}
h_\mathrm{front} = OH = \frac{b+\rho}{1-\kappa},
\label{eqn:gcs_hgt}
\end{equation}
where $b = OB $ and $\rho = BD $.
At the same time, from geometrical considerations we observed that 
\begin{equation}
OH =  {OC_1} +  R(\beta=\pi/2),
\end{equation}
and that the total speed of the CME apex $v_\mathrm{3D}$ can be related to the variation over time of the parameter $h_\mathrm{front}$, 
the expansion speed $v_\mathrm{exp}$ can be related to the variation in time of $R(\beta=\pi/2)$, 
and the radial speed $v_\mathrm{rad}$ can be related to that of $ OC_1$ (Figure~\ref{fig:gcs_sketch}).
Therefore, the radial and expansion speed can be calculated based on the standard GCS output parameters as:
\begin{equation}
\begin{cases} 
v_\mathrm{rad} = \frac{d OC_1}{d t}  \\
v_\mathrm{exp} = \frac{d R(\beta=\pi/2)}{d t}.
\end{cases}
\end{equation}
The heliocentric distance of the apex centre, $OC_1$, and the cross section radius of the apex, $R(\beta=\pi/2)$,
are in turn related to the leading edge height $h_\mathrm{front}$ by the following relations \citep{thernisien:2011}:
\begin{equation}
\begin{cases} 
OC_1 			= \frac{b+\rho}{1-\kappa^2} =  \frac{1}{1+\kappa} \, h_\mathrm{front} \\
R(\beta=\pi/2) = \frac{b+\rho}{1-\kappa^2}  \,  \kappa  =  \frac{\kappa}{1+\kappa}  \, h_\mathrm{front}.
\end{cases}
\end{equation}
so that $OC_1 + R(\beta = \pi/2) = OH$ (as shown in Figure~\ref{fig:gcs_sketch}).
Combining these results and remembering that all the GCS parameters are in principle time-dependent, one obtains
\begin{align}
v_\mathrm{rad} &= \frac{d}{d t} \left( \frac{h_\mathrm{front}}{1+\kappa} \right) 
                = \frac{1}{1+\kappa} \frac{d h_\mathrm{front}}{dt} 
                - h_\mathrm{front} \frac{1}{(1+\kappa)^2} \frac{d \kappa}{dt},
\label{eqn:gcs_vrad_full}
\end{align}
and 
\begin{align}
v_\mathrm{exp} &= \frac{d}{d t} \left( \frac{\kappa}{1+\kappa}h_\mathrm{front} \right) = \nonumber \\
        &= \frac{\kappa}{1+\kappa} \frac{d h_\mathrm{front}}{dt} 
        + h_\mathrm{front} \, \left( \frac{1}{1+\kappa} - \frac{\kappa}{(1+\kappa)^2} \right) \frac{d \kappa}{dt}.
\label{eqn:gcs_vexp_full}
\end{align}
For CMEs where $\kappa$ can be kept fixed in time, the above equations simplify to 
\begin{equation}
\boxed{v_\mathrm{rad} = \frac{1}{1+\kappa} \frac{d h_\mathrm{front}}{dt},}
\label{eqn:gcs_vrad_full_02}
\end{equation}
and 
\begin{equation}
\boxed{v_\mathrm{exp} = \frac{\kappa}{1+\kappa} \frac{d h_\mathrm{front}}{dt}.}
\label{eqn:gcs_vexp_full_02}
\end{equation}

% ==================================================================================================
\section{EUHFORIA results on the ecliptic and meridional planes}
\label{sec:appendixB}

\begin{sidewaysfigure}
\centering
         \includegraphics[width=0.98\hsize, trim={0mm 0mm 0mm 0mm},clip]{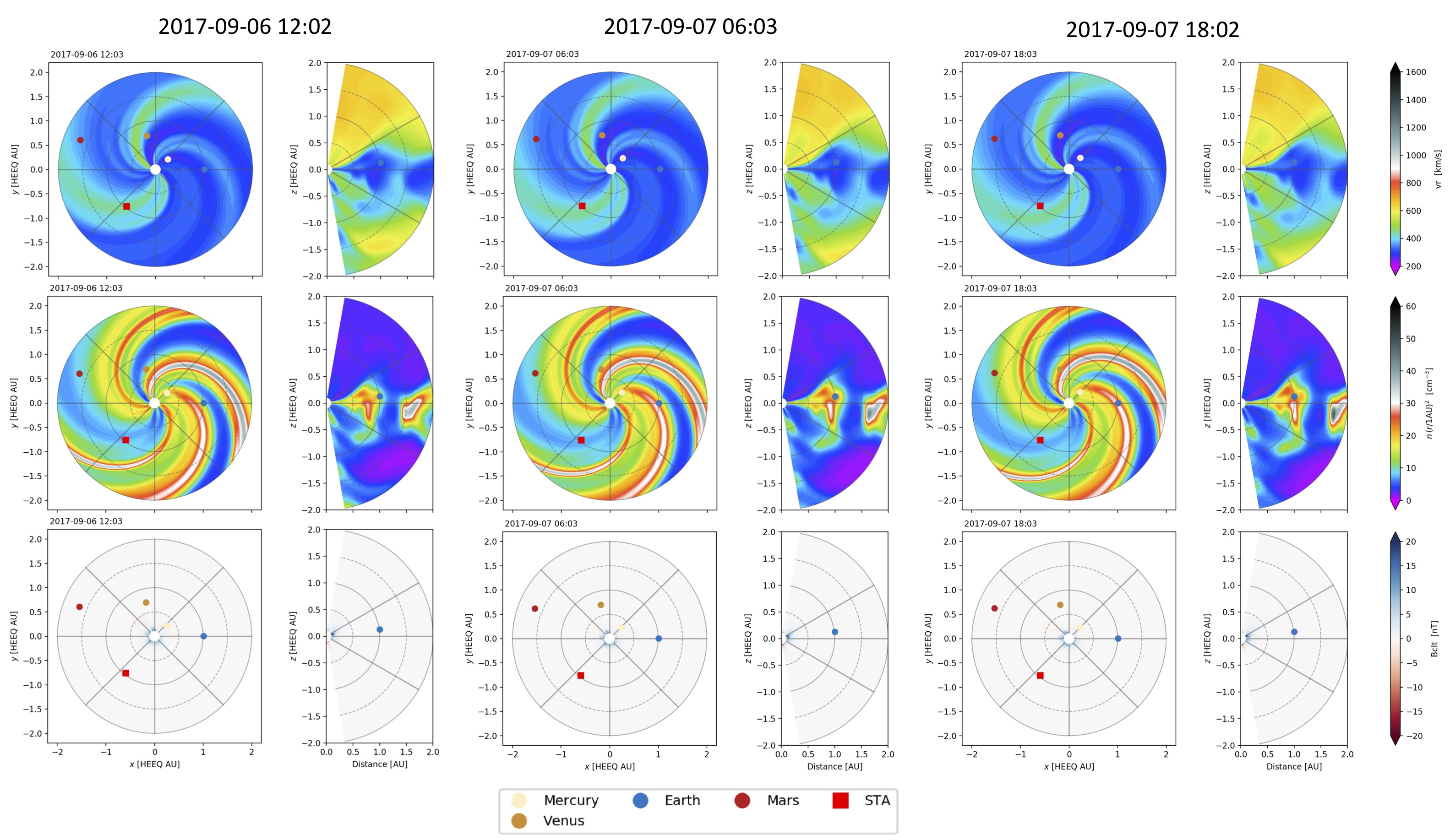} 
\caption{Snapshots from EUHFORIA run 00-00-00 showing the radial speed ($v_r$, top row), scaled number density ($n (D/1~\mathrm{AU})^2$, middle row), and co-latitudinal magnetic field ($B_\mathrm{clt}$, bottom row) in the heliographic equatorial plane and in the meridional plane that includes Earth (which is indicated by solid blue circles). 
Left column: September 6, 2017 at 12:02~UT. 
Middle column: September 7, 2017 at 06:03~UT.
Right column: September 7, 2017 at 18:02~UT. }
\label{fig:euhforia_000000}
\end{sidewaysfigure}

\begin{sidewaysfigure}
\centering
         \includegraphics[width=0.98\hsize, trim={0mm 0mm 0mm 0mm},clip]{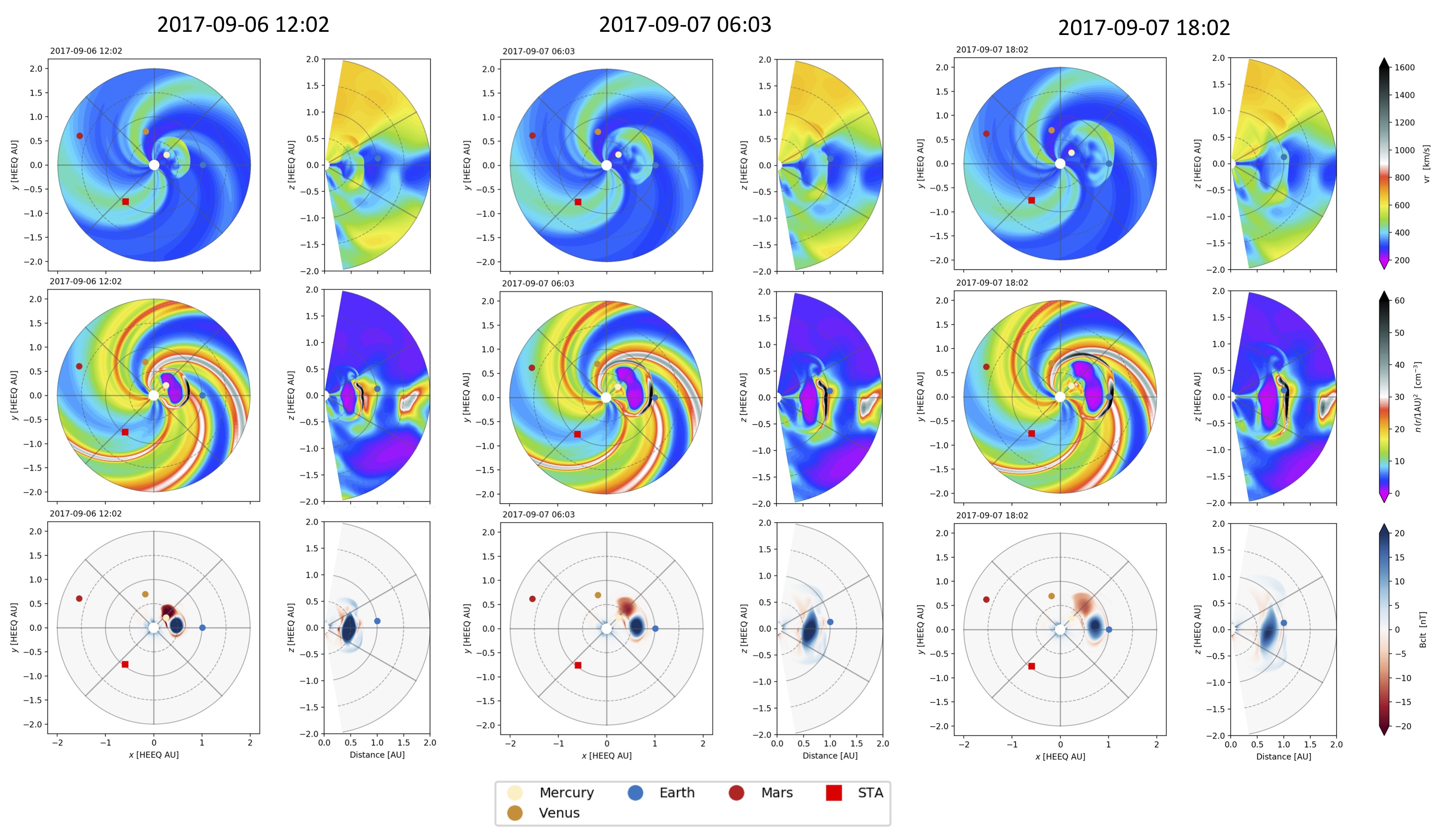} 
\caption{Snapshots from EUHFORIA run 01-00-00 showing the radial speed ($v_r$, top row), scaled number density ($n (D/1~\mathrm{AU})^2$, middle row), and co-latitudinal magnetic field ($B_\mathrm{clt}$, bottom row) in the heliographic equatorial plane and in the meridional plane that includes Earth (which is indicated by solid blue circles). 
Left column: September 6, 2017 at 12:02~UT. 
Middle column: September 7, 2017 at 06:03~UT.
Right column: September 7, 2017 at 18:02~UT. }
\label{fig:euhforia_010000}
\end{sidewaysfigure}

\begin{sidewaysfigure}
\centering
         \includegraphics[width=0.98\hsize, trim={0mm 0mm 0mm 0mm},clip]{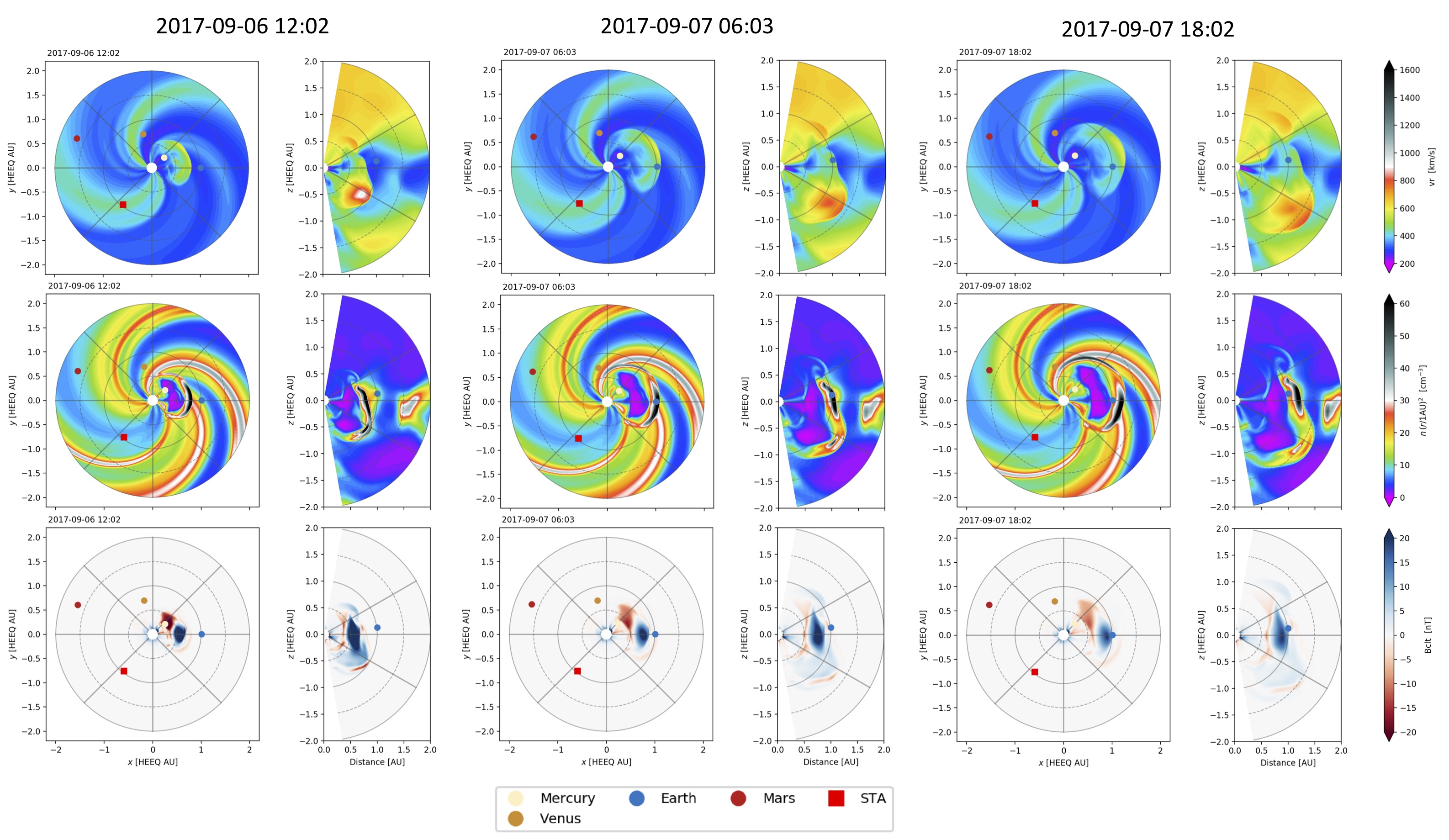} 
\caption{Snapshots from EUHFORIA run 01-01-00 showing the radial speed ($v_r$, top row), scaled number density ($n (D/1~\mathrm{AU})^2$, middle row), and co-latitudinal magnetic field ($B_\mathrm{clt}$, bottom row) in the heliographic equatorial plane and in the meridional plane that includes Earth (which is indicated by solid blue circles). 
Left column: September 6, 2017 at 12:02~UT. 
Middle column: September 7, 2017 at 06:03~UT.
Right column: September 7, 2017 at 18:02~UT. }
\label{fig:euhforia_010100}
\end{sidewaysfigure}

\begin{sidewaysfigure}
\centering
         \includegraphics[width=0.98\hsize, trim={0mm 0mm 0mm 0mm},clip]{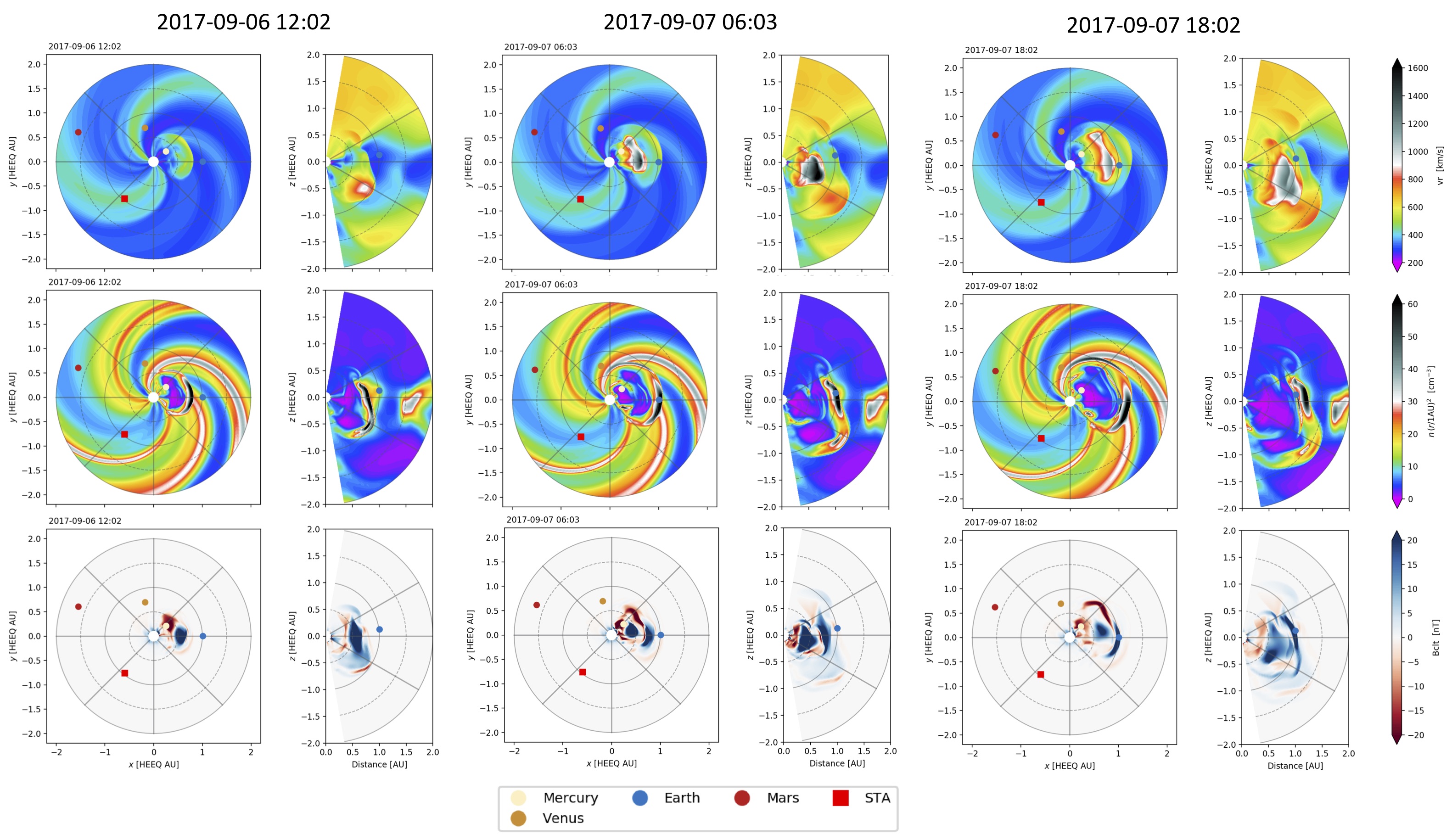} 
\caption{Snapshots from EUHFORIA run 01-01-01 showing the radial speed ($v_r$, top row), scaled number density ($n (D/1~\mathrm{AU})^2$, middle row), and co-latitudinal magnetic field ($B_\mathrm{clt}$, bottom row) in the heliographic equatorial plane and in the meridional plane that includes Earth (which is indicated by solid blue circles). 
Left column: September 6, 2017 at 12:02~UT. 
Middle column: September 7, 2017 at 06:03~UT.
Right column: September 7, 2017 at 18:02~UT. }
\label{fig:euhforia_010101}
\end{sidewaysfigure}

\begin{sidewaysfigure}
\centering
         \includegraphics[width=0.98\hsize, trim={0mm 0mm 0mm 0mm},clip]{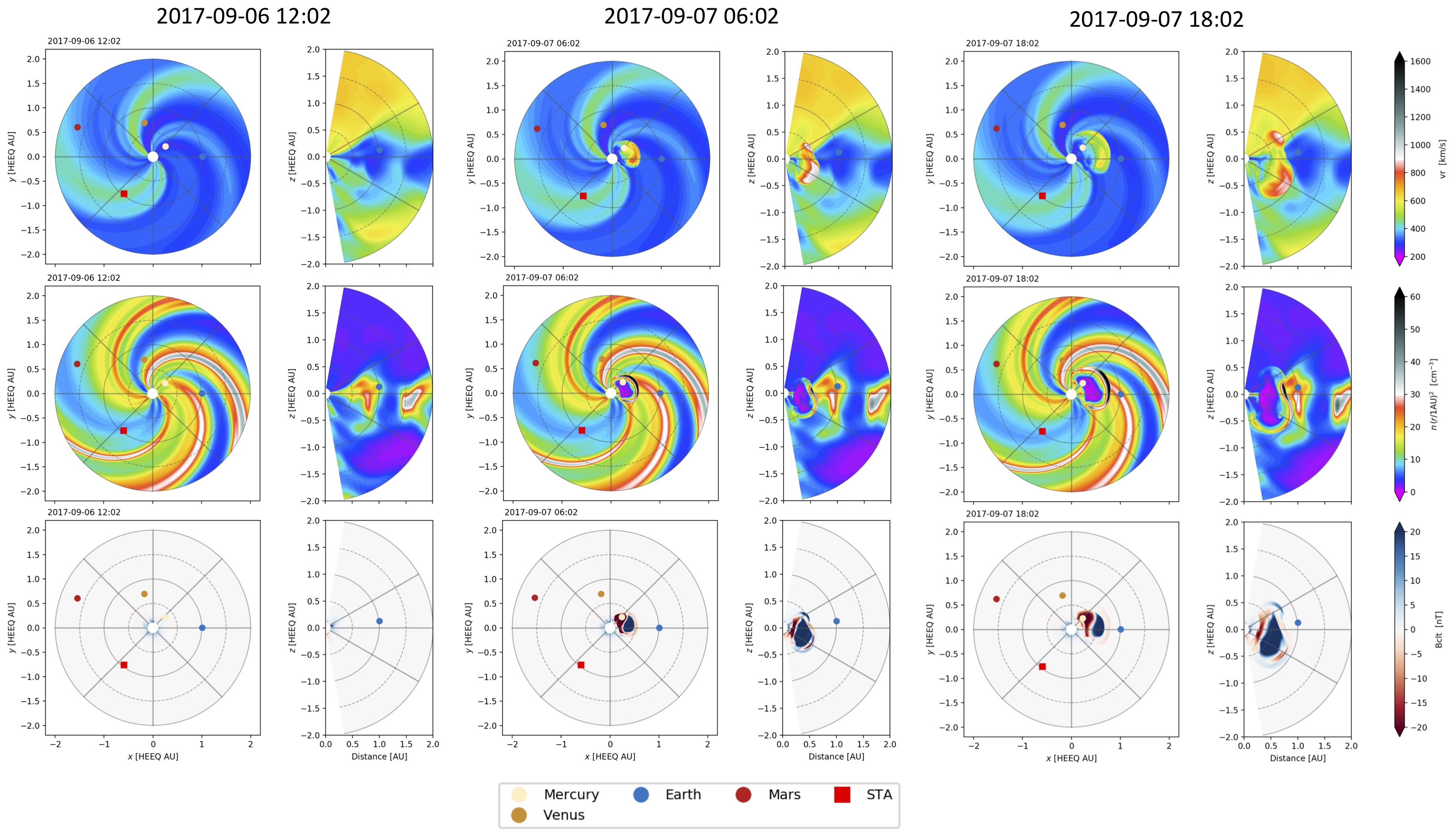} 
\caption{Snapshots from EUHFORIA run 00-00-01 showing the radial speed ($v_r$, top row), scaled number density ($n (D/1~\mathrm{AU})^2$, middle row), and co-latitudinal magnetic field ($B_\mathrm{clt}$, bottom row) in the heliographic equatorial plane and in the meridional plane that includes Earth (which is indicated by solid blue circles). 
Left column: September 6, 2017 at 12:02~UT. 
Middle column: September 7, 2017 at 06:03~UT.
Right column: September 7, 2017 at 18:02~UT. }
\label{fig:euhforia_000001}
\end{sidewaysfigure}

Figures~\ref{fig:euhforia_000000},  \ref{fig:euhforia_010000}, \ref{fig:euhforia_010100}, \ref{fig:euhforia_010101}, and \ref{fig:euhforia_000001} show the results of EUHFORIA simulations on the ecliptic plane and on the meridional plane containing Earth, for each of the 5 runs performed (Table~\ref{tab:euhforia_runs}).

% ==================================================================================================

\section{EUHFORIA predictions at Earth}
\label{sec:appendixC}

\begin{sidewaysfigure}
\centering
         \includegraphics[width=0.49\hsize, trim={0mm 0mm 0mm 0mm},clip]{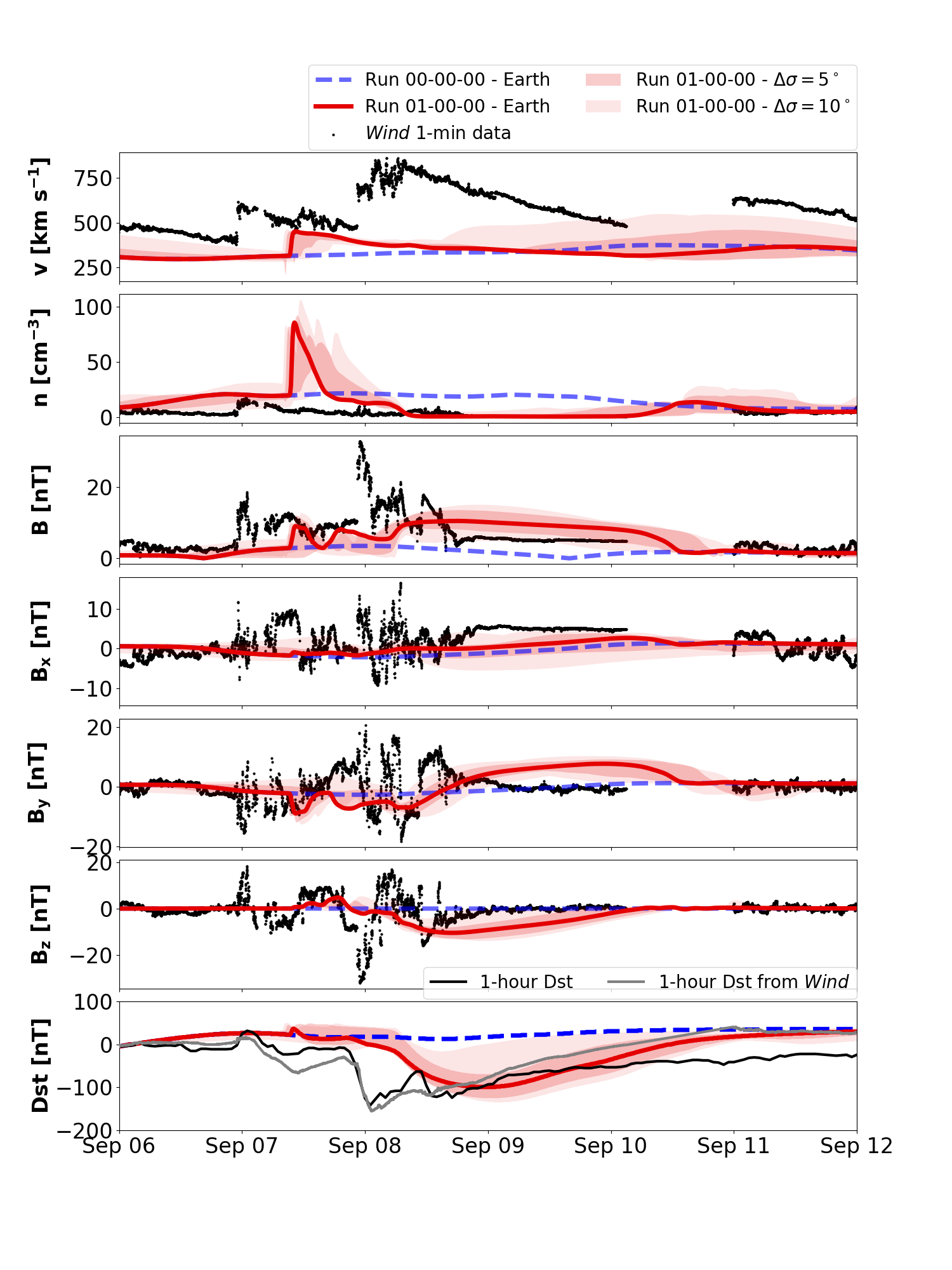} 
         \includegraphics[width=0.49\hsize, trim={0mm 0mm 0mm 0mm},clip]{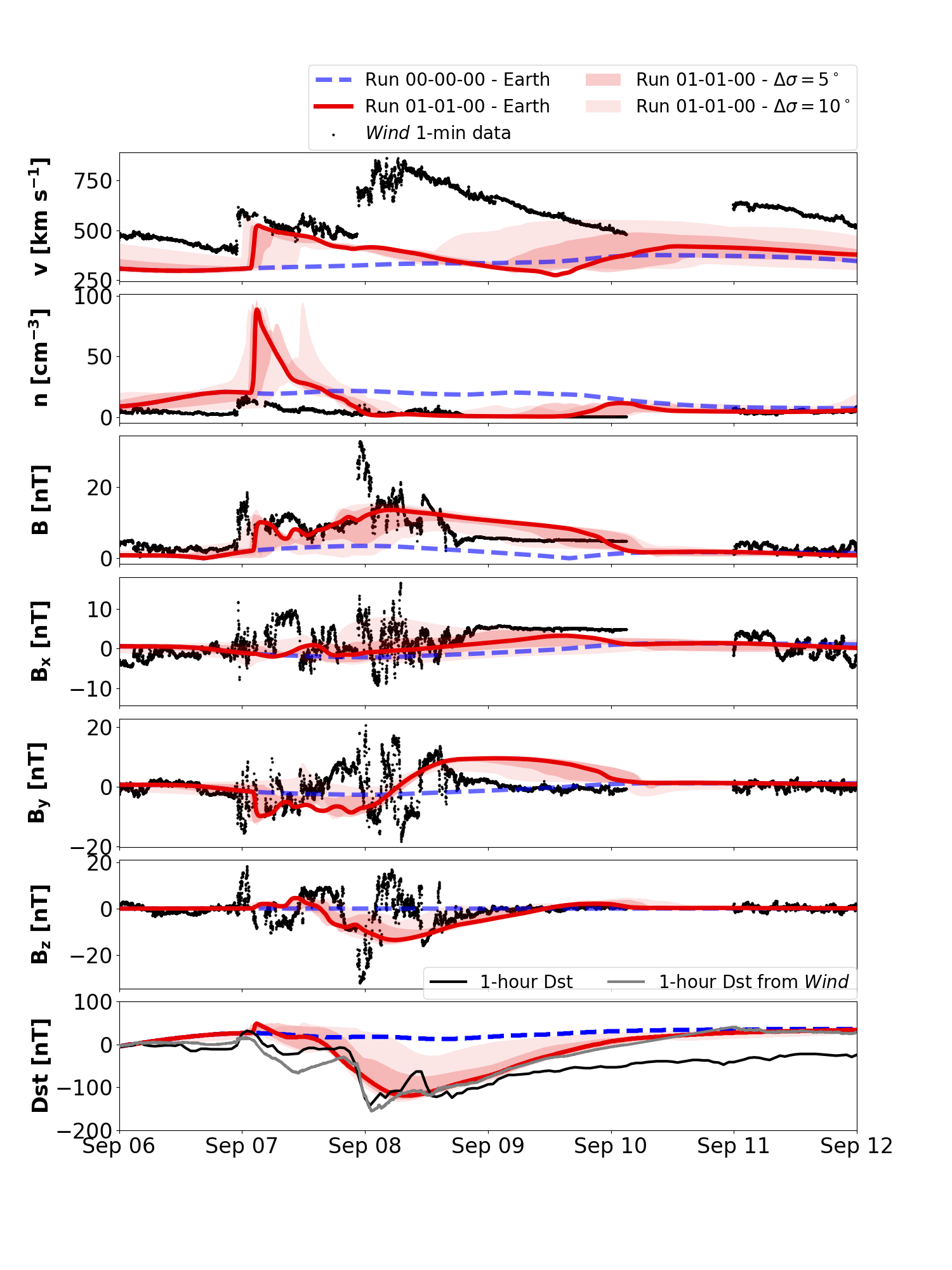}
\caption{Comparison of EUHFORIA time series (red and blue) with in-situ measurements from $Wind$ (black) for the whole temporal computational domain (both in GSE coordinates). The bottom panel shows a comparison of the measured Dst index (black) with predictions obtained from solar wind measurements at $Wind$ (grey), and from EUHFORIA time series after conversion into GSM coordinates (red and blue).
}
\label{fig:euhforia_Earth_a}
\end{sidewaysfigure}

\begin{sidewaysfigure}
\centering
         \includegraphics[width=0.49\hsize, trim={0mm 0mm 0mm 0mm},clip]{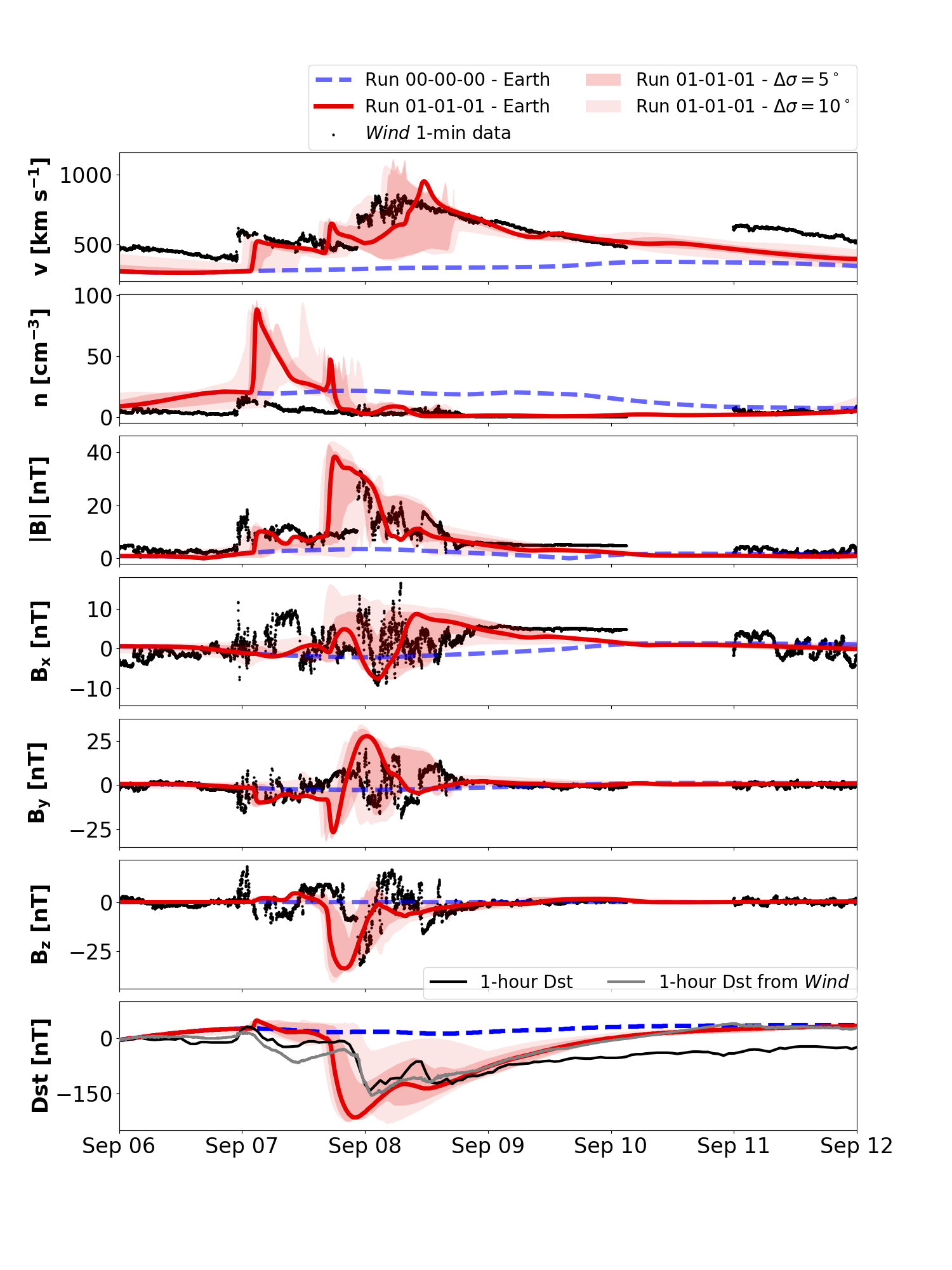} 
         \includegraphics[width=0.49\hsize, trim={0mm 0mm 0mm 0mm},clip]{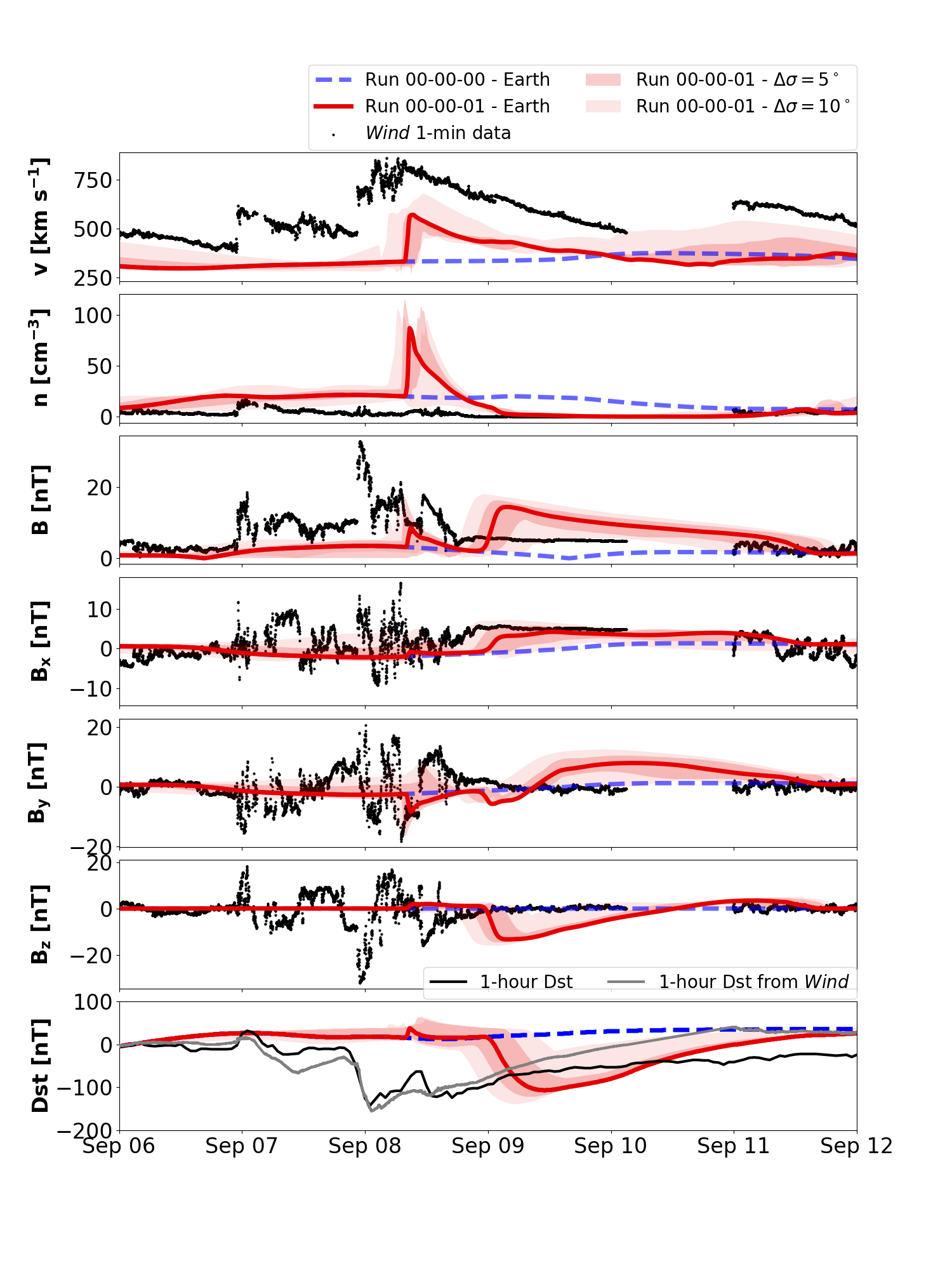} 
\caption{Comparison of EUHFORIA time series (red and blue) with in-situ measurements from $Wind$ (black) for the whole temporal computational domain (both in GSE coordinates). The bottom panel shows a comparison of the measured Dst index (black) with predictions obtained from solar wind measurements at $Wind$ (grey), and from EUHFORIA time series after conversion into GSM coordinates (red and blue).
}
\label{fig:euhforia_Earth_b}
\end{sidewaysfigure}

Figures~\ref{fig:euhforia_Earth_a} and \ref{fig:euhforia_Earth_b} show the time series extracted from EUHFORIA at Earth and surrounding virtual spacecraft, together with the predicted Dst index, for each of the 5 runs performed (Table~\ref{tab:euhforia_runs}).
$Wind$ in-situ measurements at L1 and Dst measurements at Earth are included for comparison.

%
%______________________________________________________________
% BIBLIOGRAPHY
%\bibliographystyle{aasjournal} % style aasjournal.bst
%\bibliography{Bibliography} % your references Yourfile.bib

%\begin{thebibliography}{}

%\end{thebibliography}

\end{document}